\documentclass[twocolumn,tighten,times,trackchanges,fleqn]{aastex701}

\usepackage{graphicx}
\usepackage{amsmath}
\usepackage{amssymb}
\usepackage{natbib}
\usepackage{color}
\usepackage{enumitem}

\begin{document}

\title{Global Disc Simulations of the Magnetorotational Instability using Smoothed Particle Magnetohydrodynamics}

\author[0009-0004-0131-6370,sname=Narvaez Coral,gname=Jacksen]{Jacksen Narvaez Coral}
\affiliation{Memorial University of Newfoundland, Department of Computer Science, St. John’s, NL A1B 3X5, Canada}
\affiliation{Memorial University of Newfoundland, Department of Mathematics and Statistics, St. John's, NL A1C 5S7, Canada}
\email{jjnarvaezcor@mun.ca}

\author[0000-0002-6238-9096]{Terrence S. Tricco}
\affiliation{Memorial University of Newfoundland, Department of Computer Science, St. John’s, NL A1B 3X5, Canada}
\affiliation{Memorial University of Newfoundland, Department of Physics and Physical Oceanography, St. John’s, NL A1B 3X7, Canada}
\email{tstricco@mun.ca}

\author[0009-0006-5777-3989, sname=B. Lapeyre, gname=Yona]
{Yona B. Lapeyre}
\affiliation{ENS de Lyon, Université Lyon 1,  CNRS, CRAL, UMR 5574, Lyon, France}
\email{yona.lapeyre@ens-lyon.fr}

\begin{abstract}

The magnetorotational instability (MRI) has been investigated extensively with grid-based simulations, in both local shearing-box and global models, but has received little attention with particle-based methods. The few studies of the MRI with smoothed particle magnetohydrodynamics (SPMHD) have used local models, which have successfully reproduced the instability, but global disc models remain largely unexplored. We perform three-dimensional ideal MHD simulations of global accretion discs with the \textsc{Phantom} and \textsc{Shamrock} SPMHD codes with resolutions up to 100M particles. We start with a purely toroidal magnetic field with plasma $\beta=25$. Our highest resolution calculation reproduces the main signatures of MRI-driven turbulence. The mean toroidal magnetic field undergoes periodic reversals producing the characteristic butterfly diagram, the normalized Maxwell stress is $\alpha_{\rm M} \sim$ 0.38, and angular momentum transport is dominated by the Maxwell stress, which is 4--6 times higher than the Reynolds stress. Turbulence is not sustained over long timescales due to the effective magnetic Reynolds number ($R_\mathrm{m}\sim1000$). We find that using a geometric density weighting in the gradient discretizations in the magnetic solver avoids excessive dissipation in the presence of steep density gradients. Our resolution study suggests SPMHD can model MRI-driven turbulence in magnetized accretion discs, provided the numerical dissipation is low and the MRI quality factor is high enough (occurring for approximately $\gtrsim$20M particles).

\end{abstract}


\section{Introduction}\label{sec:Introduction}

In sufficiently ionized regions of accretion discs, the magnetorotational instability (MRI) is a leading mechanism for driving turbulence \citep{balbus_powerful_1991, hawley_powerful_1991, balbus_instability_1998}. This turbulence acts as an effective viscosity that transports angular momentum, setting the rate of mass accretion onto the central object \citep{shakura_black_1973}. The efficiency of the MRI, however, depends strongly on the environment. In protoplanetary discs, where the gas can be very weakly ionized, non-ideal effects suppress the instability in extended dead zones \citep[][see \citealt{lesur_magnetohydrodynamics_2021} for a review]{gammie_layered_1996}. In AGN discs, by contrast, the gas is expected to stay well coupled to the magnetic field, so non-ideal effects are less important. This good coupling may extend even to the colder outer parts if X-rays and cosmic rays provide enough ionization, although their contribution is still uncertain \citep{menou_ionization_2001}.

The MRI has been studied extensively with numerical simulations, both in local shearing-box models \citep{brandenburg_dynamo-generated_1995, hawley_local_1995, stone_three-dimensional_1996, fleming_effect_2000, miller_formation_2000, guan_locality_2009, davis_sustained_2010, bai_local_2013, salvesen_accretion_2016, deng_local_2019, wissing_magnetorotational_2022, suzuki_mhd_2023, reboul-salze_mri-driven_2025, squire_rapid_2025, jannaud_numerical_2026} and in global models \citep{hawley_global_2000, arlt_r_global_2001, hawley_global_2001, fromang_s_global_2006, flock_turbulence_2011, flock_large-scale_2012, sorathia_global_2012, parkin_equilibrium_2013, flock_radiation_2017, cui_turbulence_2022, rodman_evolution_2024, roberts_global_2025}. Local models have been essential for understanding the instability. By simulating only a small patch, they can resolve much smaller scales. Their diagnostics, such as the saturation predictors, are also similar to those of global simulations at comparable resolution \citep{sorathia_global_2012, simon_emergent_2012}. However, local models cannot capture the full dynamics of an accretion disc. The limited size and periodic boundaries of the shearing box restrict the largest scales the turbulence can reach. In addition, processes that depend on the large-scale structure of the disc, such as outflows and the transport of magnetic flux, can be studied reliably only in global models \citep{regev_viability_2008, fromang_local_2013, bai_local_2013, zhu_global_2018}.

Most numerical studies of the MRI have used Eulerian grid-based methods, with only a few based on Lagrangian meshless schemes \citep{deng_local_2019, wissing_magnetorotational_2022}. Each approach introduces its own errors and challenges, so agreement between different methods on the same system increases reliability of the results. It is therefore important to simulate the MRI with meshless schemes as well.

Among these, one of the most widely used is smoothed particle hydrodynamics (SPH; \citealt{lucy_numerical_1977, gingold_smoothed_1977}). It offers excellent conservation properties, Galilean invariance, flexibility with complex geometries, and resolution that naturally follows the mass. These advantages have made SPH popular across cosmology and astrophysics. One of its most common applications is accretion discs, where angular momentum transport is usually modelled by calibrating the artificial viscosity to an $\alpha$-viscosity prescription \citep{lodato_diffusive_2010}.

Smoothed particle magnetohydrodynamics (SPMHD; \citealt{price_smoothed_2004a, price_smoothed_2004b}; see \citealt{price_smoothed_2012} and \citealt{tricco_smoothed_2023} for reviews), the magnetic extension of SPH, inherits its advantages and applies them to magnetized flows. SPMHD has since matured considerably, with improved schemes that reduce excess numerical dissipation \citep{cullen_inviscid_2010, price_phantom_2018}, the effects of the tensile instability \citep{borve_regularized_2001}, and divergence errors in the magnetic field \citep{tricco_constrained_2012, tricco_constrained_2016}. As a result, it has been applied to a wide range of problems, including star formation \citep{wurster_can_2016, wurster_non-ideal_2020a, wurster_non-ideal_2020b}, the evolution of galactic magnetic fields \citep{dobbs_magnetic_2016, steinwandel_origin_2020}, dynamo amplification \citep{tricco_comparison_2016, wissing_numerical_2023}, and tidal disruption events \citep{bonnerot_magnetic_2017}.

One application that still requires further exploration, however, is MRI-driven turbulence in accretion discs and its associated dynamo. Previous studies have mostly been restricted to local models. In 2D unstratified shearing sheets, SPMHD has recovered the linear growth rate of the MRI and a structure consistent with grid-based results \citep{gaburov_astrophysical_2011, vanaverbeke_gradspmhd_2014, tricco_simulating_2015, hopkins_accurate_2016}. \footnote{\citet{gaburov_astrophysical_2011} used a meshless weighted particle MHD scheme.} \citet{gaburov_astrophysical_2011} and \citet{vanaverbeke_gradspmhd_2014} also presented 3D global disc simulations, but these lacked the resolution to capture the characteristic MRI wavelength. \citet{forgan_fragmentation_2016} performed MHD simulations of self-gravitating global discs, although the MRI remained unresolved. \citet{deng_local_2019} studied unstratified and vertically stratified 3D shearing boxes with meshless finite-mass (MFM) and SPMHD. In the unstratified case with a net vertical flux, both methods maintained MRI-driven turbulence. With zero net flux, tested only with MFM, turbulence decayed within about 40 orbits, most likely because numerical resistivity was too high at the resolutions used. In the stratified case, SPMHD failed to sustain turbulence and instead built up strong, unphysical toroidal magnetic fields. Using a geometric density average force formulation (GDSPH; \citealt{wadsley_gasoline2_2017, wissing_smoothed_2020}), \citet{wissing_magnetorotational_2022} sustained MRI-driven turbulence for at least 100 orbits in a 3D stratified shearing box. They used resolutions in the range $(1.6$--$12.8) \times 10^6$ particles. In the unstratified case with zero net flux, turbulence required a magnetic Prandtl number above a critical value of $\sim 2.5$, in good agreement with the results from grid-based codes.

Although SPMHD has been shown to capture the MRI in local models, its ability to do so in global discs remains largely untested. In this paper, we assess the suitability of SPMHD for simulating the MRI in global accretion discs. To this end, we perform three-dimensional ideal MHD simulations of locally isothermal discs over a wide range of resolutions and dissipation parameters.

The paper is organized as follows. Section~\ref{sec:Methods} describes the numerical method, including the continuum governing equations and their SPMHD discretization, the initial conditions, codes, and the diagnostics used in our analysis. Section~\ref{sec:Results} presents results on the evolution of the disc, including the dependence on resolution and artificial resistivity. Section~\ref{sec:Conclusions} summarizes the study and states our conclusions.

\section{Methods}\label{sec:Methods}

This section describes the numerical method and the physical model adopted in our disc simulations. Section~\ref{subsec:MHD_Equations} summarizes the MHD equations in continuum form, and Section~\ref{subsec:_initial_conditions} sets out the disc model and initial conditions. The SPMHD discretization is presented in Section~\ref{subsec:SPMHD}, followed by the simulation suite (Section~\ref{subsec:simulations}), the code units (Section~\ref{subsec:code_units}), and a resolution criterion based on the quality factor (Section~\ref{subsec:resolution}). Finally, we define the diagnostics used in the analysis: averages (Section~\ref{subsec:averages}), dissipation parameters (Section~\ref{subsec:dissipation}), stresses (Section~\ref{subsec:stresses}), and the azimuthal spectrum (Section~\ref{subsec:spectrum}).

Throughout this paper, $\{R, z, \phi\}$ denote the cylindrical radius, height, and azimuthal angle, respectively, with the coordinate system centred at the origin. We also use the spherical radius $r=\sqrt{R^2+z^2}$.

\subsection{Ideal MHD equations}\label{subsec:MHD_Equations}

The set of ideal MHD equations that we solve is given by
\begin{gather}
    \frac{{\rm d} \rho}{{\rm d} t} = -\rho \nabla \cdot \mathbf{v}, \label{MHDEqs:mass}\\
    \frac{{\rm d}\mathbf{v}}{{\rm d} t} = \frac{1}{\rho}\nabla \cdot \mathbf{S} - \nabla\Phi, \label{MHDEqs:momentum}\\
    \frac{{\rm d}}{{\rm d} t}\left(\frac{\mathbf{B}}{\rho}\right) = \frac{1}{\rho}\left(\mathbf{B}\cdot \nabla\right)\mathbf{v}, \label{MHDEqs:induction}\\
    \nabla \cdot \mathbf{B} = 0, \label{MHDEqs:divB}
\end{gather}
which are the continuity, momentum, and induction equations, and the divergence-free constraint, respectively. Here, ${\rm d}/ {\rm d}t = \partial/\partial t + \mathbf{v}\cdot \nabla$ is the material time derivative, $\rho$ is the gas density, $\mathbf{v}$ is the velocity, $\mathbf{B}$ is the magnetic field, $\Phi$ is the gravitational potential, and $\mathbf{S}$ is the MHD stress tensor, with components
\begin{equation}\label{MHDEqs:stress_tensor}
    S^{ij} = -\left(P+\frac{1}{2\mu_0}B^2\right)\delta^{ij}+\frac{1}{\mu_0}B^i B^j,
\end{equation}
where $P$ is the gas pressure, $B = |\mathbf{B}|$, $\mu_0$ is the permeability of free space, and $\delta^{ij}$ is the Kronecker delta. The self-gravity of the disc is neglected, so that $\Phi$ corresponds only to the potential of the central point mass.

The system of equations \eqref{MHDEqs:mass}--\eqref{MHDEqs:divB} is closed by a locally isothermal equation of state,
\begin{equation}\label{MHDEqs:IsoEOS}
    P = c_\mathrm{s}^2(r) \rho,
\end{equation}
where $c_\mathrm{s}$ is the speed of sound.

\subsection{Initial conditions} \label{subsec:_initial_conditions}

The disc model follows radial power-law profiles in surface density and sound speed, with all quantities in code units (Section~\ref{subsec:code_units}). The initial particle distribution is generated by Monte Carlo sampling between $R_\mathrm{in} = 1$ and $R_\mathrm{out} = 10$, following the surface density profile
\begin{equation}\label{ICs:density}
    \Sigma(R) = \Sigma_0 \left(1-\sqrt{\frac{R_\mathrm{in}}{R}}\right)\left(\frac{R}{R_\mathrm{ref}}\right)^{-p},
\end{equation}
with $R_\mathrm{ref}=1$, $p=1$, and $\Sigma_0$ set by the disc mass $M_\mathrm{d} = 0.05$. The central object is modelled as a sink particle of mass $M = 1$ and accretion radius $R_\mathrm{acc} = 0.8$. The sound speed is
\begin{equation}\label{ICs:cs}
    c_\mathrm{s}(r) = c_\mathrm{s,0}\left(\frac{r}{R_\mathrm{in}}\right)^{-q},
\end{equation}
with $c_\mathrm{s,0}=0.05$ and $q=0.25$. The azimuthal velocity includes a correction for the radial pressure gradient,
\begin{equation}\label{ICs:velocityphantom}
    v_\phi^2 = v_\mathrm{k}^2 - c_\mathrm{s}^2\left(\frac{3}{2}+q+p\right),
\end{equation}
where $v_\mathrm{k}$ is the Keplerian velocity.

The particle distribution is first relaxed by performing purely hydrodynamic simulations for 100 orbits at $R_\mathrm{ref}$. This step reduces numerical artefacts from the random Monte Carlo placement and from the density discontinuity at the outer edge of the disc. The relaxed positions and velocities then serve as the starting point for the MHD simulations, to which a purely toroidal magnetic field is added according to
\begin{equation}\label{ICs:Bfield}
    B_\phi = \sqrt{\frac{2\mu_0P}{\beta_0}}, \quad B_R = B_z = 0 .
\end{equation}
The plasma beta, defined as $\beta = 2\mu_0 P/B^2$, is initially set to
$\beta_0 = 25$, as in \citet{fromang_s_global_2006, flock_turbulence_2011}. To balance the magnetic pressure and tension, we also apply a small correction to the azimuthal velocity,
\begin{equation}\label{ICs:velocitymag}
    v_{\phi,\mathrm{mhd}}^2 = v_{\phi,\mathrm{hydro}}^2 + v_\mathrm{mag}^2,
\end{equation}
where $v_\mathrm{mag}^2 = -(c_\mathrm{s}^2/\beta_0)\left(p+q-\tfrac{1}{2}\right)$.

\subsection{Smoothed particle magnetohydrodynamics} \label{subsec:SPMHD}

Smoothed particle magnetohydrodynamics is a Lagrangian method that discretizes the dynamical equations \eqref{MHDEqs:mass}--\eqref{MHDEqs:induction} on a set of interpolation points, referred to as particles. Fluid properties at each particle are estimated by weighting contributions of its neighbours with a smoothing kernel. For a detailed description of the method, see \citet{price_smoothed_2012, price_phantom_2018, tricco_smoothed_2023}.

The density and smoothing length of particle $a$ are computed by solving simultaneously
\begin{gather}
    \rho_{\mathrm{sum},a} = \sum_b m_b W_{ab}(h_a), \label{SPMHD:density_sum}\\
    \rho_a = m_a \left(\frac{\eta_\mathrm{fact}}{h_a}\right)^{3}, \label{SPMHD:density_h}
\end{gather}
where $m$ is the particle mass, $h$ is the smoothing length, and $\eta_\mathrm{fact}$ is a dimensionless parameter that sets the ratio of the smoothing length to the mean local particle spacing. The kernel is written as $W_{ab}(h_a) \equiv W(|\mathbf{r}_a - \mathbf{r}_b|, h_a)$, where $\mathbf{r}$ is the particle position.

Two general forms for discretizing the gradient of a quantity $f$ \citep{price_smoothed_2012} are obtained by rewriting $\nabla f$ in terms of an arbitrary differentiable scalar $\xi$. The symmetric form is
\begin{align} \notag
\nabla f &= \xi
\left[ \frac{f}{\xi^{2}}\nabla\xi + \nabla\!\left(\frac{f}{\xi}\right)\right] \\ \label{gradient:symmetric}
\left(\nabla f\right)_a&\approx \sum_{b}
\frac{m_b}{\rho_b}
\left[ f_a \frac{\xi_b}{\xi_a}\nabla_a W_{ab}(h_a) \right.\\
&\hspace{18mm} \left.+ f_b \frac{\xi_a}{\xi_b}\nabla_a W_{ab}(h_b) \right]
,
\end{align}
and the antisymmetric form is
\begin{align} \notag
\nabla f & = \frac{1}{\xi}
\left[ \nabla(\xi f)-f\nabla\xi \right]\\ \label{gradient:antisymmetric}
\left(\nabla f\right)_a&\approx \sum_{b}
\frac{m_b}{\rho_b}
\frac{\xi_b}{\xi_a} (f_b-f_a) \nabla_a W_{ab}(h_a).
\end{align}
With $\xi=\rho$, these expressions reduce to the standard SPMHD gradient operators used in \citet{price_phantom_2018}. With $\xi=1$, the densities of each particle pair enter the momentum and induction equations through their geometric mean \citep{wadsley_gasoline2_2017, wissing_smoothed_2020}. We refer to this choice as the geometric density weighting.

Recent studies have reported benefits of the geometric density weighting, especially for MHD problems \citep{wissing_smoothed_2020, wissing_magnetorotational_2022}. The gains are largest in systems with steep density gradients that span a wide range of scales. In the collapse of a magnetized cloud, for example, it produces collimated jets with fewer particles and achieves faster convergence with resolution \citep{wissing_smoothed_2020}. Motivated by these findings, we apply the geometric density weighting to the magnetic tension, the induction equation, and the $\nabla\cdot\mathbf{B}$ terms, while the pressure gradients keep the standard weighting. To do so, we separate the MHD stress tensor, $S^{ij} = S^{ij}_{\rm iso} + S^{ij}_{\rm anis}$, into its isotropic and anisotropic parts,
\begin{equation}\label{SPMHD:stress_split}
    S^{ij}_{\rm iso} = -\left(P + \frac{B^2}{2\mu_0}\right)\delta^{ij}, \qquad
    S^{ij}_{\rm anis} = \frac{B^iB^j}{\mu_0}.
\end{equation}

The momentum equation, Equation~\eqref{MHDEqs:momentum}, is then discretized as
\begin{equation} \label{SPMHD:momentum}
\begin{split}
    \frac{{\rm d}v^i_a}{{\rm d}t} = &\sum_b m_b \left[\frac{S^{ij}_{{\rm iso},a}}{\rho_a^2}\frac{\nabla^j_a W_{ab}(h_a)}{\Omega_a} + \frac{S^{ij}_{{\rm iso},b}}{\rho_b^2}\frac{\nabla^j_a W_{ab}(h_b)}{\Omega_b} \right]\\
    &+\sum_b \frac{m_b}{\rho_a\rho_b} \left[S^{ij}_{{\rm anis},a}\frac{\nabla^j_a W_{ab}(h_a)}{\Omega_a} + S^{ij}_{{\rm anis},b}\frac{\nabla^j_a W_{ab}(h_b)}{\Omega_b} \right]\\
    &+\Pi^i_{\mathrm{shock}, a} + f^i_{\mathrm{divB}, a} + a^i_\mathrm{sink\text{-}gas},
\end{split}
\end{equation}
where
\begin{equation}\label{SPMHD:Omega_factor}
    \Omega_a = 1 + \frac{h_a}{3\rho_a}\sum_b m_b \frac{\partial W_{ab}(h_a)}{\partial h_a}
\end{equation}
is the correction for the spatial variation of the smoothing length (the grad-$h$ term). The stress-tensor terms in the first two lines are derived from a discrete Lagrangian (Appendix~\ref{app:lagrangian_formulation}). The last three terms in Equation~\eqref{SPMHD:momentum} are the shock-capturing term $\Pi^i_{\mathrm{shock},a}$, the correction $f^i_{\mathrm{divB},a}$ that removes the tensile instability, and the gravitational acceleration $a^i_{\mathrm{sink\text{-}gas}}$ from the sink particle.

Advection is computed exactly in SPMHD because the method is Lagrangian. Thus, dissipation arises only from the shock-capturing terms added to stabilize the method. In the momentum equation, they act as an artificial viscosity (AV) \citep{monaghan_sph_1997, lodato_diffusive_2010},
\begin{equation}\label{SPMHD:shock_AV}
    \Pi^i_{\mathrm{shock}, a} = -\sum_b m_b \left[\frac{q^a_{ab}}{\rho_a^2}\frac{\nabla^i_{a} W_{ab}(h_a)}{\Omega_a} + \frac{q^b_{ab}}{\rho_b^2}\frac{\nabla^i_{a} W_{ab}(h_b)}{\Omega_b} \right],
\end{equation}
with
\begin{equation}\label{SPMHD:q_AV}
    q^a_{ab} =
    \begin{cases}
        -\frac{1}{2}\rho_a v_{{\mathrm{sig},a}}\, \mathbf{v}_{ab}\cdot \hat{\mathbf{r}}_{ab}, & \mathbf{v}_{ab}\cdot \hat{\mathbf{r}}_{ab} < 0, \\
        0, & \mathrm{otherwise},
    \end{cases}
\end{equation}
where $\mathbf{v}_{ab} = \mathbf{v}_a - \mathbf{v}_b$, $\hat{\mathbf{r}}_{ab} = (\mathbf{r}_a - \mathbf{r}_b)/|\mathbf{r}_a - \mathbf{r}_b|$, and the maximal signal speed is
\begin{equation}\label{SPMHD:v_sig}
    v_{\mathrm{sig},a} = \alpha_a^{\mathrm{AV}} \sqrt{c^2_{\mathrm{s},a} + v^2_{\mathrm{A},a}} + \beta^{\mathrm{AV}}|\mathbf{v}_{ab}\cdot \hat{\mathbf{r}}_{ab}|.
\end{equation}
Here, $v_\mathrm{A}$ is the Alfv\'en speed, and $\alpha_a^{\mathrm{AV}}$ and $\beta^{\mathrm{AV}}$ are dimensionless parameters. The linear coefficient $\alpha_a^{\mathrm{AV}}$ varies between 0 and 1 according to the shock detector of \citet{cullen_inviscid_2010}, which reduces unwanted dissipation away from shocks. We set $\beta^{\mathrm{AV}}=2$ to avoid particle interpenetration.

To address the tensile instability, we follow the correction from \citet{borve_regularized_2001},
\begin{equation}\label{SPMHD:tensile_instability}
\begin{split}
    f^i_{\mathrm{divB}, a} = -\frac{\hat{\beta}}{\mu_0}B^i_a \sum_b \frac{m_b}{\rho_a\rho_b} &\left[\mathbf{B}_a\cdot\frac{\nabla_{a} W_{ab}(h_a)}{\Omega_a} \right.\\
    & + \left.\mathbf{B}_b\cdot \frac{\nabla_{a} W_{ab}(h_b)}{\Omega_b} \right],
\end{split}
\end{equation}
where $\hat{\beta}$ introduces a linear transition that switches the correction off as the plasma beta increases \citep{borve_two-dimensional_2004, tricco_constrained_2012},
\begin{equation}\label{SPMHD:beta_fdivB}
    \hat{\beta} =
    \begin{cases}
        1, & \beta \leq 2, \\
        (10-\beta)/8, & 2 < \beta < 10, \\
        0, & \mathrm{otherwise}.
    \end{cases}
\end{equation}

The discretized form of the induction equation, Equation~\eqref{MHDEqs:induction}, reads
\begin{equation}\label{SPMHD:Induction_Eq}
\begin{split}
    \frac{{\rm d}}{{\rm d}t}\left(\frac{\mathbf{B}}{\rho}\right)_a = &-\sum_b \frac{m_b}{\rho_a\rho_b} \mathbf{v}_{ab} \left[\mathbf{B}_a \cdot \frac{\nabla_a W_{ab} (h_a)}{\Omega_a}\right] \\
    &-\sum_b \frac{m_b}{\rho_a\rho_b} \left[ \psi_a \frac{\nabla_a W_{ab} (h_a)}{\Omega_a} + \psi_b\frac{\nabla_a W_{ab} (h_b)}{\Omega_b}\right]\\
    &+ \mathcal{D}_{\mathrm{diss}, a},
\end{split}
\end{equation}
where $\psi$ is the divergence-cleaning field. The artificial resistivity (AR) used to capture shocks and discontinuities is \citep{price_phantom_2018}
\begin{equation}\label{SPMHD: art resistivity}
\begin{split}
    \mathcal{D}_{\mathrm{diss}, a} = \frac{1}{2}\sum_b &\frac{m_b}{\rho_a\rho_b}\, v^B_{\mathrm{sig}, ab}\, \mathbf{B}_{ab} \\
    &\hat{\mathbf{r}}_{ab}\cdot\left[\frac{\nabla_a W_{ab} (h_a)}{\Omega_a} + \frac{\nabla_a W_{ab} (h_b)}{\Omega_b}\right],
\end{split}
\end{equation}
where $\mathbf{B}_{ab} = \mathbf{B}_a - \mathbf{B}_b$, and the signal speed is
\begin{equation}
    v^B_{\mathrm{sig}, ab} = \alpha_\mathrm{B}|\mathbf{v}_{ab}\times \hat{\mathbf{r}}_{ab}|,
\end{equation}
with $\alpha_\mathrm{B}$ a dimensionless constant parameter.

To constrain the divergence of the magnetic field, Equation~\eqref{MHDEqs:divB}, we use the hyperbolic/parabolic divergence cleaning scheme of \citet{tricco_constrained_2012} and \citet{tricco_constrained_2016}. The divergence-cleaning field evolves according to
\begin{equation}\label{SPMHD:psi}
\begin{split}
    \frac{{\rm d}}{{\rm d}t}\left(\frac{\psi}{c_\mathrm{h}}\right)_a = ~& c_{\mathrm{h},a}\sum_b \frac{m_b}{\rho_b} \mathbf{B}_{ab} \cdot \frac{\nabla_a W_{ab} (h_a)}{\Omega_a} \\
    &+ \frac{\psi_a}{2c_{\mathrm{h},a}\rho_a}\sum_b m_b \mathbf{v}_{ab} \cdot \frac{\nabla_a W_{ab} (h_a)}{\Omega_a}\\
    &- \frac{\psi_a}{c_{\mathrm{h},a}\tau_{\mathrm{c},a}},
\end{split}
\end{equation}
where $c_\mathrm{h} = f_\mathrm{clean}\sqrt{c_\mathrm{s}^2 + v_\mathrm{A}^2}$ is the cleaning speed, with $f_\mathrm{clean} = 1$, and $\tau_\mathrm{c}$ is the decay timescale,
\begin{equation}
    \tau_{\mathrm{c},a} = \frac{h_a}{\sigma_\mathrm{c}\, c_{\mathrm{h},a}}.
\end{equation}
with $\sigma_c=1$ a dimensionless factor.

\subsection{Codes and simulations} \label{subsec:simulations}

We use the two codes \textsc{Phantom} \citep{price_phantom_2018} and \textsc{Shamrock} \citep{david-cleris_shamrock_2025}. \textsc{Phantom} has had a robust public SPMHD implementation since 2018 and is widely used for simulations of accretion discs. \textsc{Shamrock} is built on the same SPH methods. Both codes were modified to include the formulation of Section~\ref{subsec:SPMHD}, so they share the same discretization but differ in their parallelization and time integration. In Appendix~\ref{Append_MHD_testing}, we validate that the \textsc{Shamrock} SPMHD implementation is in agreement with \textsc{Phantom}.

Table~\ref{tab:sims} lists our simulations, which span a range of resolutions and dissipation parameters. Each run is labelled \texttt{CNpM\_2aBx}, where \texttt{C} denotes the code (\texttt{S} for \textsc{Shamrock}, \texttt{P} for \textsc{Phantom}), \texttt{Np} is the number of particles in millions, and \texttt{x} sets the artificial resistivity parameter through $\alpha_\mathrm{B} = 2^{-\texttt{x}}$. For example, \texttt{S102M\_2aB2} uses \textsc{Shamrock} with $102\times10^6$ particles and $\alpha_\mathrm{B} = 0.25$. Every run uses the quintic spline kernel ($M_6$).

\textsc{Phantom} is parallelized for shared memory with OpenMP. The MPI implementation in this version of \textsc{Phantom} was not found to scale well for our problem. \textsc{Shamrock}, on the other hand, relies on the SYCL standard to achieve performance portability across hardware. It combines two levels of parallelism: an MPI decomposition of the domain into patches, and device-level kernels within each patch. Using \textsc{Shamrock}, our simulations can scale efficiently up to 1920 cores.

\textsc{Shamrock} simulations were run on AMD EPYC 9655 CPUs, corresponding to the Fir cluster of the Digital Research Alliance of Canada. \textsc{Phantom} simulations were run on Intel 6972P CPUs (the Nibi cluster).

For 300 orbits of model \texttt{102M\_2aB2}, \textsc{Phantom} required $\approx$ 203k CPU-hours (${\sim}$132 days on 64 cores), compared with $\approx$ 1.2M CPU-hours (${\sim}$32 days on 1536 cores) for \textsc{Shamrock}. The reason for such a large discrepancy between running times is due to how timesteps are handled. \textsc{Phantom} evolves the particles on individual timesteps, whereas \textsc{Shamrock} uses a single global timestep.

To improve parallelization efficiency, \textsc{Shamrock} uses a density-based smoothing length with a limit of 500 neighbours, and it also discards any particle that moves beyond a kill radius $r_\mathrm{kill}=20\,\mathrm{au}$.

\begin{deluxetable}{lccccccc}
\tabletypesize{\footnotesize}
\tablecaption{Summary of the MHD simulations. From left to right: label, number of particles, artificial resistivity parameter, initial quality factor, magnetic Reynolds number, magnetic Prandtl number, normalized Maxwell stress, and simulated time. Time averages are computed over 200--300\,yr. For $R_\mathrm{m}$, we report the geometric mean. \label{tab:sims}}
\tablewidth{0pt}
\tablehead{
\colhead{Label} & \colhead{$N_\mathrm{P}$} & \colhead{$\alpha_\mathrm{B}$} & \colhead{$\langle \mathrm{Q}_{\phi,0}\rangle^{\rm V}$} & $\overline{\langle R_{\rm m}\rangle}$ & $\overline{\langle P_{\rm m}\rangle}$ & \colhead{$\overline{\alpha_{\rm M}}$} & \colhead{$t_\mathrm{max}$}\\[-0.2cm]
\colhead{} & \colhead{($10^6$)} & \colhead{} & \colhead{} &
\colhead{} & \colhead{} & \colhead{} & \colhead{(yr)}
}
\startdata
\texttt{S102M\_2aB2}  & 102 & 0.25  & 30.3 & 1041 & 1.83 & 0.35 & 1000 \\
\texttt{P102M\_2aB1} & 102 & 0.5 & 30.3 & 724 & 1.16 & 0.30 & 315 \\
\texttt{P60M\_2aB2} & 60 & 0.25 & 25.4 & 822 & 1.72 & 0.32 & 353 \\
\texttt{P60M\_2aB1} & 60 & 0.5 & 25.4 & 558 &  1.08 & 0.26 & 337 \\
\texttt{P40M\_2aB2} & 40 & 0.25 & 22.2 & 674 & 1.62 & 0.29 & 356 \\
\texttt{P40M\_2aB1} & 40 & 0.5 & 22.2 & 471 & 1.05 & 0.24 & 351 \\
\texttt{P40M\_2aB0} & 40 & 1 & 22.2 & 304 & 0.63 & 0.13 & 326 \\
\texttt{P20M\_2aB2} & 20 & 0.25 & 17.7 & 495 & 1.48 & 0.25 & 316 \\
\texttt{P20M\_2aB1} & 20 & 0.5 & 17.7 & 336 & 0.96 & 0.16 & 316 \\
\texttt{P20M\_2aB0} & 20 & 1 & 17.7 & 204 & 0.57 & 0.10 & 302 \\
\hline
\texttt{P10M\_2aB2}\tablenotemark{a} & 10 & 0.25 & 14.0 & 344 & 1.36 & 0.19 & 288 \\
\texttt{P10M\_2aB1} & 10 & 0.5 & 14.0 & 234 & 0.90 & 0.12 & 389 \\
\texttt{P5M\_2aB2} & 5 & 0.25 & 11.2 & 234 & 1.24 & 0.16 & 595 \\
\texttt{P5M\_2aB1} & 5 & 0.5 & 11.2 & 161 & 0.84 & 0.11 & 599 \\
\texttt{P5M\_2aB0} & 5 & 1 & 11.2 & 89 & 0.48 & 0.06 & 585 \\
\texttt{P2M\_2aB2} & 2 & 0.25 & 8.3 & 147 & 1.17 & 0.11 & 632 \\
\texttt{P2M\_2aB1} & 2 & 0.5 & 8.3 & 90 & 0.74 & 0.10 & 632 \\
\texttt{P2M\_2aB0} & 2 & 1 & 8.3 & 51 & 0.43 & 0.04 & 632 \\
\hline
\texttt{stdP60M\_2aB2}  & 60  & 0.25  & 25.4 & -- & -- & -- & 79 \\
\texttt{stdP60M\_2aB1}  & 60  & 0.5   & 25.4 & -- & -- & -- & 51 \\
\texttt{stdP40M\_2aB2}  & 40  & 0.25  & 22.2 & -- & -- & -- & 60 \\
\texttt{stdP20M\_2aB2}  & 20  & 0.25  & 17.7 & -- & -- & -- & 60 \\
\enddata
\tablecomments{Simulations are labelled \texttt{CNpM\_2aBx}, where \texttt{C}
is the code integrating the system (\texttt{S} for \textsc{Shamrock} and
\texttt{P} for \textsc{Phantom}), \texttt{Np} is the number of particles in millions, and \texttt{x} is related to the artificial resistivity parameter through $\alpha_\mathrm{B}=2^{-\texttt{x}}$. The prefix \texttt{std} stands for standard implementation of SPMHD (see section \ref{sub:gradient_choice}).}
\tablenotetext{a}{Time averages were taken from 200\,yr to $t_\mathrm{max}$.}
\end{deluxetable}

\subsection{Code units} \label{subsec:code_units}

The ideal MHD equations are scale-free, so physical units serve only as a reference for interpreting the results. The code uses $u_\mathrm{length} = 1~\mathrm{au}$ and $u_\mathrm{mass} = 1~\mathrm{M}_\odot$, with the gravitational constant $G = 1$ and $\mu_0 = 1$, so one orbit at $R_\mathrm{ref}$ lasts $2\pi$ in code time. For convenience, we express all times in years; one year then corresponds to one orbit at $R_\mathrm{ref}$.

\subsection{Quality factor} \label{subsec:resolution}

To measure how well the particles resolve the MRI, we follow the approach of \citet{noble_dependence_2010} and \citet{hawley_assessing_2011}. We define a quality factor as the local number of resolution elements across the characteristic MRI wavelength,
\begin{equation}\label{quality_factor_phi}
    Q_\phi = \frac{\lambda_\mathrm{MRI}}{\Delta} = \frac{2\pi|v_{\mathrm{A},\phi}|}{\Omega\,\Delta},
\end{equation}
where $v_{\mathrm{A},\phi}$ is the azimuthal component of the Alfv\'en velocity, $\Omega$ is the angular velocity, and $\Delta$ is the resolution length. Note that the wavelength $\lambda_\mathrm{MRI}$ is a factor $\sqrt{16/15}$ smaller than that of the fastest-growing mode. In grid-based codes, $\Delta$ is usually defined by the cell size. Particle methods, however, have no unique equivalent; possible choices include the smoothing length, the mean particle spacing, and the compact support radius. In this work, we adopt $\Delta = h_\mathrm{SPH} = 2\sigma$, where $\sigma$ is the standard deviation of the kernel \citep{2012MNRAS.425.1068D, deng_local_2019}. For the quintic spline, $h_\mathrm{SPH} \approx 1.37\,h$.

Although the quality factor is a common resolution diagnostic, its reliability as a convergence metric is debated. It is based on the linear growth of the MRI in local models, and may be less meaningful in global discs or once the instability saturates into nonlinear turbulence. \citet{jannaud_numerical_2026} argued that, in the nonlinear regime, a high quality factor is at most a requirement for convergence, not a guarantee of it. They found that even ideal zero-net-flux shearing boxes, which are known to be unconverged, reach values above the usual thresholds. We nevertheless use $Q_\phi$, since it facilitates the comparison of our results with previous studies. However, we treat it as a qualitative guide rather than a strict criterion for resolving the MRI.

\subsection{Averages} \label{subsec:averages}

This section defines the averages used to characterize the disc. For equal-mass particles, the mass-weighted average of a quantity $f$ over a region $\zeta$ reduces to the arithmetic mean
\begin{equation} \label{ave:_mass_weighted_sum}
    \langle f\rangle = \frac{1}{N_\zeta}\sum_b f_b,
\end{equation}
where the sum runs over the $N_\zeta$ particles in $\zeta$. To facilitate comparison with previous studies, we also use the volume-weighted average,
\begin{equation} \label{ave:_volume_weighted_sum}
    \langle f\rangle^{\mathrm{V}} = \frac{\sum_b f_b/\rho_b}{\sum_b 1/\rho_b}.
\end{equation}

These definitions extend to particles grouped in bins along selected coordinates. For instance, the azimuthal average $\langle f\rangle_\phi(R, z)$ is computed over an annular bin of radial width $\Delta R$ and height $\Delta z$, centred on $(R, z)$. The azimuthal and vertical average $\langle f\rangle_{\phi z}(R)$ is obtained in the same way, with the particles grouped instead in cylindrical shells of width $\Delta R$. Empty bins are excluded from the analysis.

These averages are calculated directly from the particles. When $\langle f\rangle_\phi$ is plotted in the meridional plane, however, few particles lie at $|z| \gtrsim 3H$, so the methods above can leave many bins empty. In that case, it is preferable to interpolate the particle data onto a three-dimensional cylindrical or spherical grid, evaluate $f$ at the cell centres, and then average over $\phi$. The result is denoted by square brackets, as in $[f]_\phi$.

Time averages over an interval $[t_0, t_1]$ are denoted by an overline,
\begin{equation}\label{ave: time}
    \overline{\langle f\rangle} = \frac{1}{t_1 - t_0}\int_{t_0}^{t_1} \langle f\rangle~dt.
\end{equation}

\subsection{Dissipation} \label{subsec:dissipation}

Dissipation in our simulations comes mainly from the shock-capturing terms. In the continuum limit, these terms act as a physical viscosity and resistivity whose strength scales with the smoothing length \citep{monaghan_smoothed_2005, price_smoothed_2012}. Following \citet{wissing_magnetorotational_2022}, we estimate the shear viscosity at each particle by matching the heating rate of the artificial viscosity, $(\mathrm{d}u/\mathrm{d}t)_\mathrm{AV}$, to the viscous heating of the Navier--Stokes equations. Assuming a bulk viscosity $\nu_\mathrm{b} = \tfrac{5}{3}\nu_\mathrm{AV}$ \citep{lodato_diffusive_2010}, this gives
\begin{equation} \label{AV dissipation}
    \nu_\mathrm{AV} = \frac{\left(\frac{{\rm d}u}{{\rm d}t}\right)_\mathrm{AV}}{\frac{1}{2}\left(\frac{\partial v^i}{\partial x^j} + \frac{\partial v^j}{\partial x^i}\right)^2+\left(\nabla\cdot \mathbf{v}\right)^2}.
\end{equation}
Similarly, matching the heating rate of the artificial resistivity, $(\mathrm{d}u/\mathrm{d}t)_\mathrm{AR}$, to Ohmic heating gives the resistivity
\begin{equation} \label{AR resistivity}
    \eta_\mathrm{AR} = \frac{\rho}{J^2} \left(\frac{{\rm d}u}{{\rm d}t}\right)_\mathrm{AR},
\end{equation}
where $J=|\nabla \times \mathbf{B}|$ is the current density.

From these quantities, we compute the magnetic Reynolds number and the Reynolds number,
\begin{equation}
    R_\mathrm{m} = \frac{c_\mathrm{s} H}{\eta_\mathrm{AR}}, \qquad
    R_\mathrm{e} = \frac{c_\mathrm{s} H}{\nu_\mathrm{AV}},
\end{equation}
where $H = c_\mathrm{s}/\Omega$ is the disc scale height. Their ratio defines the magnetic Prandtl number,
\begin{equation}
    P_\mathrm{m} = \frac{R_\mathrm{m}}{R_\mathrm{e}} = \frac{\nu_\mathrm{AV}}{\eta_\mathrm{AR}}.
\end{equation}

\subsection{Turbulence and angular momentum transport} \label{subsec:stresses}

We characterize the turbulence through the fluctuations $\delta\mathbf{v}$ and $\delta\mathbf{B}$ of the velocity and magnetic field about their mean values. The means are computed over particles binned in cylindrical shells, so that $\delta\mathbf{v} = \mathbf{v} - \langle\mathbf{v}\rangle_{\phi z}$ and $\delta\mathbf{B} = \mathbf{B} - \langle\mathbf{B}\rangle_{\phi z}$. From these fluctuations, we compute the turbulent kinetic and magnetic energy densities,
\begin{equation} \label{eq:e_turb}
    e^\mathrm{turb}_\mathrm{kin} = \frac{1}{2}\rho\,|\delta\mathbf{v}|^2, \qquad e^\mathrm{turb}_\mathrm{mag} = \frac{1}{2}|\delta\mathbf{B}|^2 .
\end{equation}
It is also important to characterize the transport of angular momentum. We measure it through the $R\phi$ component of the total stress, which is given by the sum of the Reynolds stress,
\begin{equation}
    \mathcal{R}_{R\phi} = \rho\, \delta v_R\, \delta v_\phi,
\end{equation}
and the Maxwell stress,
\begin{equation}
    \mathcal{M}_{R\phi} = -B_R B_\phi,
\end{equation}
Following the $\alpha$-prescription of \citet{shakura_black_1973}, we normalize both stresses by the thermal pressure, with all quantities volume-averaged over cylindrical shells,
\begin{equation} \label{eq. stress: alpha_tot}
    \alpha_\mathrm{tot} = \alpha_\mathrm{Rey} + \alpha_\mathrm{Max} = \frac{\langle\mathcal{R}_{R\phi}\rangle^{\mathrm{V}}_{\phi z}}{\langle P \rangle^{\mathrm{V}}_{\phi z}} + \frac{\langle \mathcal{M}_{R\phi} \rangle^{\mathrm{V}}_{\phi z}}{\langle P \rangle^{\mathrm{V}}_{\phi z}}.
\end{equation}
We estimate the contribution of the artificial viscosity to the angular momentum transport from the shear stress it exerts in a differentially rotating disc,
\begin{equation} \label{eq:T_art}
    T_\mathrm{art} = \left|\frac{\mathrm{d}\ln\Omega}{\mathrm{d}\ln R}\right|\rho\,\nu_\mathrm{AV}\Omega ,
\end{equation}
where $\nu_\mathrm{AV}$ is given by Equation~\eqref{AV dissipation}, and normalize it in the same way,
\begin{equation} \label{eq:alpha_art}
    \alpha_\mathrm{art} = \frac{\langle T_\mathrm{art}\rangle^{\rm V}_{\phi z}}{\langle P\rangle^{\rm V}_{\phi z}}.
\end{equation}
With this definition, $\alpha_\mathrm{art}$ is directly comparable with $\alpha_\mathrm{Rey}$ and $\alpha_\mathrm{Max}$, and its ratio to $\alpha_\mathrm{tot}$ reveals the regions where numerical dissipation dominates the transport. For a Keplerian disc, it is $3/2$ times the parameter $\nu_\mathrm{AV}/(c_\mathrm{s}H)$ used in some SPH studies \citep[e.g.][]{lodato_diffusive_2010, forgan_fragmentation_2016}. We also use the Maxwell stress normalized by the magnetic pressure,
\begin{equation} \label{eq. stress: alpha_M}
    \alpha_\mathrm{M} = -\frac{2\langle B_R B_\phi\rangle^{\mathrm{V}}}{\langle B^2\rangle^{\mathrm{V}}},
\end{equation}
which depends on the correlation between $B_R$ and $B_\phi$ rather than on the field strength.

\subsection{Azimuthal spectrum}\label{subsec:spectrum}

To find on which azimuthal scales the magnetic energy is concentrated, and how much of it lies in non-axisymmetric structures, we compute its azimuthal spectrum. We follow \citet{deng_global_2020}, who applied the particle-based Fourier analysis of \citet{cossins_characterizing_2009} to the magnetic energy. For a cylindrical shell of radius $R$ containing $N_\mathrm{ann}$ particles, the amplitude of the $m$-th azimuthal mode is
\begin{equation}
  B_m^2(R) = \frac{1}{N_\mathrm{ann}}
  \left| \sum_{b=1}^{N_\mathrm{ann}} \frac{B_b^2}{2}\,
  e^{-\mathrm{i} m \phi_b} \right|,
  \label{eq:spectrum}
\end{equation}
where $B_b$ and $\phi_b$ are the magnetic field strength and azimuthal angle of particle $b$. For $m = 0$, this gives the mean magnetic energy density of the shell.

\section{Results} \label{sec:Results}

Our main goal is to study how SPMHD captures the growth of the MRI in global accretion discs and whether the resulting turbulence is sustained. We first evaluate the quality factor and define the disc region used in the analysis. We then focus on the evolution of \texttt{S102M\_2aB2}, our highest-resolution model and the one with the most favourable conditions for resolving the MRI (Table~\ref{tab:sims}). This model uses $N_\mathrm{P} = 102\times10^6$ particles and $\alpha_\mathrm{B} = 0.25$, and was run with \textsc{Shamrock}. Next, we explore how the disc dynamics change with resolution, dissipation parameters, and the choice of weighting in the SPMHD equations. Finally, we discuss the lower-resolution simulations.

\subsection{Quality factor and analysis region} \label{sub:results_quality}

\begin{figure}[t!]
\centering
\includegraphics[width=\columnwidth]{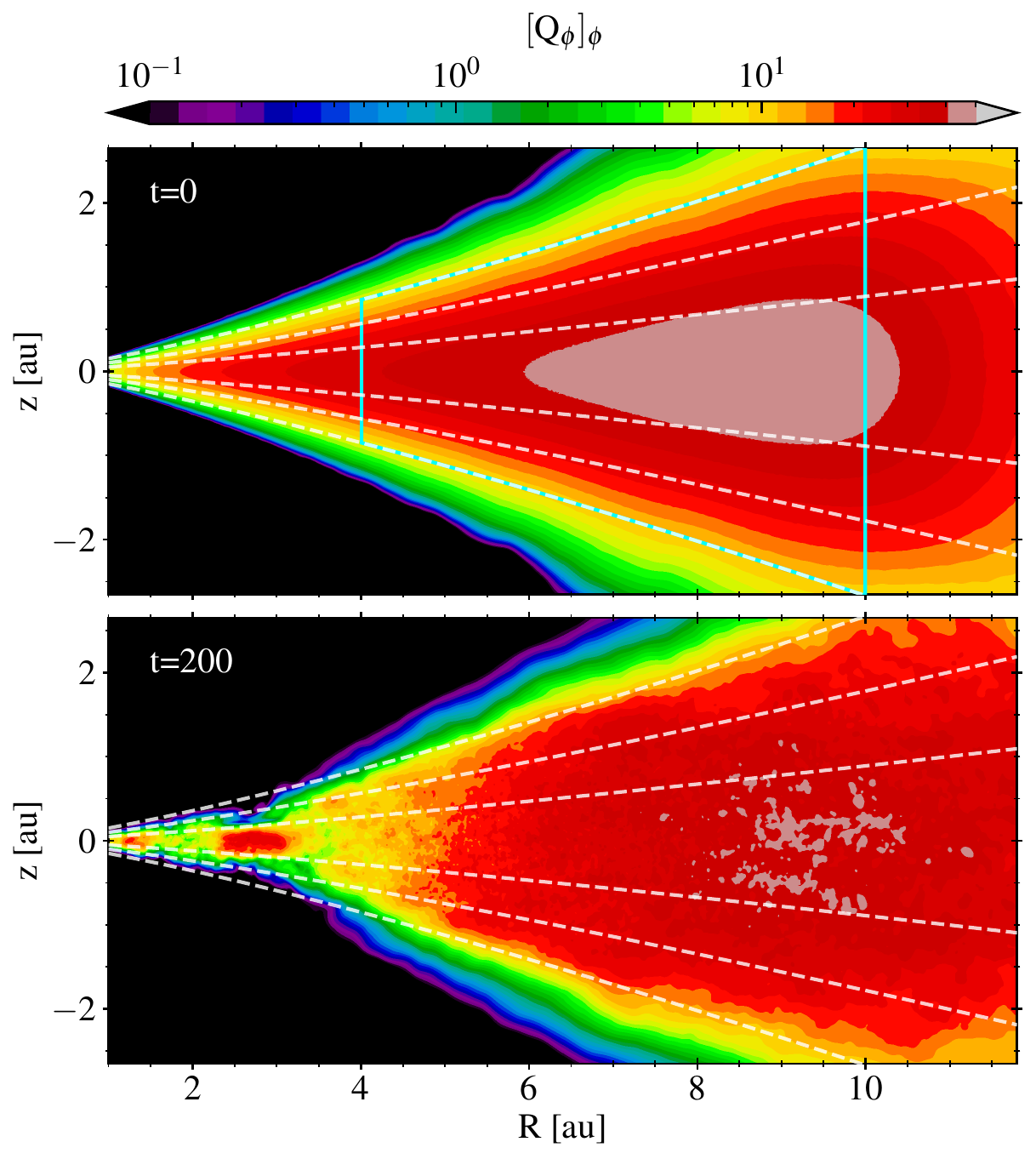}
\caption{Azimuthally averaged quality factor $[Q_\phi]_\phi$ of the \texttt{S102M\_2aB2} model at $t=0$ (top) and $200\,\mathrm{yr}$ (bottom). White dashed lines mark one, two, and three scale heights above and below the midplane. The solid cyan contour encloses the analysis region $\zeta$.}
\label{fig:quality_factor_disc}
\end{figure}

Figure~\ref{fig:quality_factor_disc} shows the azimuthally averaged quality factor of \texttt{S102M\_2aB2} at $t=0$ and $200\,\mathrm{yr}$. Initially, most of the disc within three scale heights has $Q_\phi\geq10$, with the outer disc near the midplane reaching $Q_\phi\geq30$. By $t=200\,\mathrm{yr}$, $Q_\phi$ has dropped sharply in the inner disc, especially in the atmosphere, where it falls below 1. In the outer disc, by contrast, $Q_\phi$ stays close to its initial values.

We restrict our analysis to a central part of the disc where the quality factor remains high and boundary effects are small. A minimum resolution comes from grid-based simulations, where the linear growth of the MRI needs at least six cells per wavelength \citep{hawley_local_1995, miller_formation_2000, sano_angular_2004}. Convergence studies have adopted higher thresholds, some requiring $Q_\phi\gtrsim15$ \citep{hawley_assessing_2011, sorathia_global_2012, gagnier_post-dynamical_2024, jannaud_numerical_2026}. Motivated by these criteria, we define the region $\zeta=\{4\leq R\leq 10,\ |z|\leq3H\}$, which keeps $Q_\phi\gtrsim10$ while reducing numerical artifacts from the inner boundary and from outflows in the outer disc.

\begin{figure}[t!]
\centering
\includegraphics[width=\columnwidth]{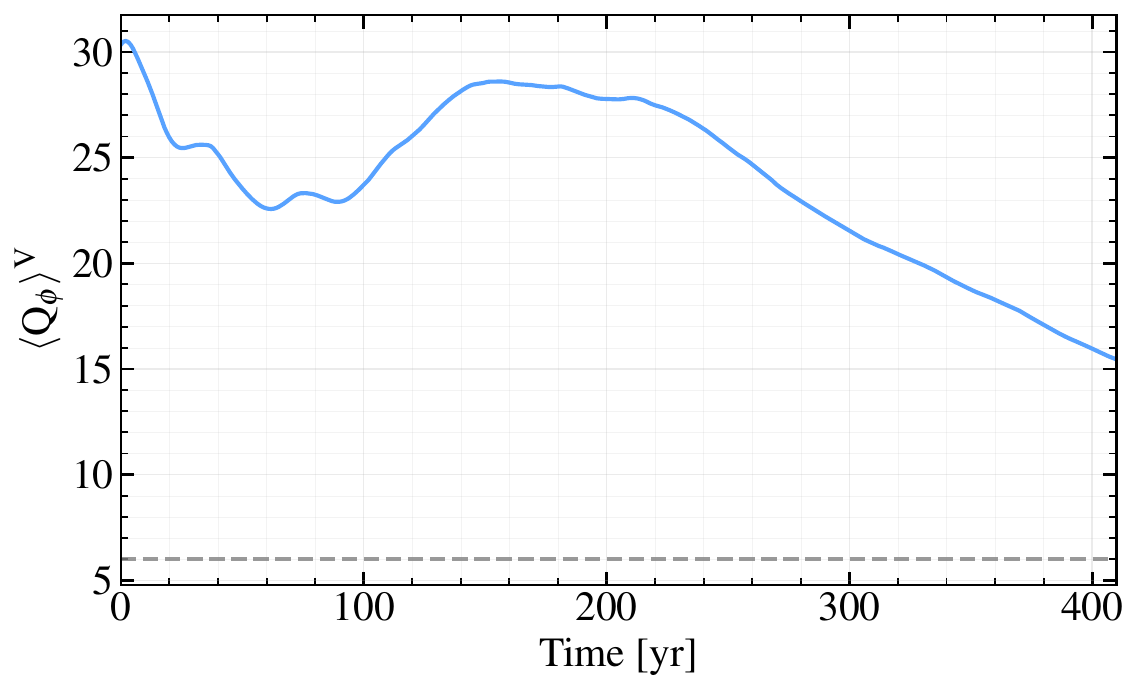}
\caption{Evolution of the azimuthal quality factor $\langle Q_\phi \rangle^{\rm V}$, volume-averaged over the analysis region $\zeta$, for \texttt{S102M\_2aB2}. The dashed grey line marks $\langle Q_\phi \rangle^{\rm V} = 6$, the threshold for resolving the linear growth of the MRI.}
\label{fig:quality_factor_evol}
\end{figure}

The volume-averaged quality factor in $\zeta$ stays above the threshold for resolving the linear growth of the MRI throughout the simulation (Figure~\ref{fig:quality_factor_evol}). After an initial drop, $\langle Q_\phi \rangle^{\rm V}$ increases as the MRI develops. As we show in the following sections, the turbulence reaches its maximum at $t\sim200\,\mathrm{yr}$, and then decays, which explains the later decrease in $\langle Q_\phi \rangle^{\rm V}$.

Unless stated otherwise, the rest of this work focuses on the region defined by $\zeta$, thus all mass- and volume-weighted averages include only the particles inside this region. Time averages are likewise computed over $t=200$--$300\,\mathrm{yr}$. This interval is chosen for three reasons: most simulations cover it, the low-dissipation discs are turbulent throughout (Section~\ref{sub:overall_disc_evolution}), and $\langle Q_\phi \rangle^{\rm V}$ stays above 20 in \texttt{S102M\_2aB2} (Figure~\ref{fig:quality_factor_evol}). A different window within this period changes the averaged values only slightly, while the trends with resolution and dissipation parameters remain the same.

\subsection{Overall disc evolution} \label{sub:overall_disc_evolution}

\begin{figure*}[t!]
\centering
\includegraphics[width=\textwidth]{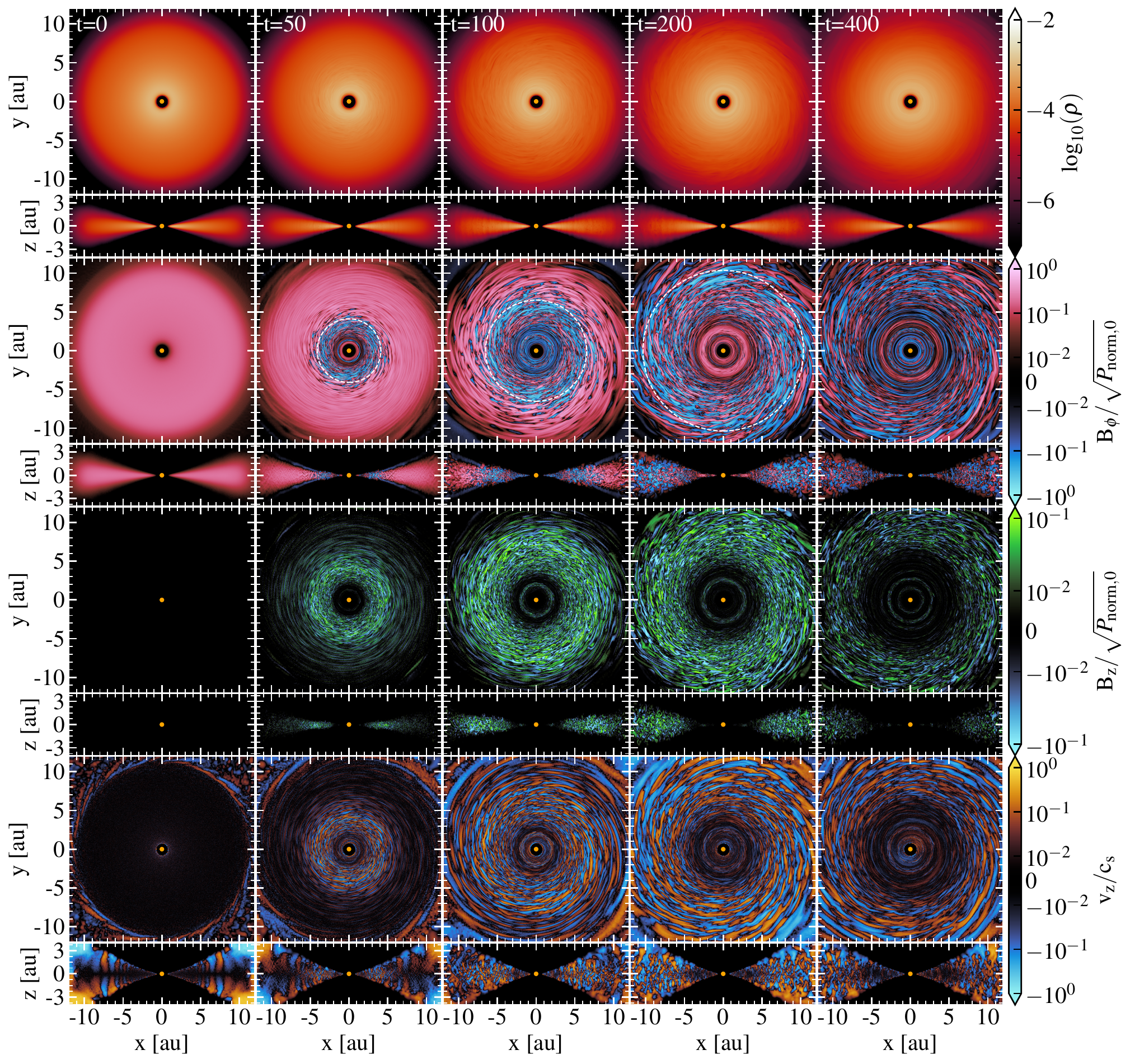}
\caption{Snapshots of the density, toroidal magnetic field, vertical magnetic field, and vertical velocity (top to bottom) at $t=0$, 50, 100, 200, and $400\,\mathrm{yr}$ (left to right) for the \texttt{S102M\_2aB2} model. For each quantity, the upper panel shows the disc midplane and the lower panel a meridional cross-section. The magnetic fields are normalized by $\sqrt{P_{\mathrm{norm},0}}$, where $P_{\mathrm{norm},0}\propto R^{-(p+q+3/2)}$ is the initial midplane pressure without the inner taper, $P(R,z=0)=\bigl(1-\sqrt{R_\mathrm{in}/R}\bigr)P_{\mathrm{norm},0}$. The orange dot marks the sink particle. In the toroidal field panels at $t=50$, 100, and $200\,\mathrm{yr}$, white dashed circles mark the radius at which the gas has completed six orbits.}
\label{fig:disc_evolution}
\end{figure*}

Figure~\ref{fig:disc_evolution} shows snapshots of the density, toroidal and vertical magnetic fields, and vertical velocity at $t=0$, 50, 100, 200, and $400\,\mathrm{yr}$ for \texttt{S102M\_2aB2}. The disc starts with a purely toroidal magnetic field, but poloidal components appear within the first few orbits and grow rapidly, reaching roughly $10\%$ of the toroidal field strength. The gas becomes turbulent after about six local orbits, as illustrated by the white dashed circles in the toroidal field panels. By $t\approx200\,\mathrm{yr}$ (around six orbits at $R=10\,\mathrm{au}$), the entire disc is turbulent, indicating that the MRI develops on the local dynamical timescale. The density has weak spiral perturbations, similar to those in previous global MRI simulations \citep{fromang_s_global_2006, deng_global_2020}. These perturbations extend over most of the disc thickness and vary little with height because of the locally isothermal equation of state.

The meridional slice of the vertical velocity at $t=0$ (bottom-left panel) displays the characteristic finger modes of the vertical shear instability. These modes begin to form during the hydrodynamical runs used to prepare the MHD initial conditions. However, they are likely poorly resolved in our models. The later snapshots show that once the MRI develops, it dominates the dynamics and replaces this vertically oriented, anisotropic pattern with more isotropic turbulence throughout the disc.

The turbulence is strongly suppressed in the inner disc and the toroidal magnetic field organizes into coherent rings. These rings are confined to a thin layer near the midplane and are weaker in the atmosphere. They are clearest at $t=200\,\mathrm{yr}$ for $R<4\,\mathrm{au}$. As the field spreads outward, it reaches regions of higher quality factor and lower dissipation, so it becomes turbulent and loses its azimuthal coherence. At $t=400\,\mathrm{yr}$, the regions where these structures persist clearly correlate with a weak poloidal field and low turbulent velocities. As a result, the spiral density perturbations disappear and are replaced by annular mass concentrations. This behaviour appears to arise from the larger dissipation in the inner disc, rather than from the inner boundary. When the sink accretion radius was varied or the density smoothing $\left(1-\sqrt{R_\mathrm{in}/R}\right)$ removed, the magnetic structure did not change significantly. In contrast, the coherent region narrows and the toroidal field weakens as the resolution increases or the resistivity decreases. This morphology is then explained by the inhibition of MRI-driven turbulence, which would otherwise regenerate the field and buoyantly transport magnetic flux away from the midplane. Consequently, the flux accumulates into the long-lived rings.

\begin{figure}[t!]
\centering
\includegraphics[width=\columnwidth]{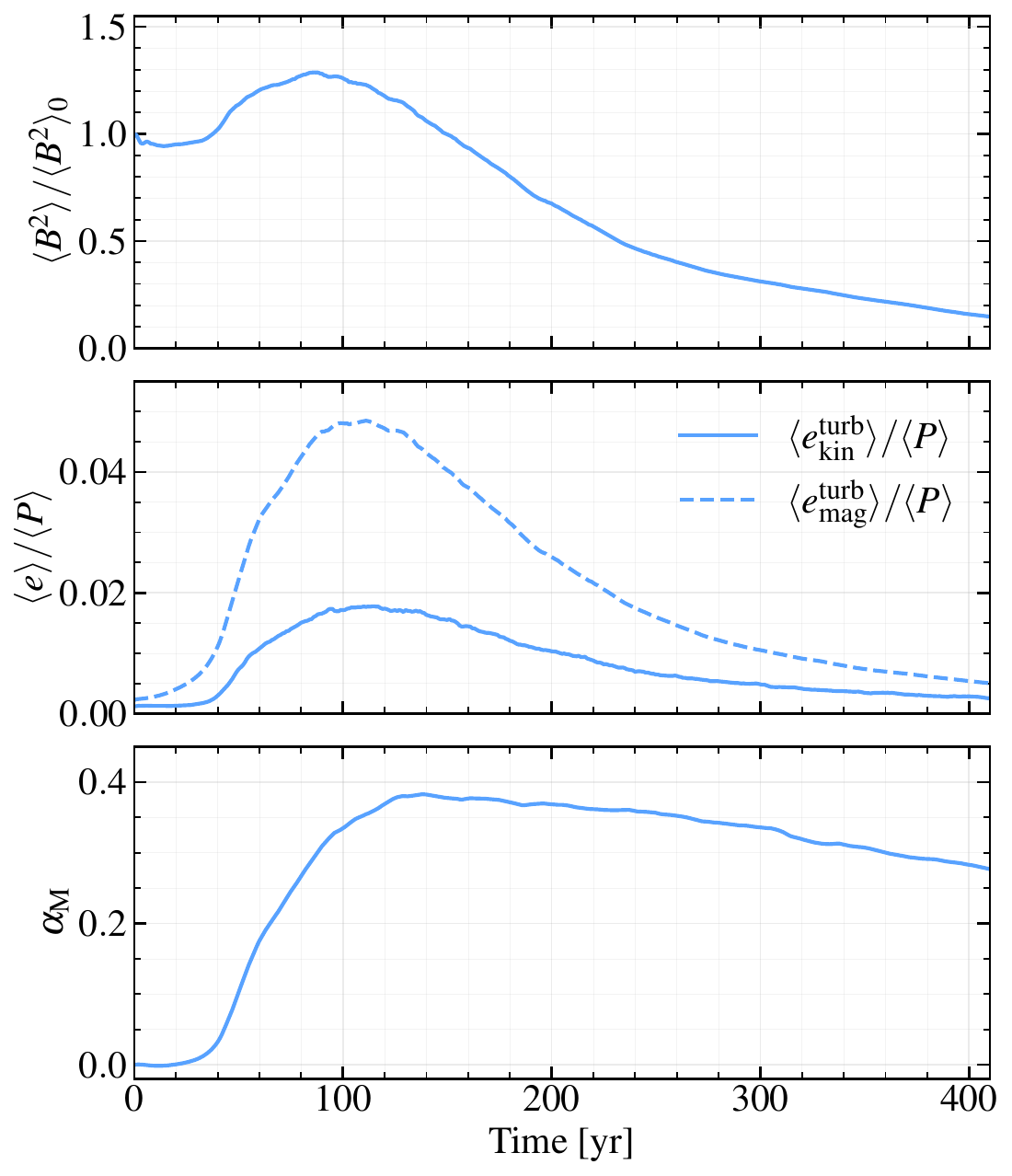}
\caption{Time evolution of the magnetic energy normalized by its initial value (top), the turbulent kinetic and magnetic energies normalized by the thermal pressure (middle), and the normalized Maxwell stress $\alpha_\mathrm{M}$ (bottom) for the \texttt{S102M\_2aB2} model. All quantities are averaged over the analysis region $\zeta$.}
\label{fig:stress_evol}
\end{figure}

Figure~\ref{fig:stress_evol} shows the time evolution of the magnetic energy, the turbulent kinetic and magnetic energies, and the normalized Maxwell stress, averaged over $\zeta$. The magnetic energy reaches a maximum at around $90\,\mathrm{yr}$ due to the amplification of the magnetic field by the linear growth of the MRI. As the instability develops, the turbulent kinetic and magnetic energies and the Maxwell stress rise sharply. The turbulent energies peak at $t\approx110\,\mathrm{yr}$ and the Maxwell stress somewhat later, at $t\approx140\,\mathrm{yr}$, with $\alpha_\mathrm{M,peak}=0.38$. This delay likely comes from the different weightings: the energies are mass-weighted and give relatively more weight to the inner radii, where the MRI saturates first, whereas $\alpha_\mathrm{M}$ is volume-weighted and favours the outer radii, where the orbital period is longer. During this phase, the turbulent magnetic energy exceeds the kinetic one by a factor of $\approx2$--3. All these quantities then decay. By $t=400\,\mathrm{yr}$, the magnetic energy has fallen to $\sim15\%$ of its initial value and both turbulent energies have dropped by almost an order of magnitude from their peaks, whereas the normalized Maxwell stress has decreased more moderately to $\alpha_\mathrm{M}\approx0.28$. These trends show that the MRI grows, reaches a maximum, and then gradually weakens as the turbulent magnetic field it draws on is dissipated. This weakening is consistent with the decline in $Q_\phi$ shown in Figure~\ref{fig:quality_factor_evol}.

\begin{figure}[t!]
\centering
\includegraphics[width=\columnwidth]{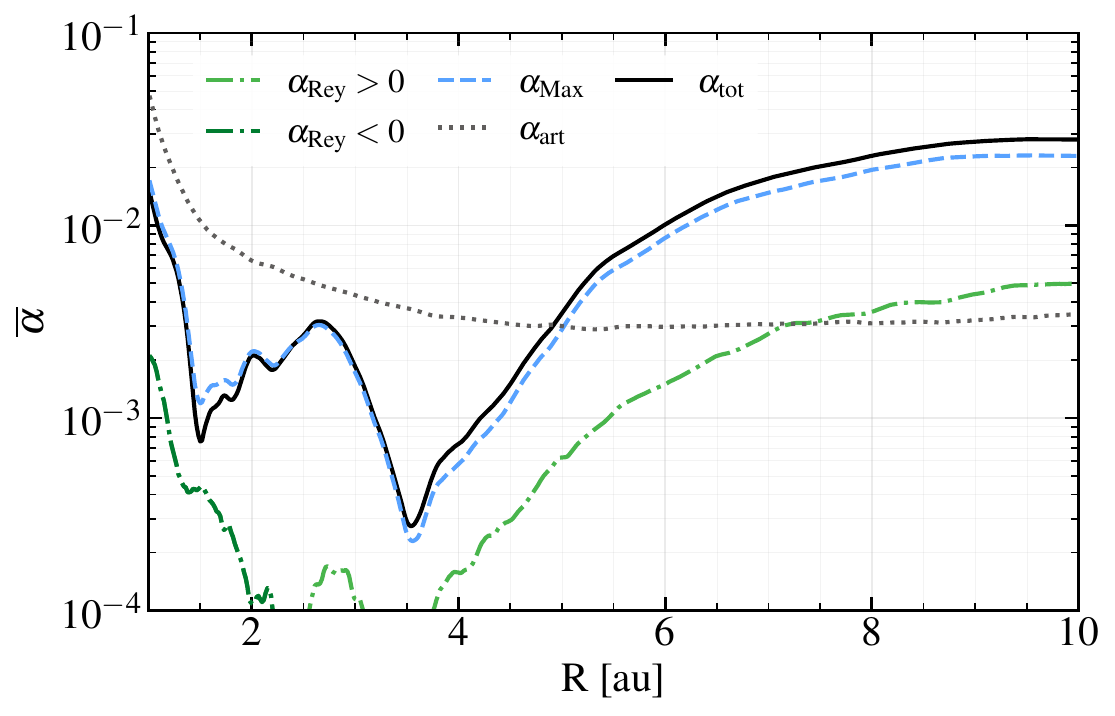}
\caption{Radial profiles of the contributions to the $\alpha$ parameter for angular momentum transport, time-averaged over $t=200$--$300\,\mathrm{yr}$.}
\label{fig:alphasR}
\end{figure}

Angular momentum transport is also quantified by the Shakura \& Sunyaev $\alpha$ prescription. Figure~\ref{fig:alphasR} shows the time-averaged radial profiles of the Reynolds and Maxwell stress contributions, $\alpha_{\rm Rey}$ and $\alpha_{\rm Max}$. In the $\zeta$ region, where the disc has already become turbulent, the Maxwell stress dominates the transport and is around 4--6 $\times$ the Reynolds stress. This ratio is consistent with the value of about 5 obtained in stratified SPMHD shearing-box simulations with net toroidal flux \citep{wissing_magnetorotational_2022}, and larger than the factor of about three found in global simulations \citep{fromang_s_global_2006,flock_turbulence_2011}.

The Reynolds and Maxwell stresses dominate the angular momentum transport in the outer disc, whereas the artificial-viscosity contribution, $\alpha_{\rm art}$, governs it in the inner disc. In the $\zeta$ region, $\alpha_{\rm tot}$ ranges from $\sim3\times10^{-3}$ to $3\times10^{-2}$ and exceeds $\alpha_{\rm art}$ by up to an order of magnitude in the outer parts. These values are comparable to the $\alpha\approx4$--$6\times10^{-3}$ reported in global simulations initialized with a toroidal field \citep{fromang_s_global_2006, flock_turbulence_2011}, although a direct comparison depends on how the stresses are normalized and averaged. In contrast, for $R\lesssim5\,\mathrm{au}$, $\alpha_{\rm art}$ is larger than $\alpha_{\rm tot}$. In this region, $\alpha_{\rm Max}$ decreases to a minimum of $\sim3\times10^{-4}$ at $R\approx3.5\,\mathrm{au}$, while the Reynolds stress becomes negligible and, inside $R\approx2\,\mathrm{au}$, negative, indicating inward transport of angular momentum by the velocity fluctuations. The low Maxwell stress in the inner disc is consistent with the suppression of MRI turbulence and the formation of coherent toroidal fields seen in Figure~\ref{fig:disc_evolution}, which we attribute to the larger dissipation.

\begin{figure}[t!]
\centering
\includegraphics[width=\columnwidth]{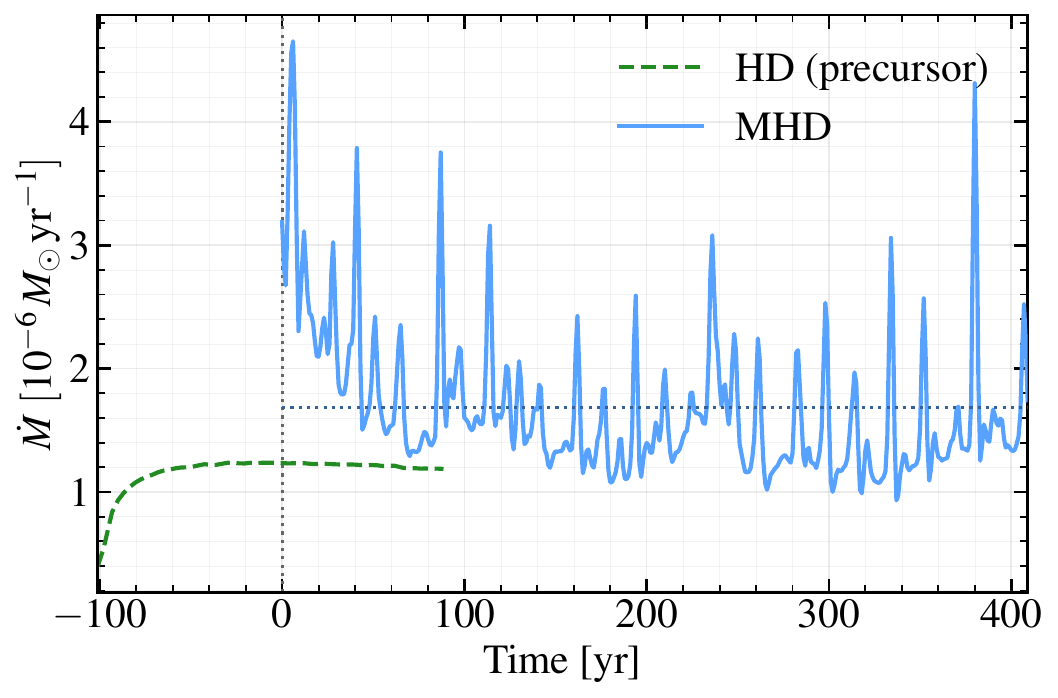}
\caption{Accretion rate $\dot{M}$ on to the central sink particle as a function of time. The dashed green line shows the purely hydrodynamical simulation used to generate the initial conditions, plotted at negative times to indicate that it precedes the insertion of the magnetic field at $t=0$ (vertical dotted line). The solid blue line shows the MHD simulation. Rates are averaged over $1\,\mathrm{yr}$ intervals. The horizontal dotted line marks the mean accretion rate over $0\leq t\leq410\,\mathrm{yr}$, $\approx1.7\times10^{-6}\,M_\odot\,\mathrm{yr}^{-1}$.}
\label{fig:accretion_rates}
\end{figure}

Figure~\ref{fig:accretion_rates} shows the accretion rate on to the central sink particle, $\dot{M}$, computed as the difference in the accumulated mass over $1\,\mathrm{yr}$ intervals. The dashed green line corresponds to the hydrodynamical model used to generate the MHD initial conditions, and the solid blue line to the subsequent MHD run, where $t=0$ corresponds to 100 inner orbits after the start of the hydrodynamical run. The hydrodynamical model has a nearly constant accretion rate of $1.2\times10^{-6}\,M_\odot\,\mathrm{yr}^{-1}$. In contrast, once the magnetic field is introduced, the accretion rate becomes highly variable, taking values between $\sim1\times10^{-6}$ and $5\times10^{-6}\,M_\odot\,\mathrm{yr}^{-1}$, with a mean of $1.7\times10^{-6}\,M_\odot\,\mathrm{yr}^{-1}$.

\begin{figure}[t!]
\centering
\includegraphics[width=\columnwidth]{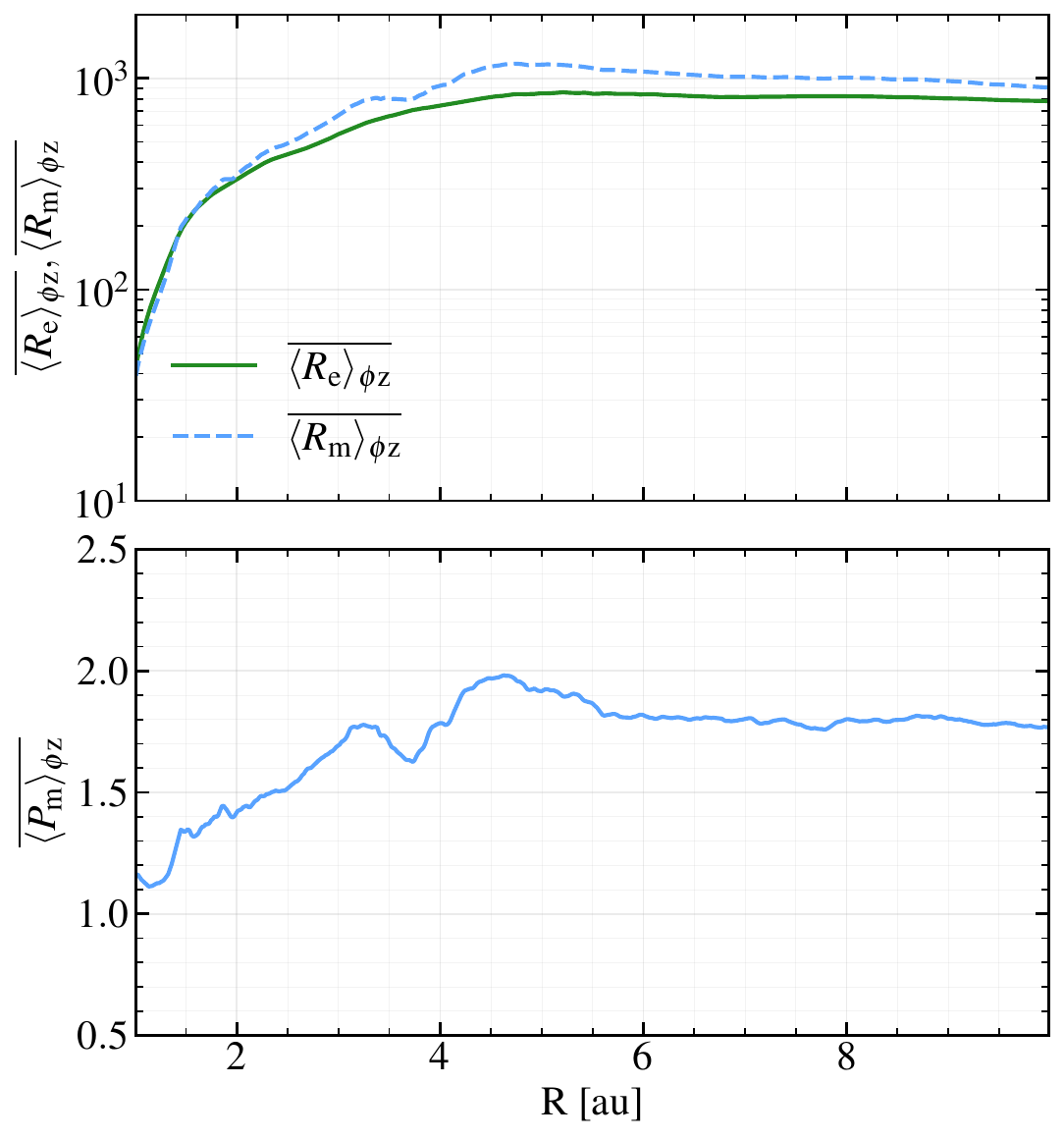}
\caption{Reynolds and magnetic Reynolds numbers (\textit{top}) and magnetic Prandtl number (\textit{bottom}) as a function of radius. All quantities are averaged azimuthally, over $|z|\leq 3H$, and in time over $t=200$--$300\,\mathrm{yr}$.}
\label{fig:numbers_r}
\end{figure}

Figure~\ref{fig:numbers_r} shows the Reynolds and magnetic Reynolds numbers (top) and the magnetic Prandtl number (bottom), averaged azimuthally, over $|z|\leq 3H$, and in time over $200$--$300\,\mathrm{yr}$. Both Reynolds numbers increase with radius in the inner disc and flatten beyond $R\approx5\,\mathrm{au}$, where $\overline{\langle R_\mathrm{e}\rangle_{\phi z}}\approx800$ and $\overline{\langle R_\mathrm{m}\rangle_{\phi z}}\approx1000$. The magnetic Prandtl number rises from $\approx1.2$ at the inner edge to a maximum of $\approx2$ at $R\approx4.5\,\mathrm{au}$, and then decreases slightly, remaining close to $1.8$ throughout the outer disc. Since $P_\mathrm{m}$ is averaged directly, it differs from the ratio of the averaged Reynolds numbers. These values of $R_\mathrm{m}$ lie in the regime where MRI turbulence decays in global simulations with explicit Ohmic resistivity, $R_\mathrm{m}\lesssim3000$ \citep{flock_turbulence_2012}. Based on this threshold, we expect the turbulence in our discs to decay in the long term rather than saturate.

\begin{figure}[t!]
\gridline{\fig{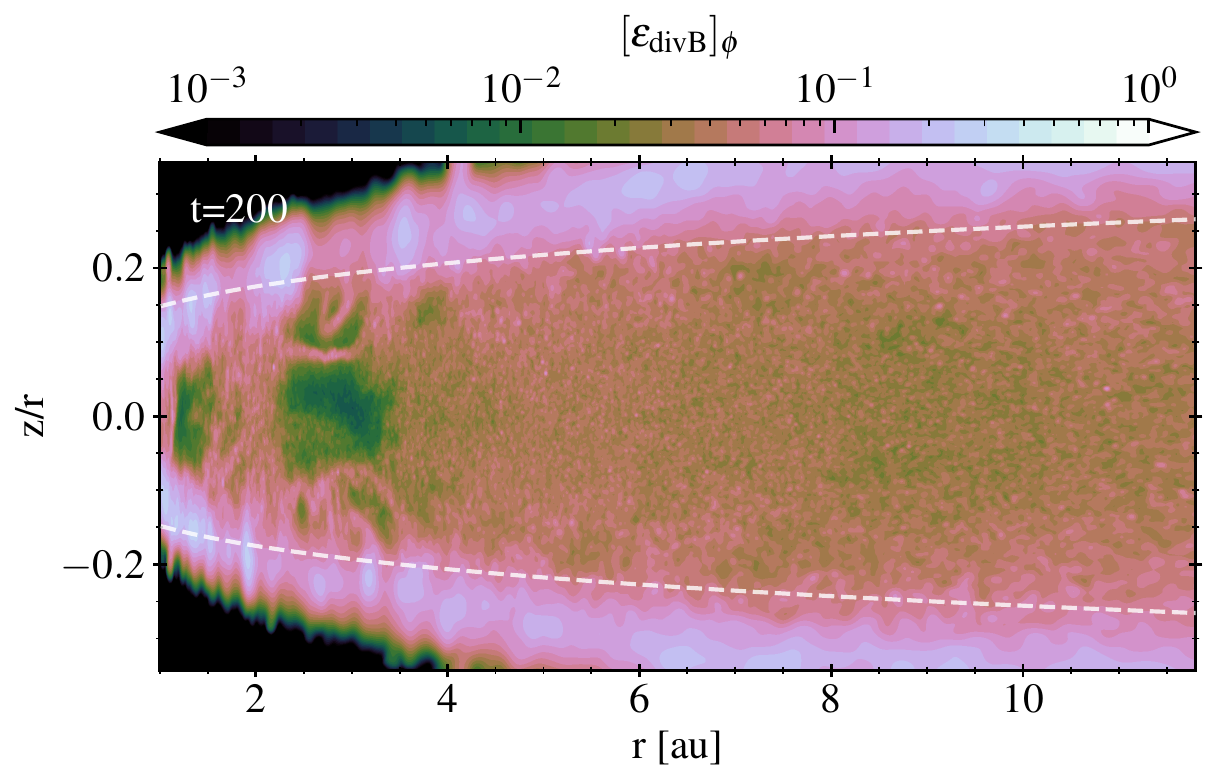}{\columnwidth}{(a)}}
\gridline{\fig{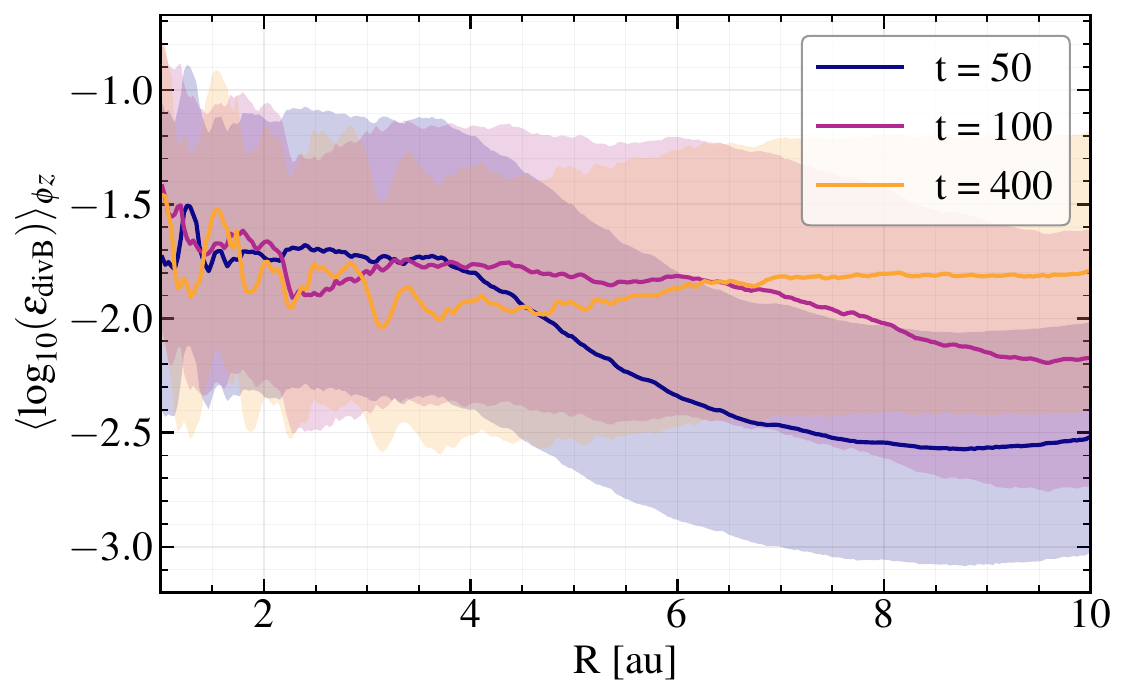}{\columnwidth}{(b)}}
\caption{(a) Azimuthally averaged divergence error in the meridional plane at $t=200\,\mathrm{yr}$. The white dashed lines mark $|z|=3H$. Inside the $\zeta$ region, the error is of order $10^{-2}$--$10^{-1}$. (b) Divergence error as a function of radius at $t=50$, 100, and $400\,\mathrm{yr}$. Solid lines show the geometric mean over cylindrical shells, and shaded regions show one standard deviation.}
\label{fig:error_divB}
\end{figure}

In SPMHD, the divergence-free condition is not satisfied to machine precision. The most common approach to date is to control the divergence error with the hyperbolic/parabolic divergence cleaning scheme \citep{tricco_constrained_2012, tricco_constrained_2016}. To ensure that this error remains small, we monitor it throughout the evolution of the disc with the standard dimensionless metric
\begin{equation}
    \epsilon_{\rm divB} = h \frac{|\nabla \cdot \mathbf{B}|}{|\mathbf{B}|},
\end{equation}
which compares the divergence of the field with the field strength on the scale of the smoothing length.

Figure~\ref{fig:error_divB}(a) shows the azimuthally averaged divergence error in the meridional plane at $t=200\,\mathrm{yr}$. Within $|z|\leq3H$, the error lies between $10^{-2}$ and $10^{-1}$, with typical values of a few $\times10^{-2}$ in the turbulent outer disc. Above $3H$, it increases to $\epsilon_{\rm divB}\gtrsim10^{-1}$, mainly because of edge effects in the disc atmosphere. Figure~\ref{fig:error_divB}(b) shows the radial profiles of the divergence error at $t=50$, 100, and $400\,\mathrm{yr}$. At $t=50$ and $100\,\mathrm{yr}$, the error drops at the radius where the gas has completed six orbits, which marks the edge of the turbulent region. Once turbulence sets in, the error rises to $\epsilon_{\rm divB}\approx2\times10^{-2}$. By $t=400\,\mathrm{yr}$, the mean error is $\epsilon_{\rm divB}\approx1$--$2\times10^{-2}$ at all radii, and in the inner disc, where turbulence developed first, it does not increase between $t=50$ and $400\,\mathrm{yr}$. Suppressing divergence error to the $\sim1\%$ level has been found sufficient for most practical applications \citep{tricco_smoothed_2023}. The divergence error grows as MRI turbulence develops and then saturates. At all times shown, the mean error stays below $10^{-1}$, as does the upper edge of the one-standard-deviation band at most radii.

Our errors are comparable to those reported in SPMHD simulations with divergence cleaning: of order $10^{-2}$ for the MRI in shearing boxes \citep{wissing_magnetorotational_2022} and $10^{-2}$--$10^{-1}$ for galactic dynamos \citep{wissing_numerical_2023}. The coherent toroidal rings in the inner disc do not coincide with an increase in the divergence error, although a more detailed study is needed to determine whether the error affects the evolution of these regions directly. Stronger cleaning lowers the error but may add numerical dissipation. In moving-mesh shearing boxes with Dedner cleaning, zero-net-flux MRI turbulence decays above a threshold cleaning strength \citep{zier_simulating_2022}. Increasing $f_\mathrm{clean}$, and hence the cleaning speed, by a factor of 10 typically leads to the divergence error being reduced by a factor of 2, but incurs $10\times$ computational expense for the simulation. This was not affordable at the resolutions used in this study.

\subsection{Dynamo signatures}\label{sub:dynamo}

\begin{figure}[t!]
\centering
\includegraphics[width=\columnwidth]{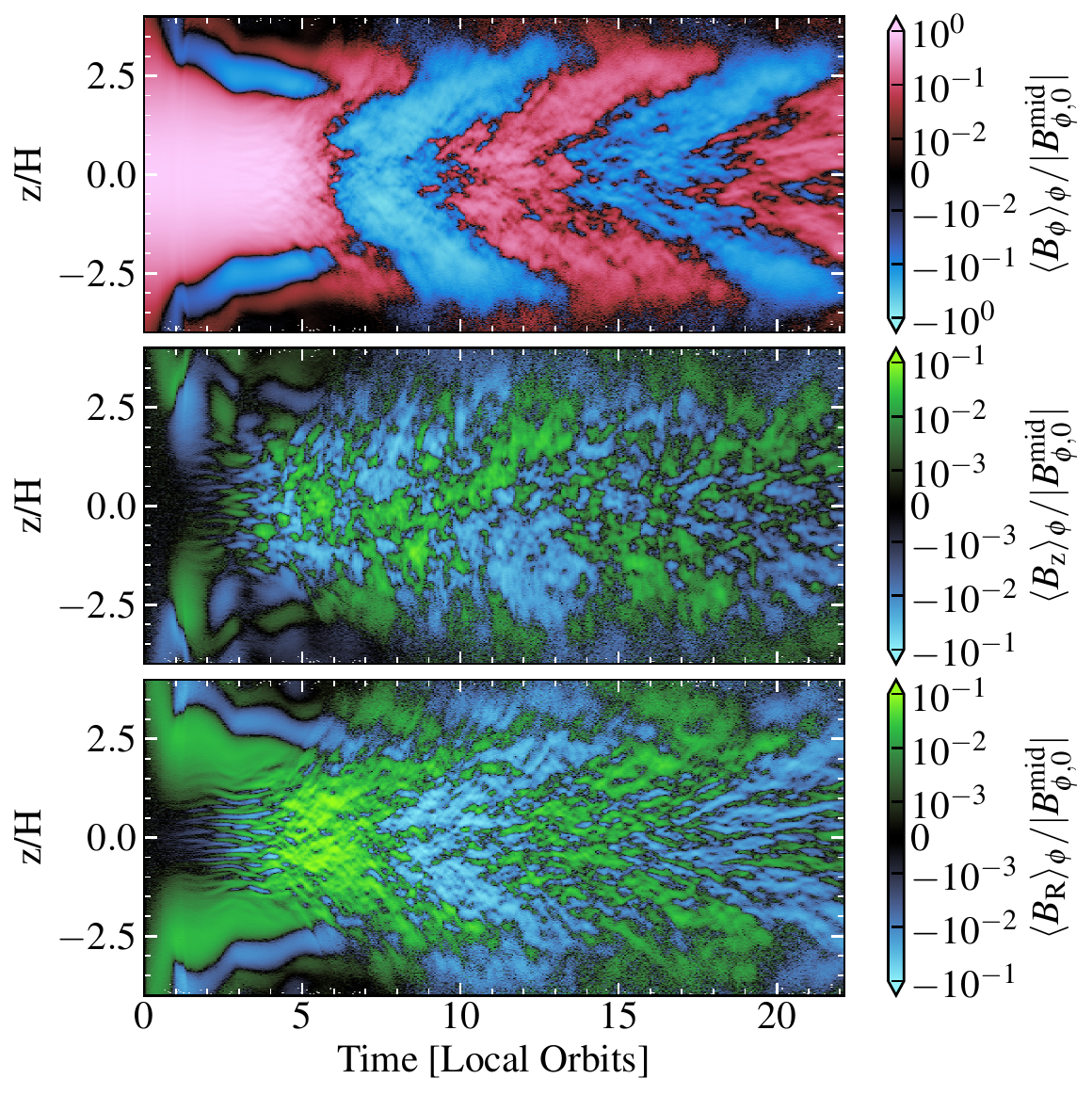}
\caption{Butterfly diagram for \texttt{S102M\_2aB2}. Azimuthally averaged magnetic field components at $R = 7\,\mathrm{au}$, within a cylindrical shell of width $\Delta R = 0.1\,\mathrm{au}$, as a function of height and time. \textit{Top}: $\langle B_\phi \rangle_\phi$. \textit{Middle}: $\langle B_z \rangle_\phi$. \textit{Bottom}: $\langle B_R \rangle_\phi$. All components are normalized by the initial midplane toroidal field $|B_{\phi, 0}^\mathrm{mid}|$, and time is given in local orbits.}
\label{fig:butterfly_diagram}
\end{figure}

In vertically stratified MRI turbulence, the mean toroidal field changes polarity quasi-periodically \citep{brandenburg_dynamo-generated_1995}, and the reversals migrate from the midplane towards the disc surfaces. In a space--time representation, this produces the characteristic \textit{butterfly diagram}, with a cycle period of $\sim10$ orbits in both local \citep{davis_sustained_2010, deng_local_2019, wissing_magnetorotational_2022} and global simulations \citep{flock_turbulence_2011, flock_large-scale_2012}.

Figure~\ref{fig:butterfly_diagram} shows this diagram at $R = 7\,\mathrm{au}$, within a thin cylindrical shell of width $\Delta R = 0.1\,\mathrm{au}$, as a function of height and time. After the first reversal at $\approx 6$ local orbits, the polarity of the toroidal field alternates with a cycle period of $\approx10$ orbits, consistent with previous work. The reversals start near the midplane and migrate towards larger $|z|$. The same cycle is present in the radial component, although less sharply defined, while the vertical one is the weakest and nearly antisymmetric about the midplane. The field strength decreases significantly above $|z|\approx3.5H$, but this may be partly a numerical effect in these low-density regions. Since the resolution follows the density, the number of particles drops steeply in the atmosphere, where the MRI is likely under-resolved ($Q_\phi\lesssim1$; Figure~\ref{fig:quality_factor_disc}) and the divergence error is larger (Figure~\ref{fig:error_divB}a). Therefore, we interpret the field evolution there with caution and base our analysis mostly on the disc body ($|z|\lesssim3H$), where the dynamo signatures are clearest.

\begin{figure}[t!]
\centering
\includegraphics[width=\columnwidth]{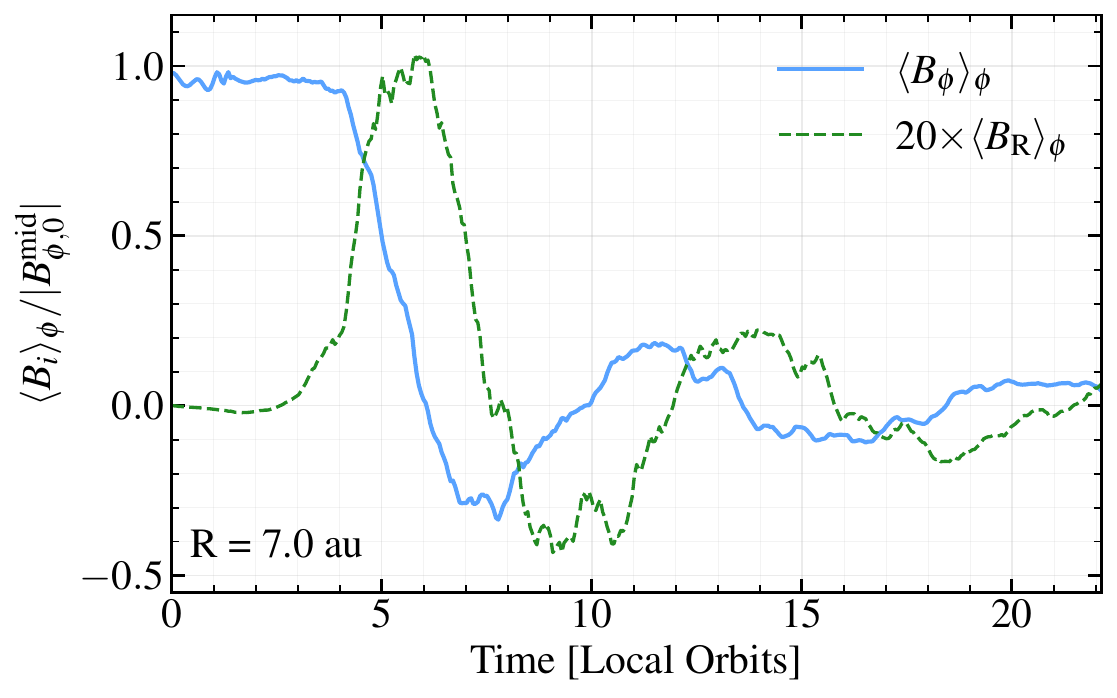}
\caption{Time evolution of the toroidal (solid line) and radial (dashed line) magnetic field components at $R = 7\,\mathrm{au}$, averaged azimuthally and over $|z| \leq H$, and normalized by $|B_{\phi, 0}^\mathrm{mid}|$. The radial component is multiplied by 20 for a better comparison with the toroidal field.}
\label{fig:bmid_field}
\end{figure}

Figure~\ref{fig:bmid_field} shows the toroidal and radial fields averaged near the midplane ($|z| \leq H$) at the same radius as in Figure~\ref{fig:butterfly_diagram}. The two components oscillate out of phase. This phase offset may indicate an $\alpha\Omega$ mean-field dynamo, in which the disc shear generates toroidal field from the radial field, while the turbulent electromotive force regenerates the radial component from the toroidal one \citep{guan_radially_2011, simon_resistivity-driven_2011}. The amplitude of the oscillations decreases over successive cycles, to $\approx10\%$ of $|B_{\phi, 0}^\mathrm{mid}|$ after $\sim20$ local orbits, consistent with the decay of the magnetic energy shown in Figure~\ref{fig:stress_evol}.

\begin{figure}[t!]
\centering
\includegraphics[width=\columnwidth]{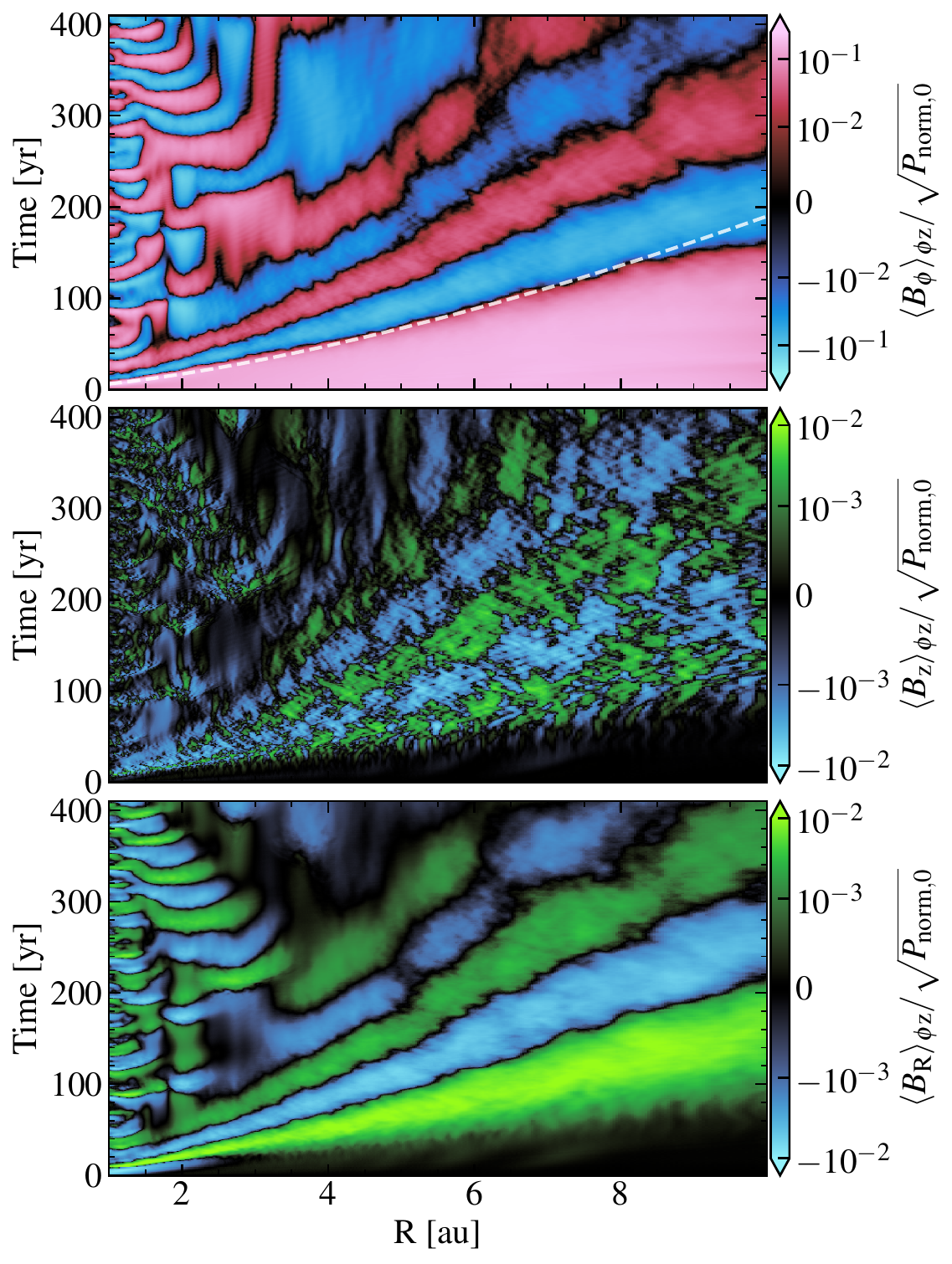}
\caption{Magnetic field components as a function of radius and time, averaged azimuthally and over $0<z<2H$, and normalized by $\sqrt{P_\mathrm{norm,0}}$. From top to bottom: $B_\phi$, $B_z$, and $B_R$. The white dashed line in the top panel marks the time at which the gas at each radius has completed six orbits.}
\label{fig:mean_field}
\end{figure}

Figure~\ref{fig:mean_field} shows the mean field components as a function of radius and time, averaged azimuthally and from the midplane up to two scale heights in the northern hemisphere ($0<z<2H$). The toroidal field first reverses after about six local orbits at all radii (white dashed line). In the outer disc, the period lengthens after $\sim3$ cycles, which coincides with a decrease in the poloidal components. Although the inner disc is dominated by artificial dissipation, all three mean field components there still change sign repeatedly. These reversals propagate outwards, but they do not extend beyond $R\approx3\,\mathrm{au}$.

\subsection{The effect of resolution} \label{subsec:the_effect_of_resolution}

In Section~\ref{sub:overall_disc_evolution}, we found that the inner region of the disc is dominated by numerical dissipation. Here, we assess how this dissipation affects the disc dynamics and how the growth of MRI depends on resolution, using simulations with $N_\mathrm{P} = 20$, 40, 60, and $102\times10^6$ particles. To first order, the numerical viscosity and resistivity are proportional to the smoothing length, and from Equation~\eqref{SPMHD:density_h}, $h \propto N_\mathrm{P}^{-1/3}$. Therefore, halving the numerical dissipation requires eight times more particles, so between our lowest- and highest-resolution models it is expected to decrease by only $\sim40\%$.

\begin{figure}[t!]
\centering
\includegraphics[width=\columnwidth]{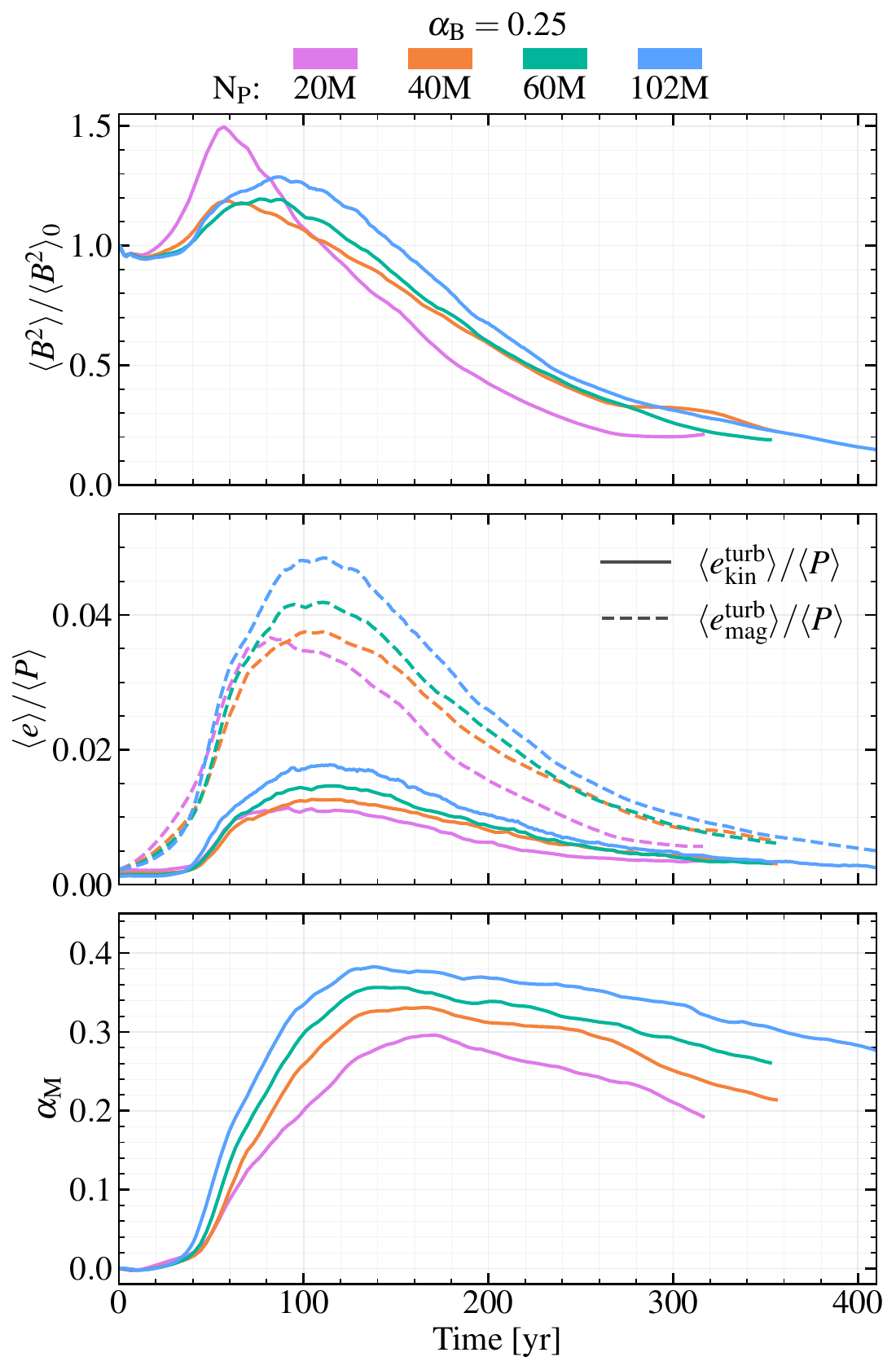}
\caption{Time evolution of the magnetic energy (top), the turbulent kinetic and magnetic energies (middle), and the normalized Maxwell stress (bottom) for $N_\mathrm{P}=20$, 40, 60, and $102\times10^6$, with $\alpha_\mathrm{B}=0.25$.}
\label{fig:_high_resolution}
\end{figure}

Figure~\ref{fig:_high_resolution} shows that the turbulent energies and the normalized Maxwell stress increase with resolution. The initial amplification of the magnetic field comes mostly from the toroidal component. The magnetic energy peaks at $\approx1.5$ times its initial value for $N_\mathrm{P}=20\times10^6$, compared with $\approx1.2$--$1.3$ at higher resolutions. The turbulent magnetic energy of this model, however, is the lowest of all, so its stronger amplification is mostly in the mean field. It also loses its magnetic energy fastest, and after $t\approx270\,\mathrm{yr}$ it levels off at $\approx0.2$, close to the values that the other models reach by the end of their runs. The normalized Maxwell stress grows faster at higher resolution, with peaks rising from $\alpha_\mathrm{M}\approx0.30$ for $N_\mathrm{P}=20\times10^6$ to $\approx0.38$ for $102\times10^6$. After reaching the maximum, $\alpha_\mathrm{M}$ decreases more steeply at lower resolution.

The largest differences during the growth phase of the MRI occur for the $N_\mathrm{P}=20\times10^6$ model, which also has $\langle Q_{\phi,0}\rangle^{\rm V}<20$. In the models with higher quality factors, zones dominated by numerical dissipation are restricted to the inner disc during the first $\sim200\,\mathrm{yr}$, which limits their influence on the growth phase within $\zeta$. However, this early stage of the disc shows no sign of convergence at the resolutions used here.

\begin{figure}[t!]
\centering
\includegraphics[width=\columnwidth]{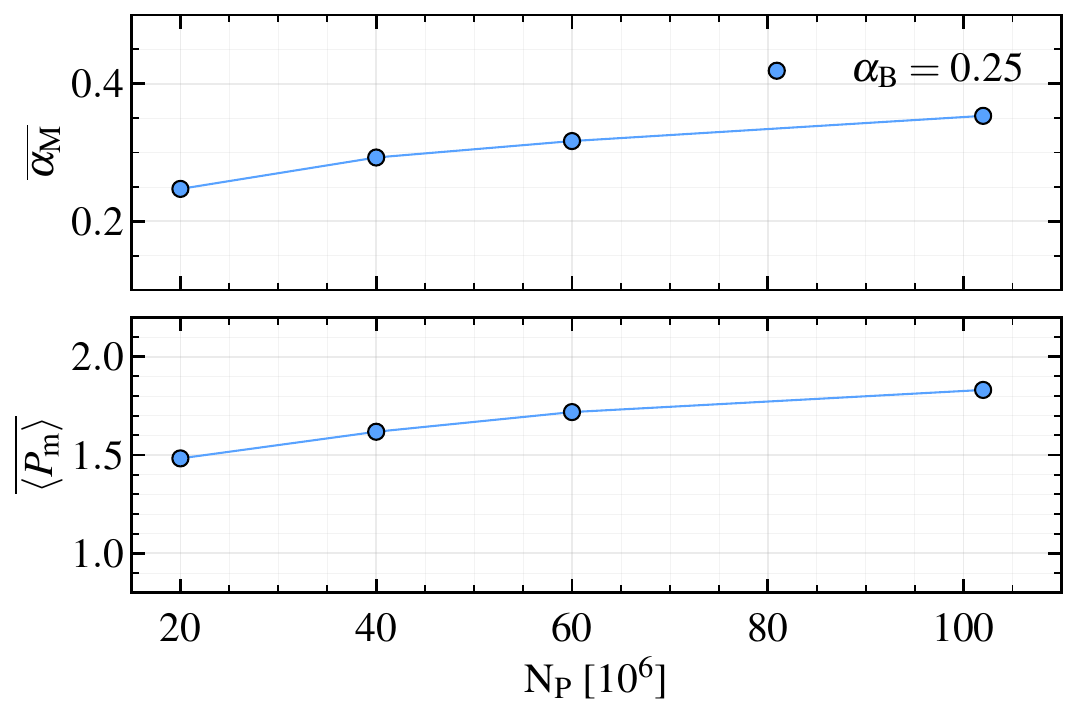}
\caption{Normalized Maxwell stress (\textit{top}) and magnetic Prandtl number (\textit{bottom}) as a function of the number of particles for $\alpha_\mathrm{B} = 0.25$, averaged between 200 and $300\,\mathrm{yr}$.}
\label{fig:am_pm_npart}
\end{figure}

Figure~\ref{fig:am_pm_npart} shows the normalized Maxwell stress and the magnetic Prandtl number as a function of resolution for $\alpha_\mathrm{B}=0.25$. From $N_\mathrm{P}=20\times10^6$ to $102\times10^6$, $\alpha_\mathrm{M}$ increases monotonically from 0.25 to 0.35, and $P_\mathrm{m}$ from 1.48 to 1.83. \citet{wissing_magnetorotational_2022} also found an initial increase of the normalized Maxwell stress with resolution in local stratified shearing boxes.

\begin{figure}[t!]
\centering
\includegraphics[width=\columnwidth]{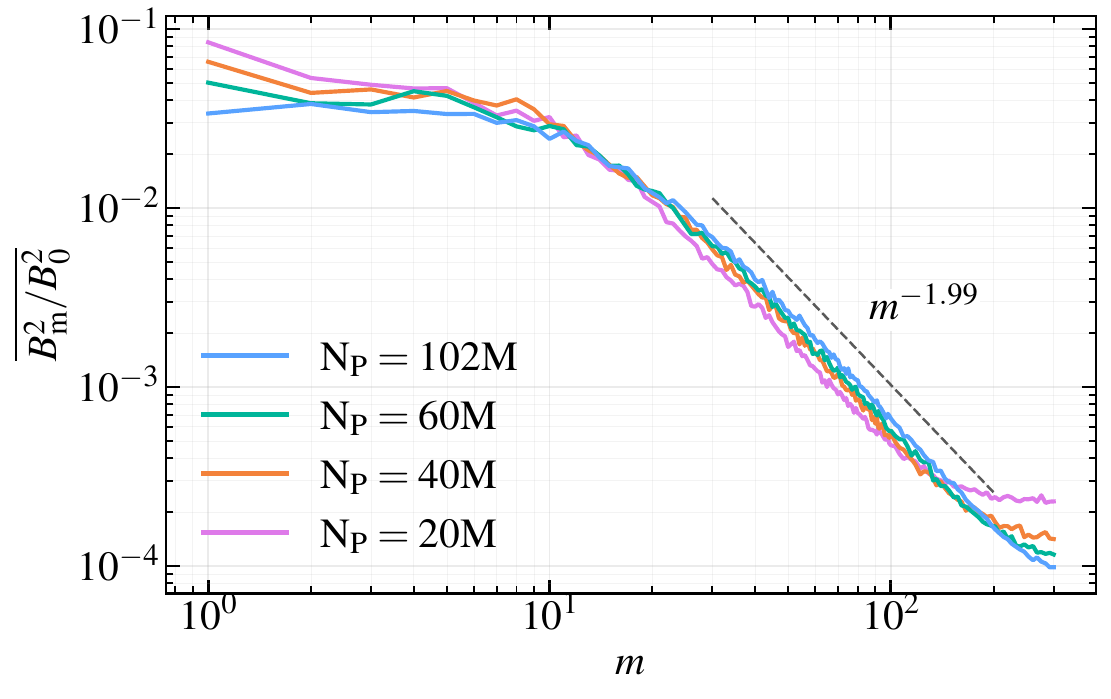}
\caption{Azimuthal spectrum of the magnetic energy density, $B_m^2/B_0^2$, for $N_\mathrm{P} = 20$, 40, 60, and $102\times10^6$ particles with $\alpha_\mathrm{B}=0.25$, averaged over annuli between $R = 5$ and $9\,\mathrm{au}$ in the $\zeta$ region and over $t = 200$--$300\,\mathrm{yr}$. The dashed line shows the $m^{-1.99}$ scaling from a power-law fit to the $N_\mathrm{P}=102\times10^6$ spectrum over $30 \leq m \leq 200$.}
\label{fig:spectrum}
\end{figure}

We also compute the azimuthal spectrum of the magnetic energy density. For this, the disc is divided into annuli of width $\Delta R = 0.5\,\mathrm{au}$ between $R = 5$ and $9\,\mathrm{au}$, using only particles within $\zeta$. In each annulus, Equation~\eqref{eq:spectrum} gives the energy in each azimuthal mode, $B_m^2$. We normalize each spectrum by its local $B_0^2$, so that $B_m^2/B_0^2$ measures the energy in mode $m$ relative to the azimuthal mean ($m=0$), independently of the radial decline of the field. Finally, we average the spectra radially and over $t = 200$--$300\,\mathrm{yr}$. Equation~\eqref{eq:spectrum} describes the azimuthal distribution of the magnetic energy density, so comparisons with the power spectra of the field components used in grid-based studies \citep[e.g.][]{flock_turbulence_2011} are only qualitative.

Figure~\ref{fig:spectrum} shows the resulting spectra. For $30 \lesssim m \lesssim 100$, the models approximately follow an $m^{-2}$ scaling. A power-law fit to the $N_\mathrm{P}=102\times10^6$ spectrum over $30 \leq m \leq 200$ gives $B_m^2 \propto m^{-1.99\pm 0.01}$. The scaling is steeper than $m^{-5/3}$, which is in agreement with \citet{flock_turbulence_2011} and \citet{deng_global_2020}. The main effect of resolution appears at high $m$, where the spectra flatten. This flattening starts at larger $m$ and at a lower amplitude as the number of particles increases: for $N_\mathrm{P} = 20\times10^6$ it appears at $m \approx 150$, while for $102\times10^6$ the spectrum keeps declining up to the highest modes shown. Higher resolution (i.e. a smaller smoothing length) extends the power-law range to smaller azimuthal scales.

\subsection{The effect of artificial dissipation} \label{subsec:the_effect_of_dissipation}

The numerical dissipation throughout the disc plays an important role in the amplification of magnetic fields and in the development of the dynamo. In ideal SPMHD, this dissipation comes mainly from the shock-capturing terms, namely the artificial viscosity and the artificial resistivity (Equations~\eqref{SPMHD:shock_AV} and \eqref{SPMHD: art resistivity}, respectively). The artificial viscosity coefficient, $\alpha_\mathrm{AV}$, is set by the switch of \citet{cullen_inviscid_2010}, which detects shocks and reduces $\alpha_\mathrm{AV}$ away from them to avoid excessive dissipation. For the artificial resistivity, the signal velocity itself plays this role, so $\alpha_\mathrm{B}$ only sets its overall amplitude. We let $\alpha_\mathrm{AV}$ vary between 0 and 1 and ran models with constant values of $\alpha_\mathrm{B} = 1$, 0.5, and $0.25$. For reference, \textsc{Phantom} \citep{price_phantom_2018} and \textsc{Shamrock} \citep{david-cleris_shamrock_2025} use $\alpha_\mathrm{B} = 1$ by default, while \textsc{Gasoline2} \citep{wadsley_gasoline2_2017, wissing_smoothed_2020} uses $\alpha_\mathrm{B} = 0.5$. Since $\eta_\mathrm{AR} \propto \alpha_\mathrm{B}$, halving $\alpha_\mathrm{B}$ halves the artificial resistivity. Values of $\alpha_\mathrm{B}$ that are too small, however, may result in a poor treatment of discontinuities and generate noise in the magnetic field. Additional standard MHD tests that we performed show that $\alpha_\mathrm{B} = 0.25$ still treats shocks and discontinuities adequately. Similar values have also been used in previous SPMHD studies \citep{wissing_magnetorotational_2022, wissing_numerical_2023}, where the lower artificial resistivity led to higher magnetic Prandtl numbers and stronger amplification of the magnetic field.

\begin{figure}[t!]
\centering
\includegraphics[width=\columnwidth]{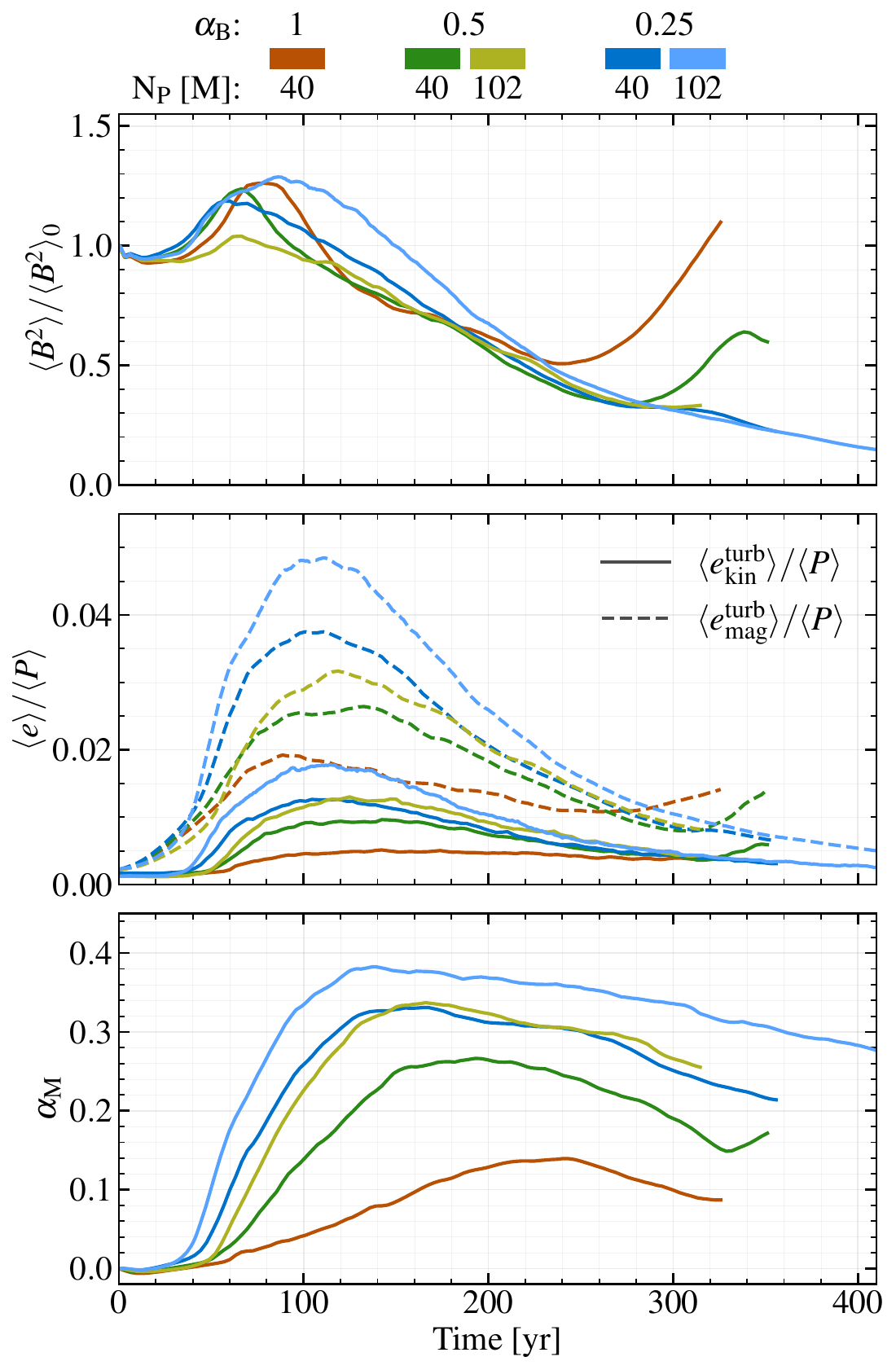}
\caption{Time evolution of the magnetic energy (top), the turbulent kinetic and magnetic energies (middle), and the normalized Maxwell stress (bottom) for different values of $\alpha_\mathrm{B}$ and resolutions of $N_\mathrm{P}=40$ and $102\times10^6$.}
\label{fig:_evol_disc_vs_alphaB}
\end{figure}

Figure~\ref{fig:_evol_disc_vs_alphaB} shows the magnetic energy, turbulent energies, and normalized Maxwell stress for the $N_\mathrm{P}=40$ and $102\times10^6$ particle calculations with $\alpha_\mathrm{B}$ varied. The turbulent energies and stresses increase as $\alpha_\mathrm{B}$ decreases. For $\alpha_\mathrm{B}=1$, the turbulent energies are the lowest of all models. The magnetic field remains largely laminar and organized in coherent rings, and the dynamo acts only through reversals of the mean field. These reversals are related to the growth of the magnetic energy at $t\gtrsim250\,\mathrm{yr}$, which occurs mostly in the mean field. The disc with $\alpha_\mathrm{B}=0.5$ and $N_\mathrm{P}=40\times10^6$ shows a similar but weaker late-time rise. For the $N_\mathrm{P}=40\times10^6$ calculations, the peak of $\alpha_\mathrm{M}$ rises from $\approx0.14$ for $\alpha_\mathrm{B}=1$ to $\approx0.33$ for $\alpha_\mathrm{B}=0.25$. The latter has a normalized Maxwell stress similar to that of the $N_\mathrm{P}=102\times10^6$ calculation with $\alpha_\mathrm{B}=0.5$ from the peak onward.

\begin{figure}[t!]
\centering
\includegraphics[width=\columnwidth]{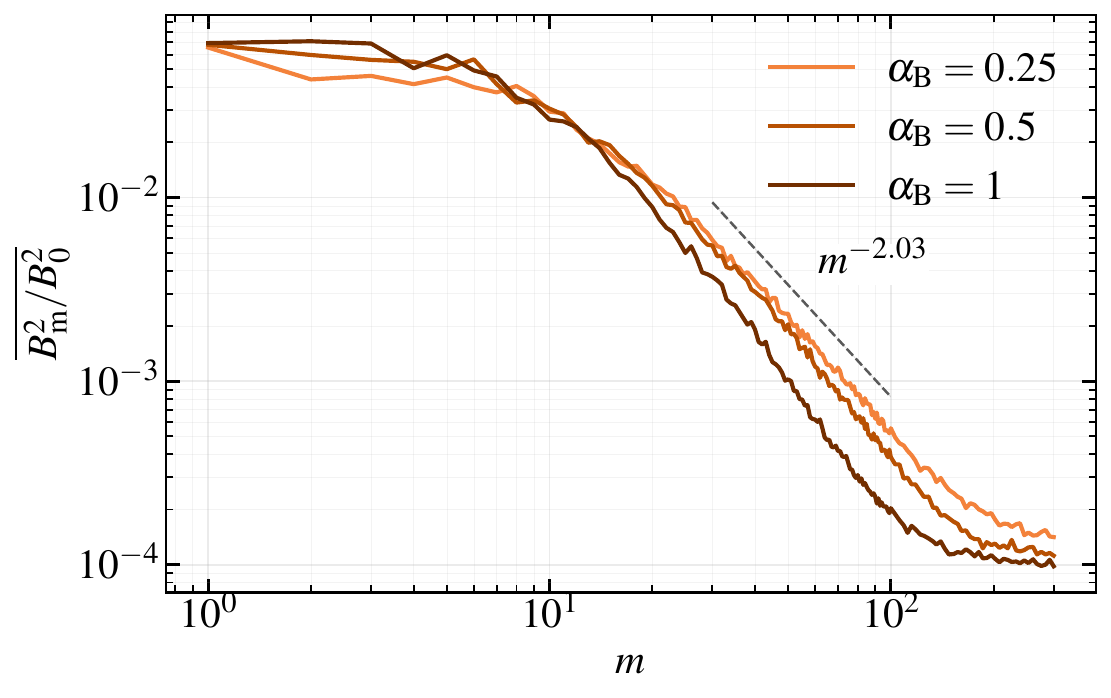}
\caption{Azimuthal spectrum of the magnetic energy density for $\alpha_\mathrm{B}=0.25$, 0.5, and 1 and $N_\mathrm{P} = 40\times10^6$. The dashed line shows the $m^{-2.03}$ scaling from a power-law fit to the $\alpha_\mathrm{B}=0.25$ spectrum over $30 \leq m \leq 100$.}
\label{fig:spectrum_alphaB}
\end{figure}

Figure~\ref{fig:spectrum_alphaB} shows the azimuthal spectra for $N_\mathrm{P}=40\times10^6$ and $\alpha_\mathrm{B}=0.25$, 0.5, and 1. As $\alpha_\mathrm{B}$ decreases, the disc becomes more turbulent and develops smaller-scale structures, and the amplitude of the high-$m$ modes increases. The spectrum also approaches an $m^{-2}$ scaling and follows it over a wider range of $m$. A power-law fit to the $\alpha_\mathrm{B}=0.25$ spectrum over $30\leq m \leq 100$ gives $B_m^2 \propto m^{-2.03\pm 0.02}$.

Table~\ref{tab:sims} lists additional diagnostic quantities for the different disc models. From these values, we observe that reducing the magnetic dissipation through $\alpha_\mathrm{B}$ is a more efficient way to reach higher $\alpha_\mathrm{M}$ and magnetic Prandtl numbers than increasing the resolution. This holds as long as $\alpha_\mathrm{B}$ remains large enough to treat shocks and discontinuities adequately and the characteristic MRI wavelength is well resolved ($\langle Q_{\phi,0}\rangle^{\rm V}\gtrsim20$). On the other hand, $\alpha_\mathrm{B} = 1$ is too dissipative for MRI turbulence to develop at $N_\mathrm{P}\lesssim40\times10^6$.

\subsection{Long-term disc dynamics}\label{sub:long_term_dynamics}

\begin{figure}[t!]
\centering
\includegraphics[width=\columnwidth]{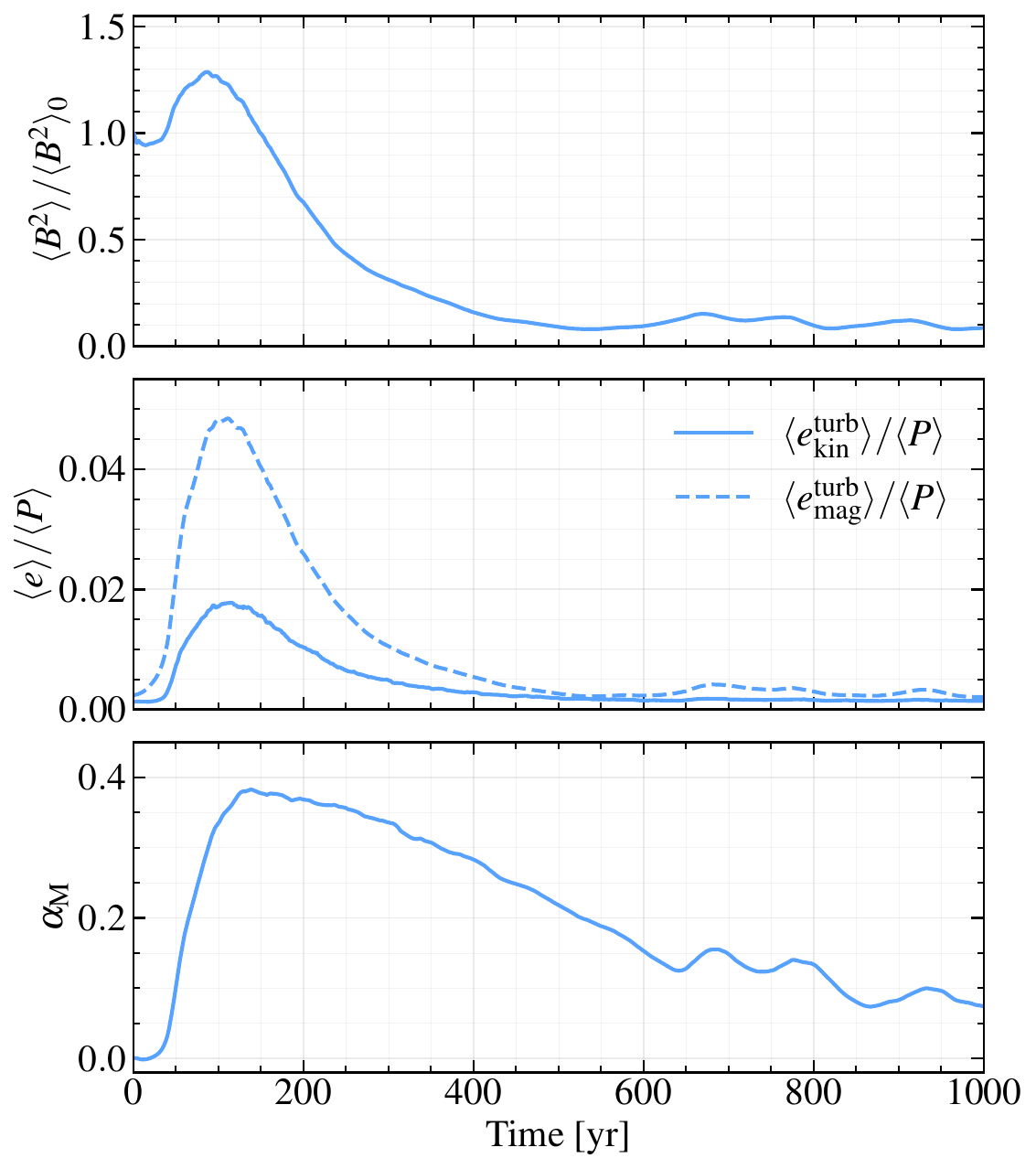}
\caption{Long-term evolution of the magnetic energy (top), the turbulent kinetic and magnetic energies (middle), and the normalized Maxwell stress (bottom) for model \texttt{S102M\_2aB2} up to $t = 1000\,\mathrm{yr}$.}
\label{fig:long_term_evolution}
\end{figure}

Figure~\ref{fig:long_term_evolution} shows the evolution of the magnetic energy, the turbulent energies, and the normalized Maxwell stress of model \texttt{S102M\_2aB2} up to $t=1000\,\mathrm{yr}$. The magnetic energy continues the decay described in Section~\ref{sub:overall_disc_evolution} until $t\sim500\,\mathrm{yr}$, and then fluctuates around $10\%$ of its initial value. The turbulent energies drop over the same period, to values close to those at $t=0$. The normalized Maxwell stress then oscillates between $\approx0.07$ and $0.15$, while still decreasing slowly. At late times, the dynamo acts only through reversals of the coherent mean field, rather than through the turbulent cycle observed during the first $400\,\mathrm{yr}$. These reversals also occur with a longer period.

\begin{figure}[t!]
\centering
\includegraphics[width=\columnwidth]{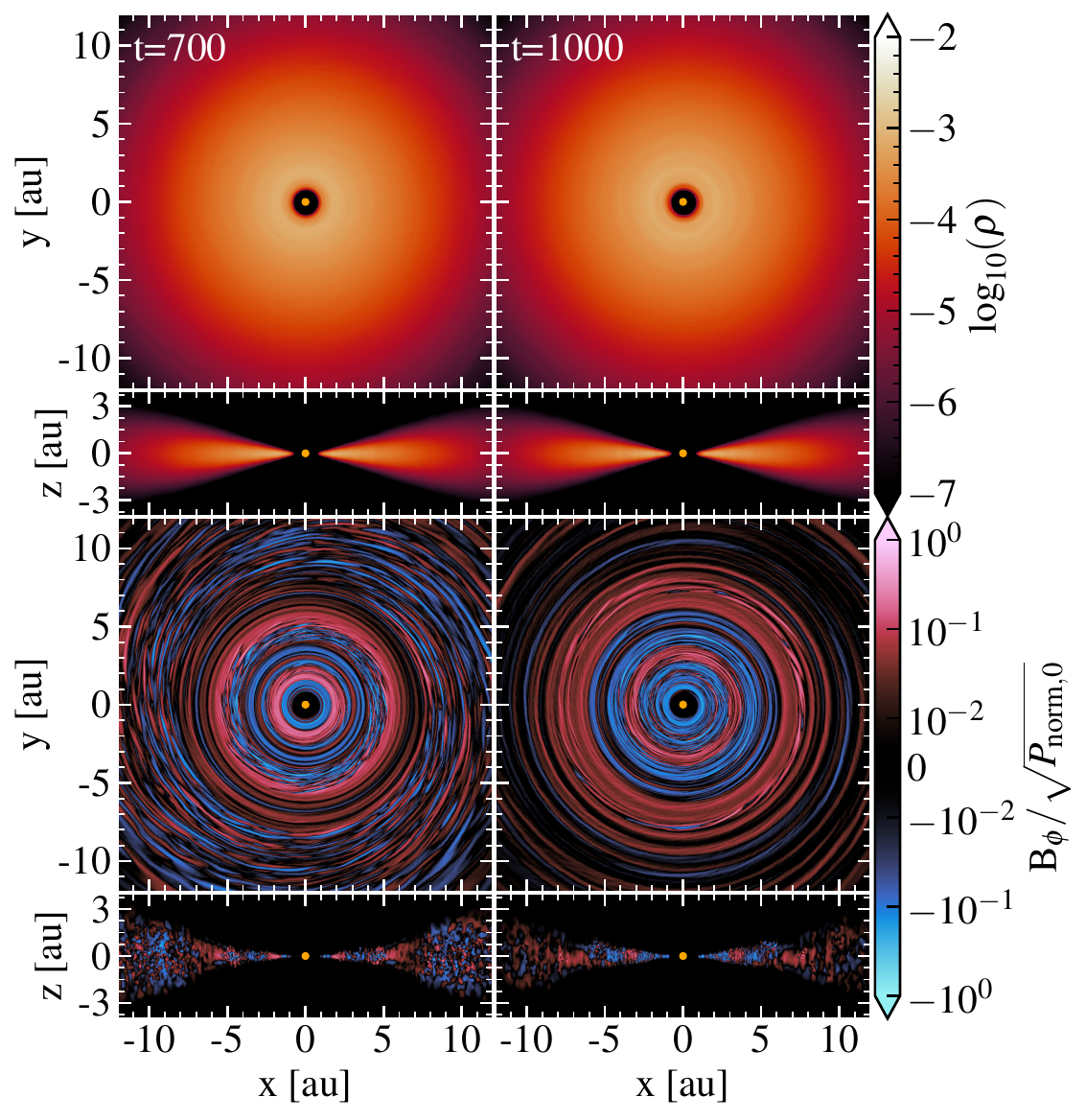}
\caption{Snapshots of the density (top) and toroidal magnetic field (bottom) at $t=700$ (left) and $1000\,\mathrm{yr}$ (right) for model \texttt{S102M\_2aB2}. For each quantity, the upper panel shows the disc midplane and the lower panel a meridional cross-section.}
\label{fig:long_term_mid}
\end{figure}

Figure~\ref{fig:long_term_mid} shows the density and toroidal magnetic field at $t=700$ and $1000\,\mathrm{yr}$, when the disc has largely lost its MRI-driven turbulence. By this time, the spiral perturbations in the density have been replaced by annular mass accumulations in the inner disc and a smooth profile in the outer regions. The toroidal field is mostly organized in coherent bands of alternating polarity, which are wider than at earlier times (Figure~\ref{fig:disc_evolution}), and weak small-scale fluctuations remain only at the outer radii.

The loss of turbulence is consistent with the low magnetic Reynolds number of the disc, $R_\mathrm{m} \sim 1000$ (Figure~\ref{fig:numbers_r}). In global simulations with an initial toroidal field and explicit Ohmic resistivity, \citet{flock_turbulence_2012} found saturated MRI turbulence only for $R_\mathrm{m} \gtrsim 5000$, damped turbulence for $3300 \lesssim R_\mathrm{m} \lesssim 5000$, and no sustained MRI below $R_\mathrm{m} \approx 3000$. Close to this critical value, the accretion stress showed long-period oscillations, which they tentatively linked to a dynamo process. In our model, the Maxwell stress also oscillates at late times, and the reversals of the coherent toroidal field are correlated with those of the radial field. In contrast, \citet{parkin_global_2013} obtained sustained turbulence in global simulations with numerical magnetic Reynolds numbers between $\approx 770$ and $2100$, as long as the azimuthal field was resolved, although the magnetic energy decreased slowly. They attributed the difference to two factors. First, their resistivity is numerical and was estimated assuming an Ohmic form, rather than being an explicit Ohmic term. Second, their third-order spatial scheme may allow turbulence to survive at lower $R_\mathrm{m}$ than the second-order scheme of \citet{flock_turbulence_2012}. Since our $R_\mathrm{m}$ is also set by numerical dissipation, a direct comparison with either study is restricted. Our results nonetheless share features with both. In the outer disc, where $R_\mathrm{m}\sim1000$, the turbulence weakens slowly and lasts longer, as in \citet{parkin_global_2013}. In the inner disc, where $R_\mathrm{m}\lesssim700$, it fades more quickly. Eventually, the turbulence decays at all radii, as expected from the threshold of \citet{flock_turbulence_2012}.

\subsection{The effect of gradient discretization choice} \label{sub:gradient_choice}

We also study the evolution of the disc under the standard SPMHD implementation in \textsc{Phantom}, which sets $\xi = \rho$ in Equations~\eqref{gradient:symmetric} and \eqref{gradient:antisymmetric} for the gradient discretization in the magnetic terms. For reference, we include model \texttt{P60M\_2aB2}, which uses $\xi = 1$, corresponding to the geometric density weighting adopted in the rest of this work.

\begin{figure}[t!]
\centering
\includegraphics[width=\columnwidth]{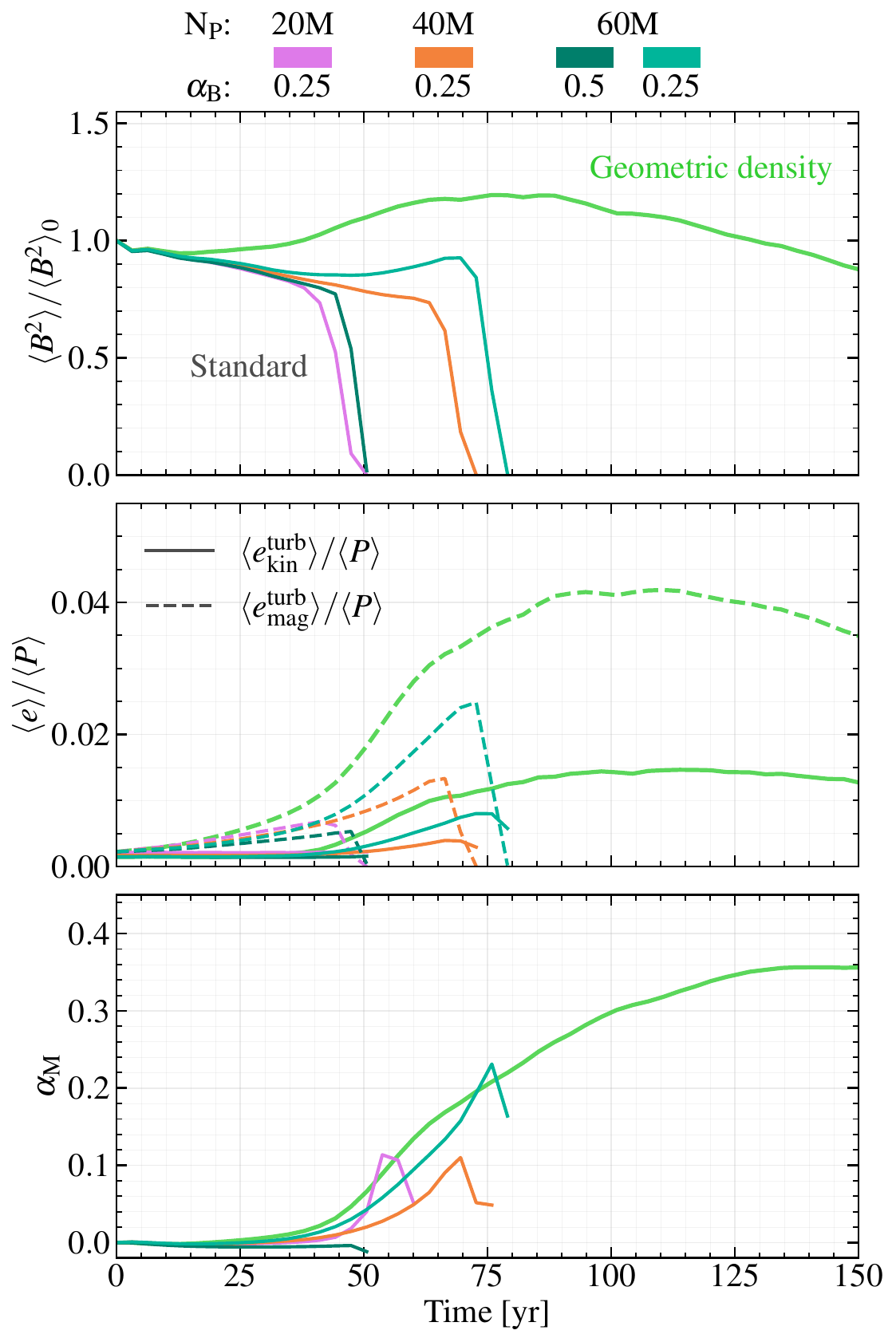}
\caption{Time evolution of the magnetic energy (top), the turbulent kinetic and magnetic energies (middle), and the normalized Maxwell stress (bottom) for models with the standard weighting ($\xi=\rho$). The light green line shows model \texttt{P60M\_2aB2}, which uses the geometric density weighting ($\xi=1$).}
\label{fig:Standard_SPMHD_evolution}
\end{figure}

Figure~\ref{fig:Standard_SPMHD_evolution} shows that, in all the standard models, the magnetic energy is abruptly dissipated within the first $\approx80\,\mathrm{yr}$, and the turbulent energies and the Maxwell stress drop with it. This sudden loss occurs later for lower numerical dissipation, from $\approx50\,\mathrm{yr}$ in the most dissipative models to $\approx80\,\mathrm{yr}$ for $N_\mathrm{P}=60\times10^6$ with $\alpha_\mathrm{B}=0.25$. Before then, the magnetic energy decreases almost steadily from the start, although the turbulent energies and the Maxwell stress start to rise, indicating that the MRI begins to develop before the field is lost. In contrast, the magnetic energy in \texttt{P60M\_2aB2} first increases by $\approx20\%$ and stays close to its initial value over the time shown, while the turbulence grows as in previous models presented in Sections~\ref{sub:overall_disc_evolution}, \ref{subsec:the_effect_of_resolution}, and \ref{subsec:the_effect_of_dissipation}.

\begin{figure}[t!]
\centering
\includegraphics[width=\columnwidth]{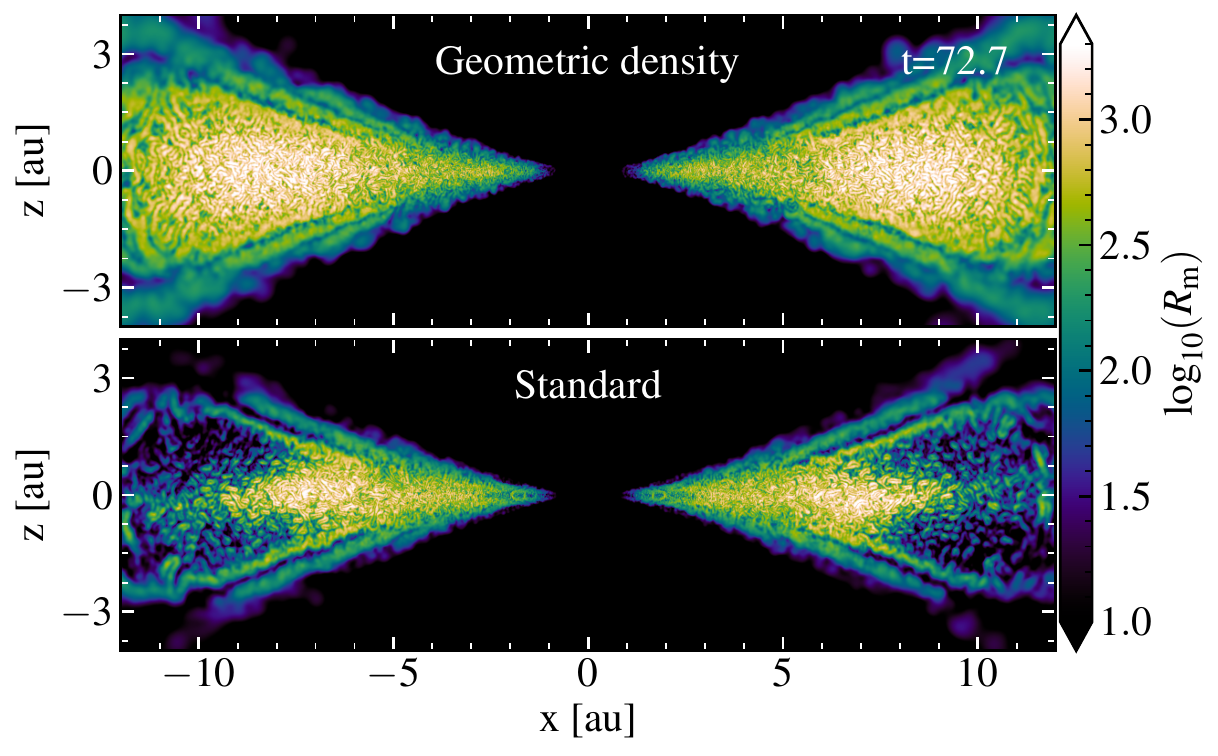}
\caption{Magnetic Reynolds number in vertical slices at $t = 72.7\,\mathrm{yr}$ with the geometric density weighting (\texttt{P60M\_2aB2}, \textit{top}) and the standard weighting (\texttt{stdP60M\_2aB2}, \textit{bottom}).}
\label{Fig:Reynolds_StdSPMHD}
\end{figure}

Figure~\ref{Fig:Reynolds_StdSPMHD} shows vertical slices of the magnetic Reynolds number at $t=72.7\,\mathrm{yr}$ for \texttt{P60M\_2aB2} and for its counterpart with the standard weighting, \texttt{stdP60M\_2aB2}. In \texttt{stdP60M\_2aB2}, which loses its field shortly afterwards, $R_\mathrm{m}$ is much smaller in the outer regions and in the atmosphere, where the density is orders of magnitude lower than in the inner disc.

We attribute the low $R_\mathrm{m}$ and the loss of magnetic energy in the standard models to the higher dissipation of this weighting across steep density gradients. To illustrate this, consider two nearby particles $a$ and $b$ with $\rho_a \ll \rho_b$, so that $h_a \gg h_b$. If $b$ lies within the compact support of $a$ but $a$ lies outside that of $b$, only the $W_{ab}(h_a)$ term is non-zero. The contribution of $b$ to the artificial resistivity term of $a$ (Equation~\eqref{SPMHD: art resistivity}) becomes
\begin{equation}
\mathcal{D}_{\mathrm{diss},a} = \frac{1}{2}\frac{m_b}{\rho_a\rho_b}\, v^B_{\mathrm{sig},ab}\, \mathbf{B}_{ab}\, \hat{\mathbf{r}}_{ab}\cdot\frac{\nabla_a W_{ab}(h_a)}{\Omega_a}
\end{equation}
for the geometric density weighting of $\xi=1$, and
\begin{equation}
\mathcal{D}^{\rm std}_{\mathrm{diss},a} = \frac{1}{2}\frac{m_b}{\rho_a^2}\, v^B_{\mathrm{sig},ab}\, \mathbf{B}_{ab}\, \hat{\mathbf{r}}_{ab}\cdot\frac{\nabla_a W_{ab}(h_a)}{\Omega_a}
\end{equation}
for the standard weighting of $\xi=\rho$. The ratio between the two is $\mathcal{D}^{\rm std}_{\mathrm{diss},a}/\mathcal{D}_{\mathrm{diss},a} = \rho_b/\rho_a \gg 1$, so the standard weighting dissipates the field of the low-density particle much faster. Moreover, although the kernel decreases monotonically with distance, its gradient does not: for the quintic kernel, $|\nabla W|$ peaks at $\approx 0.76h$, so neighbours at intermediate distances carry the largest weight. In the outer regions of our discs, the density may change by more than an order of magnitude over such distances. This effect is highly problematic in astrophysical discs, which have steep radial and vertical density gradients.

We also ran standard SPMHD simulations starting directly from the Monte Carlo particle distribution, without the hydrodynamical relaxation. In this case, the disc retains its magnetic field for up to $200\,\mathrm{yr}$ before the magnetic energy is abruptly dissipated by several orders of magnitude. We attribute this longer survival to the initial particle distribution. At $t=0$, the particles are confined to $1 \le R \le 10\,\mathrm{au}$ and $|z| \le 3\,\mathrm{au}$, so the density gradients are moderate. As the disc spreads outwards, particles reach lower densities and steeper gradients, which increases the dissipation until the magnetic energy drops abruptly. In the relaxed initial conditions, by contrast, the particle distribution already contains these low-density regions, so the field is lost before the disc becomes turbulent. The noise of the random distribution may also contribute, by generating spurious magnetic fields.

\subsection{Low Resolution}

Many SPH studies of accretion discs use fewer than $10^7$ particles \citep[e.g.][]{forgan_fragmentation_2016, ilee_observing_2020, chen_minimizing_2025}, as did early particle-based simulations of the MRI in global discs \citep{gaburov_astrophysical_2011, vanaverbeke_gradspmhd_2014}. We therefore also followed the evolution of magnetized discs at low resolutions, with $N_\mathrm{P} = 2$, 5, and $10\times10^6$. Their initial quality factors, $\langle Q_{\phi,0}\rangle^{\rm V}\approx8$--14, are above the threshold for resolving the linear growth of the MRI.

\begin{figure}[t!]
\centering
\includegraphics[width=\columnwidth]{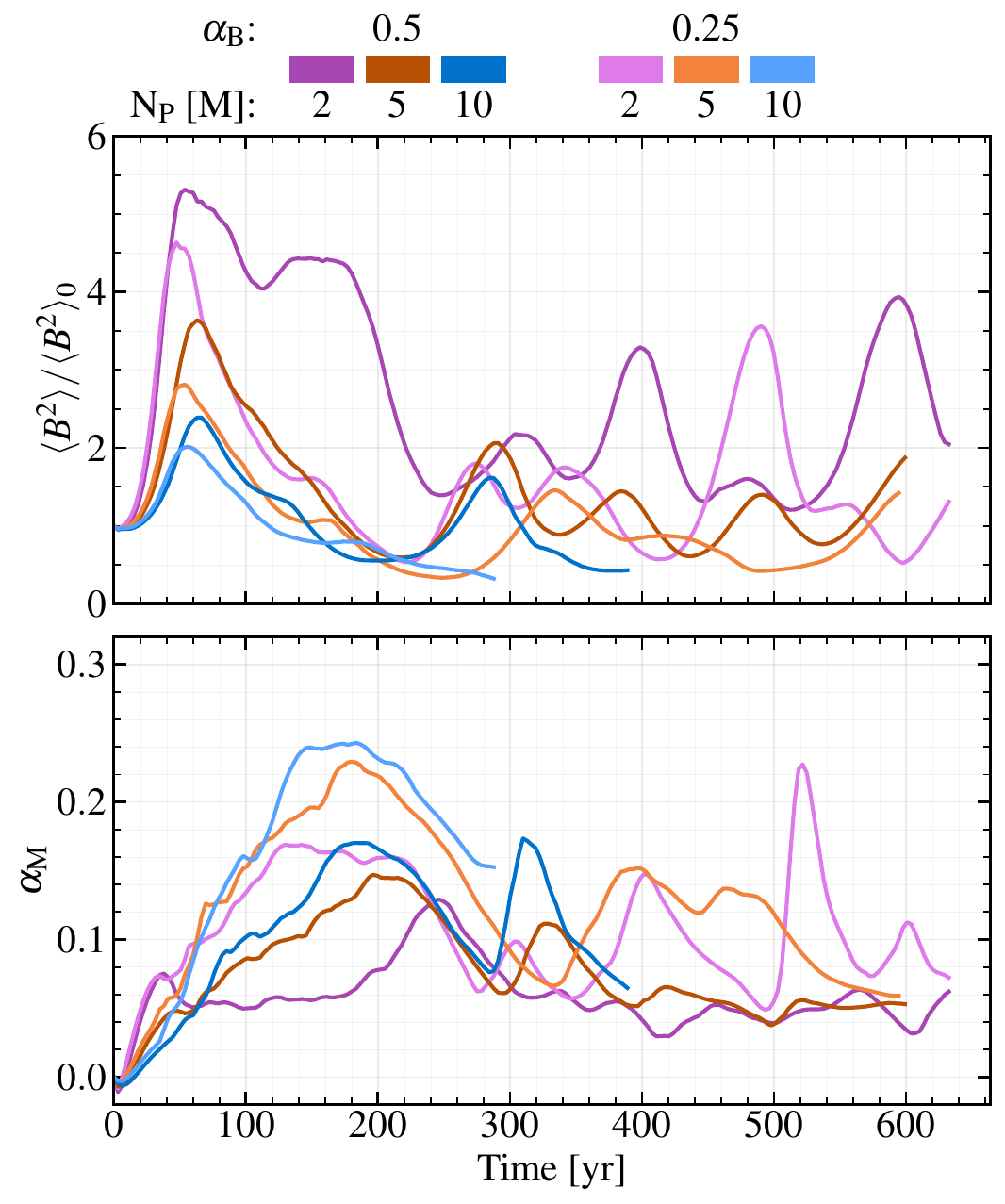}
\caption{Time evolution of the magnetic energy (top) and the normalized Maxwell stress (bottom) for the low-resolution models with $N_\mathrm{P} = 2$, 5, and $10\times10^6$. Darker lines correspond to $\alpha_\mathrm{B} = 0.5$ and lighter lines to $\alpha_\mathrm{B} = 0.25$.}
\label{fig:_low_resolution}
\end{figure}

Figure~\ref{fig:_low_resolution} shows the evolution of the magnetic energy (top) and the normalized Maxwell stress (bottom) for these models. We omit the turbulent energies here, since in these discs the shell-averaged fluctuations mostly reflect the vertical structure of the strong mean field rather than turbulence. The initial amplification of the magnetic energy is weaker at higher resolution and lower $\alpha_\mathrm{B}$, ranging from a factor of $\approx5$ for \texttt{P2M\_2aB1} to $\approx2$ for \texttt{P10M\_2aB2}. As in the high-resolution models, the Maxwell stress rises with resolution and as $\alpha_\mathrm{B}$ decreases, reaching $\alpha_\mathrm{M}\approx0.24$ for \texttt{P10M\_2aB2}. After these peaks, both quantities decrease until $t\approx250\,\mathrm{yr}$ and then oscillate around a roughly constant level, with the magnetic energy remaining of the order of its initial value. The oscillations of the magnetic energy are smaller at higher resolution.

\begin{figure}[t!]
\centering
\includegraphics[width=\columnwidth]{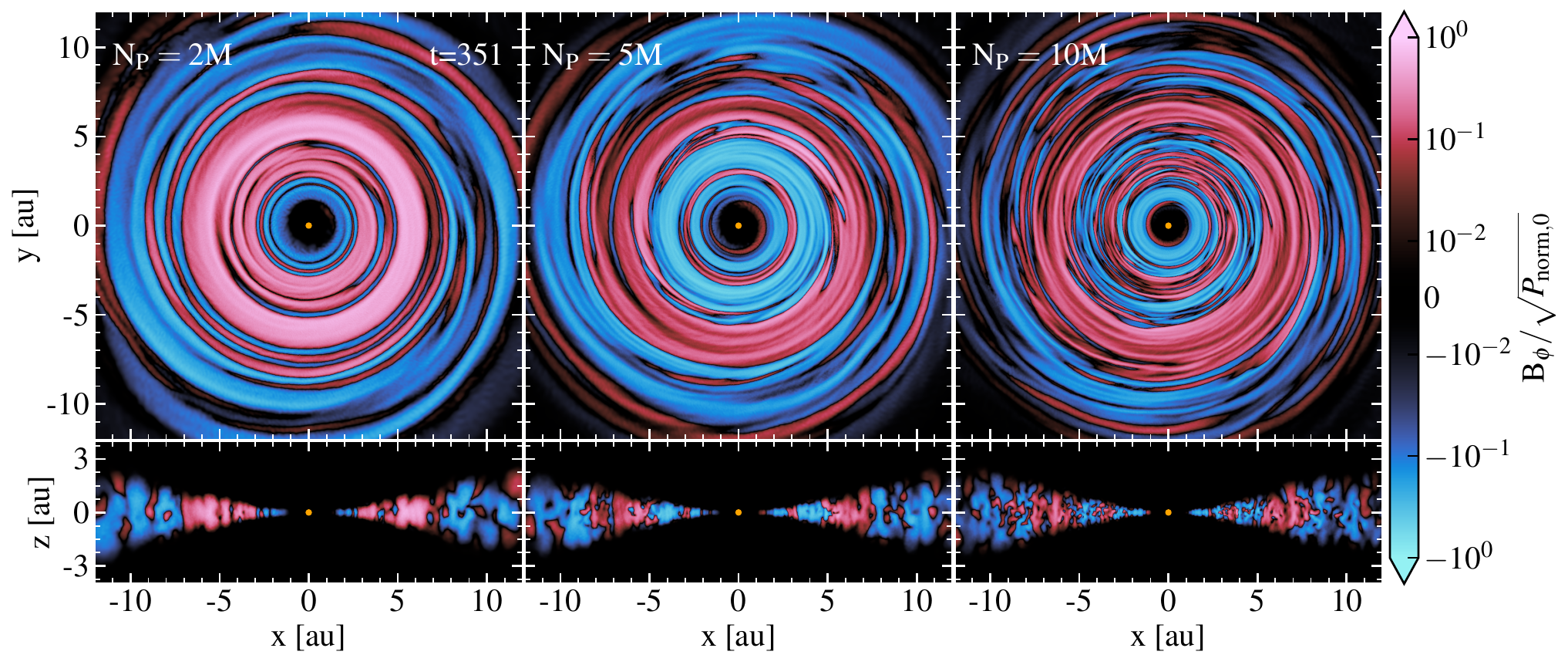}
\caption{Toroidal magnetic field, normalized by $\sqrt{P_\mathrm{norm,0}}$, in the disc midplane (\textit{top}) and in a vertical slice (\textit{bottom}) at $t = 351\,\mathrm{yr}$, for $N_\mathrm{P} = 2$, 5, and $10\times10^6$ (from left to right) with $\alpha_\mathrm{B} = 0.5$.}
\label{fig:Bphi_mid_low}
\end{figure}

Figure~\ref{fig:Bphi_mid_low} shows the toroidal field at $t=351\,\mathrm{yr}$ for the three resolutions with $\alpha_\mathrm{B}=0.5$. The field is organized in coherent bands of alternating polarity, which dominate the oscillations of the magnetic energy. These bands become narrower as the resolution increases, but they remain wider and stronger than in the high-resolution models.

None of these models sustain significant MRI-driven turbulence, as expected for $P_\mathrm{m}\lesssim 1$ and $R_\mathrm{m} \lesssim 350$, well below the values at which sustained turbulence has been reported (Section~\ref{sub:long_term_dynamics}). Even so, the coherent toroidal field reverses its polarity, so the signature of the mean-field dynamo is retained, as in the model with $\alpha_\mathrm{B}=1$ (Section~\ref{subsec:the_effect_of_dissipation}). In addition, the field strength persists with the geometric density weighting, whereas in test runs with the standard form, we found it decays even faster than in the high-resolution models of Section~\ref{sub:gradient_choice}. Therefore, we recommend the geometric density weighting for magnetized disc simulations at these resolutions as well.

\section{Summary and Conclusions} \label{sec:Conclusions}

In this study, we presented three-dimensional global ideal MHD simulations of accretion discs, to assess whether SPMHD reproduces MRI-driven turbulence and which numerical choices this requires. The discs start with a purely toroidal magnetic field with $\beta=25$. We varied the resolution by almost two orders of magnitude, from $N_\mathrm{P}=2\times10^6$ to $\approx10^8$, and the artificial resistivity parameter, with $\alpha_\mathrm{B}=0.25$, 0.5, and 1. Here, we summarize our main conclusions:

\begin{itemize}
    \item In our highest-resolution model, with $102\times10^6$ and $\alpha_\mathrm{B}=0.25$, we reach $R_\mathrm{m}\approx1000$. The MRI develops after about six local orbits, and the whole disc is turbulent by $t\approx200\,\mathrm{yr}$. The turbulence persists until $t\approx500\,\mathrm{yr}$ while its strength slowly decreases, and the magnetic energy drops to $\approx10\%$ of its initial value. This slow decay resembles that found by \citet{parkin_global_2013} at similar $R_\mathrm{m}$, and it is expected below the critical value of $\approx3000$ found by \citet{flock_turbulence_2012}.

    \item During the turbulent phase, the main MRI signatures agree with grid-based simulations. The Maxwell stress dominates the angular momentum transport and is $4$--$6$ times the Reynolds stress, and the total stress reaches $\alpha_\mathrm{tot}\sim10^{-2}$. The mean toroidal field shows the characteristic butterfly pattern, reversing direction with a period of about 10 local orbits.

    \item Where numerical dissipation is the dominant driver of angular momentum transport, MRI turbulence is suppressed and the magnetic field organizes into coherent toroidal bands. This occurs in the inner disc ($R\lesssim4\,\mathrm{au}$), where artificial viscosity transports most of the angular momentum. The same happens at most radii with $\alpha_\mathrm{B}=1$ or $N_\mathrm{P}\lesssim10^7$, and at late times ($t\gtrsim500\,\mathrm{yr}$) in the highest-resolution models. In these cases, the dynamo acts only through polarity reversals of the mean field, rather than through the turbulent cycle seen when the MRI is active.
        
    \item The gradient weighting in the magnetic terms strongly affects how long the magnetic field survives in the disc. The standard form, $\xi=\rho$, is much more dissipative across steep density gradients, which are inherent to astrophysical discs. In all our models with the standard weighting, the magnetic energy is lost within $\sim80\,\mathrm{yr}$. With the geometric density choice, $\xi=1$, the field is retained, and MRI turbulence develops when the resolution is high enough. We recommend the geometric density weighting for SPMHD simulations of magnetized discs.

    \item Whether MRI turbulence develops depends mainly on the numerical dissipation. All our models have $\langle Q_{\phi,0}\rangle^{\rm V}\gtrsim8$, above the threshold for resolving the linear growth of the MRI, but turbulence develops only for $P_\mathrm{m}\gtrsim1$, which required $\alpha_\mathrm{B}=0.5$ or 0.25, depending on the resolution. Both $\alpha_\mathrm{M}$ and $P_\mathrm{m}$ increase with resolution and with decreasing $\alpha_\mathrm{B}$. This trend continues up to $N_\mathrm{P}=102\times10^6$, so the turbulence has yet to converge.

\end{itemize}

Our results support the use of SPMHD for magnetized discs. We found that SPMHD captures the development of MRI-driven turbulence in global discs and reproduces its main signatures, provided that the resolution is high enough and the numerical dissipation is low. Our results are in agreement with previous grid-based studies \citep{flock_turbulence_2011, parkin_global_2013}. For the maximum magnetic Reynolds numbers reached here ($R_\mathrm{m}\sim1000$), however, the turbulence is not sustained over long timescales.

As it is important to keep numerical dissipation low, we recommend using $\alpha_{\rm B}$=0.25--0.5 when using the $|\mathbf{v}_{ab}\times \hat{\mathbf{r}}_{ab}|$ signal velocity in the artificial resistivity. Values below 0.25 led to unwanted numerical noise. Furthermore, we also recommend the geometric density weighting ($\xi = 1$) in the gradient discretizations used in the magnetic solver, similar to \citet{wissing_magnetorotational_2022}. The standard weighting scheme ($\xi = \rho$) incurs excessive numerical dissipation in the outer layers of the disc where there are steep density gradients. It would also be of interest to test the magnetic divergence projection method of \citet{tsukamoto_adjoint_2026}, which should be able to reduce divergence errors well below that of divergence cleaning.

For simulations using fewer than 10 million particles, the magnetic field remains laminar organized in alternating concentric rings. This occurs because the magnetic Reynolds numbers are $<$ 350 and the quality factor, $Q_\phi$, is $\lesssim$ 14. Our recommendation is that simulations should use 20 million or more particles, which provides $R_{\rm m}$ $>$ 500 and $Q_\phi >$ 15, in order to observe turbulence generated by the MRI.

It would be interesting to follow up our work with simulations of magnetized, self-gravitating discs. The spiral waves of the gravitational instability (GI) drive a dynamo that amplifies magnetic fields even when the MRI is suppressed \citep{riols_magnetorotational_2018}. \citet{deng_global_2020} recovered this dynamo in global simulations using meshless finite mass (\textsc{Gizmo} code) and found mass accretion rates several times higher than in pure GI or pure MRI discs. This dynamo also operates at low magnetic Reynolds numbers, $R_\mathrm{m}\sim1$--100, where the MRI fails to activate \citep{riols_gravitoturbulent_2019, deng_global_2020, bethune_gravitoturbulent_2022}. This may be accessible for the 1--10 million particle simulations commonly used for self-gravitating discs.

It would also be valuable to explore discs with non-ideal MHD, which is needed to model weakly ionized systems such as protoplanetary discs. Non-ideal effects often damp the MRI, but they may also favour other ways of amplifying the magnetic field. For example, an optimal Ohmic resistivity aids the GI dynamo \citep{bethune_gravitoturbulent_2022}, and the Hall effect produces a strong azimuthal field and a large-scale Maxwell stress in otherwise dead zones when the field is aligned with the rotation axis \citep{lesur_thanatology_2014}. \textsc{Phantom} contains implementations of the three non-ideal effects of Ohmic dissipation, ambipolar diffusion, and the Hall effect.

\newpage

\begin{acknowledgments}

The authors acknowledge discussions with Daniel Price and Guillaume Laibe that helped improve this paper. JNC thanks Mario Flock and David Melon Fuksman for helpful discussions on the MRI and turbulence in protoplanetary discs. The authors acknowledge the support of the Natural Sciences and Engineering Research Council of Canada (NSERC), grant number RGPIN/003786-2022. This research was enabled in part by support provided by ACENET, ace-net.ca, and the Digital Research Alliance of Canada, alliancecan.ca, by computational resources made available through their Resources for Research Groups 2026 competition.
\end{acknowledgments}

\software{We ran the simulations with \textsc{Phantom} \citep{price_phantom_2018} and \textsc{Shamrock} \citep{david-cleris_shamrock_2025}. For the analysis and visualization, we used \textsc{Sarracen} \citep{harris_sarracen_2023}, \textsc{NumPy} \citep{harris_array_2020}, \textsc{Matplotlib} \citep{hunter_matplotlib_2007}, and some colour maps from \textsc{CMasher} \citep{van_der_velden_cmasher_2021}.}

\appendix

\section{Lagrangian formulation}
\label{app:lagrangian_formulation}

The stress terms of the momentum equation, Equation~\eqref{SPMHD:momentum}, can be derived from a discrete Lagrangian, following \citet{price_smoothed_2004b}. We take
\begin{equation}\label{SPMHD:lagrangian}
    \mathcal{L} = \sum_a m_a\left[\frac{1}{2}v_a^2 - u_a - \frac{\rho_a}{2\mu_0}\left(\frac{\mathbf{B}_a}{\rho_a}\right)^2\right],
\end{equation}
and require the action $\mathcal{S} = \int\mathcal{L}\,{\rm d}t$ to be stationary under a small, arbitrary perturbation $\delta\mathbf{r}_a$ of the particle positions, at constant entropy. From the first law of thermodynamics, $(\partial u/\partial\rho)_s = P/\rho^2$, the resulting variation of the Lagrangian is
\begin{equation}\label{SPMHD:deltaL}
    \delta\mathcal{L}_a = m_a\,\mathbf{v}_a\cdot\delta\mathbf{v}_a - \sum_b \frac{m_b}{\rho_b^2}\left(P_b + \frac{B_b^2}{2\mu_0}\right)\delta\rho_b - \sum_b \frac{m_b}{\mu_0}\,\mathbf{B}_b\cdot\delta\left(\frac{\mathbf{B}}{\rho}\right)_b.
\end{equation}
The variation of the density follows from the summation density, Equation~\eqref{SPMHD:density_sum}, while the variation of $\mathbf{B}/\rho$ is imposed as a constraint from the induction equation, Equation~\eqref{SPMHD:Induction_Eq}, with $\xi = 1$:
\begin{align}
    \delta\rho_b &= \sum_c m_c\left(\delta\mathbf{r}_b - \delta\mathbf{r}_c\right)\cdot\frac{\nabla_b W_{bc}(h_b)}{\Omega_b}, \label{SPMHD:delta_rho}\\
    \delta\left(\frac{\mathbf{B}}{\rho}\right)_b &= -\sum_c \frac{m_c}{\rho_b\rho_c}\left(\delta\mathbf{r}_b - \delta\mathbf{r}_c\right)\left[\mathbf{B}_b\cdot\frac{\nabla_b W_{bc}(h_b)}{\Omega_b}\right]. \label{SPMHD:delta_Brho}
\end{align}
Expressing the perturbations in terms of $\delta\mathbf{r}_a$ with the Kronecker delta $\delta_{ab}$ and substituting into Equation~\eqref{SPMHD:deltaL} gives
\begin{equation}\label{SPMHD:deltaL_kronecker}
\begin{split}
    \delta\mathcal{L}_a = m_a\,\mathbf{v}_a\cdot\delta\mathbf{v}_a &- \delta\mathbf{r}_a\cdot\Bigg\{\sum_b \frac{m_b}{\rho_b^2}\left(P_b + \frac{B_b^2}{2\mu_0}\right)\sum_c m_c\left(\delta_{ab} - \delta_{ac}\right)\frac{\nabla_b W_{bc}(h_b)}{\Omega_b}\\
    &\qquad - \frac{1}{\mu_0}\sum_b m_b\,\mathbf{B}_b\sum_c \frac{m_c}{\rho_b\rho_c}\left(\delta_{ab} - \delta_{ac}\right)\left[\mathbf{B}_b\cdot\frac{\nabla_b W_{bc}(h_b)}{\Omega_b}\right]\Bigg\}.
\end{split}
\end{equation}

The equation of motion follows from $\delta \mathcal{S} = \int\delta\mathcal{L}_a\,{\rm d}t = 0$. Integrating the velocity term by parts,
\begin{equation}\label{SPMHD:action_parts}
\begin{split}
    \delta \mathcal{S} = -\int \delta\mathbf{r}_a\cdot\Bigg[m_a\frac{{\rm d}\mathbf{v}_a}{{\rm d}t} &+ \sum_b \frac{m_b}{\rho_b^2}\left(P_b + \frac{B_b^2}{2\mu_0}\right)\sum_c m_c\left(\delta_{ab} - \delta_{ac}\right)\frac{\nabla_b W_{bc}(h_b)}{\Omega_b}\\
    &- \frac{1}{\mu_0}\sum_b m_b\,\mathbf{B}_b\sum_c \frac{m_c}{\rho_b\rho_c}\left(\delta_{ab} - \delta_{ac}\right)\left[\mathbf{B}_b\cdot\frac{\nabla_b W_{bc}(h_b)}{\Omega_b}\right]\Bigg]{\rm d}t = 0.
\end{split}
\end{equation}
Since $\delta\mathbf{r}_a$ is arbitrary, the square bracket vanishes. Evaluating the Kronecker deltas, relabelling the summation indices, and using $\nabla_b W_{ba}(h_b) = -\nabla_a W_{ab}(h_b)$ yields
\begin{equation}\label{SPMHD:momentum_lagrangian}
\begin{split}
    \frac{{\rm d}v^i_a}{{\rm d}t} = &\sum_b m_b \left[\frac{S^{ij}_{{\rm iso},a}}{\rho_a^2}\frac{\nabla^j_a W_{ab}(h_a)}{\Omega_a} + \frac{S^{ij}_{{\rm iso},b}}{\rho_b^2}\frac{\nabla^j_a W_{ab}(h_b)}{\Omega_b} \right]\\
    &+\sum_b \frac{m_b}{\rho_a\rho_b} \left[S^{ij}_{{\rm anis},a}\frac{\nabla^j_a W_{ab}(h_a)}{\Omega_a} + S^{ij}_{{\rm anis},b}\frac{\nabla^j_a W_{ab}(h_b)}{\Omega_b} \right],
\end{split}
\end{equation}
where $S^{ij}_{\rm iso}$ and $S^{ij}_{\rm anis}$ are given by Equation~\eqref{SPMHD:stress_split}. This recovers the stress terms of Equation~\eqref{SPMHD:momentum}.

\section{MHD test problems}
\label{Append_MHD_testing}

To verify the SPMHD implementation in \textsc{Shamrock}, we compare it with \textsc{Phantom} on four standard MHD test problems: the circularly polarised Alfv\'en wave, the MHD rotor, the MHD blast wave, and the Orszag--Tang vortex. The setups follow \citet{price_phantom_2018}, to which we refer the reader for further details about the initial conditions. The initial setups were generated with \textsc{Phantom} and then read by \textsc{Shamrock}, so that both codes start from identical particle distributions. Both codes use the same numerical parameters as the accretion disc simulations presented in this study. We ran the tests at several resolutions and values of $\alpha_\mathrm{B}$, but for the sake of illustration, we show only the case $\alpha_\mathrm{B}=0.25$, which actually do not differ significantly from those obtained with $\alpha_\mathrm{B}=0.5$. Finally, we compare the two codes directly on the accretion disc problem, using model \texttt{P102M\_2aB2}, run with \textsc{Phantom}, and model \texttt{S102M\_2aB2}, run with \textsc{Shamrock}.

\subsection{Circularly polarized Alfv\'en wave}

The circularly polarized Alfv\'en wave is an exact non-linear solution of the MHD equations \citep{toth__2000}. Due to its characteristics of smoothness and being an analytical solution, it has been a standard test of the accuracy and numerical dissipation of SPMHD codes \citep{price_phantom_2018, wissing_smoothed_2020, karapiperis_smoothed_2026}.

Here, we use the 3D travelling wave setup of \citet{gardiner_unsplit_2008}, as in \citet{price_phantom_2018}. The domain is a periodic box of size $L \times L/2 \times L/2$, with $L=3$. The wave propagates along $\hat{\mathbf{x}}_1 = (\cos\alpha\cos\beta,\, \cos\alpha\sin\beta,\, \sin\alpha)$, with $\sin\alpha = 2/3$ and $\sin\beta = 2/\sqrt{5}$. The gas is initially uniform, with $\rho = 1$, $P = 0.1$, and follows an adiabatic EOS with $\gamma = 5/3$. In the frame $(x_1, x_2, x_3)$ aligned with the wave, the velocity and magnetic field are
\begin{equation}
\mathbf{v} = A\,(0,\, \sin kx_1,\, \cos kx_1), \qquad \mathbf{B} = (1,\, A\sin kx_1,\, A\cos kx_1),
\end{equation}
with $A = 0.1$ and $k = 2\pi/\lambda$, where $\lambda = 1$. Both vectors are then rotated into the box frame. The particles are placed on a close-packed lattice with $n_x = 128$ particles along the $x$-axis. Since $|\mathbf{B}|$ is uniform, there is no magnetic pressure gradient, and the wave travels at the Alfv\'en speed $v_\mathrm{A} = 1$ without changing its shape. The solution therefore returns to its initial profile after each period, $t = \lambda / v_\mathrm{A} = 1$.

\begin{figure}[t!]
\centering
\includegraphics[width=0.5\textwidth]{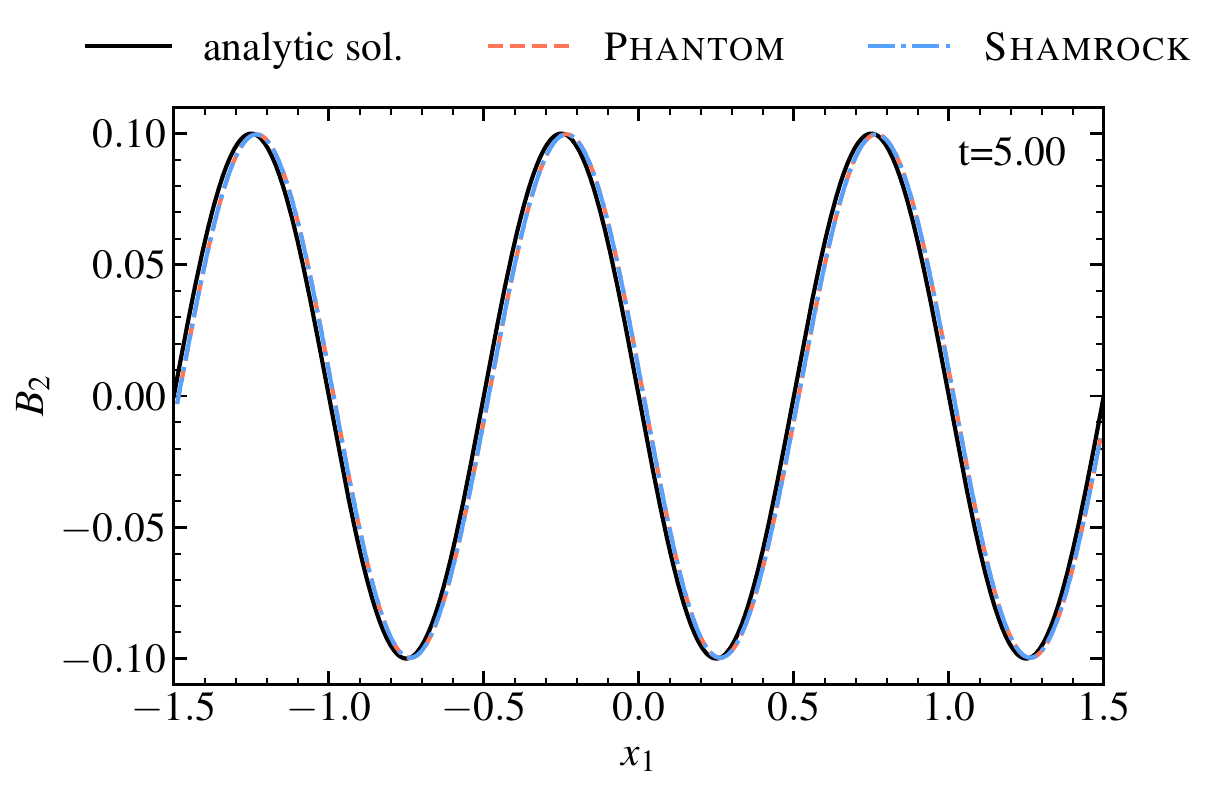}
\caption{Circularly polarized Alfv\'en wave after five periods ($t = 5$). The component of the magnetic field perpendicular to the wave vector, $B_2$, is shown as a function of $x_1$, computed with \textsc{Phantom} (red dashed line) and \textsc{Shamrock} (blue dash-dotted line), compared with the analytic solution (black solid line).}
\label{fig:mhdtest_alfven}
\end{figure}

Figure~\ref{fig:mhdtest_alfven} shows $B_2$ as a function of $x_1$ after five periods. Both codes follow the analytic solution, with no noticeable damping of the wave. Similar results are shown in figure~26 of \citet{price_phantom_2018}, figure~1 of \citet{wissing_smoothed_2020}, and figure~2 of \citet{karapiperis_smoothed_2026}.

\subsection{MHD rotor}

The MHD rotor problem \citep{balsara_staggered_1999} tests the propagation of strong torsional Alfv\'en waves. Here we use the parameters of the first rotor problem of \citet{toth__2000}, as in \citet{price_phantom_2018}. The box has periodic boundaries, a unit size in the $x$--$y$ plane, and a thin slab of 12 particle layers in $z$, with $n_x = 256$ particles along the $x$-axis. A dense cylinder of radius $R_{\rm cyl} = 0.1$ and density $\rho_{\rm cyl} = 10$, centred at the origin, rotates rigidly with velocity
\begin{equation}
\mathbf{v} = \frac{v_{\rm cyl}}{R_{\rm cyl}}\,(-y,\, x,\, 0), \qquad R < R_{\rm cyl},
\end{equation}
where $R = \sqrt{x^2 + y^2}$ and $v_{\rm cyl} = 2$ is the rotation speed at the edge of the cylinder. The surrounding medium is at rest, with density $\rho_{\rm m} = 1$. The pressure is uniform, $P = 1$, with $\gamma = 1.4$, and the whole domain is threaded by a uniform magnetic field $\mathbf{B} = (5/\sqrt{4\pi},\, 0,\, 0)$. Since all particles have equal mass, the density contrast is obtained by using close-packed lattices with different spacings inside and outside the cylinder. As the cylinder rotates, it winds up the field lines and launches torsional Alfv\'en waves into the surrounding medium, which brake the rotation.

\begin{figure}[t!]
\centering
\includegraphics[width=\textwidth]{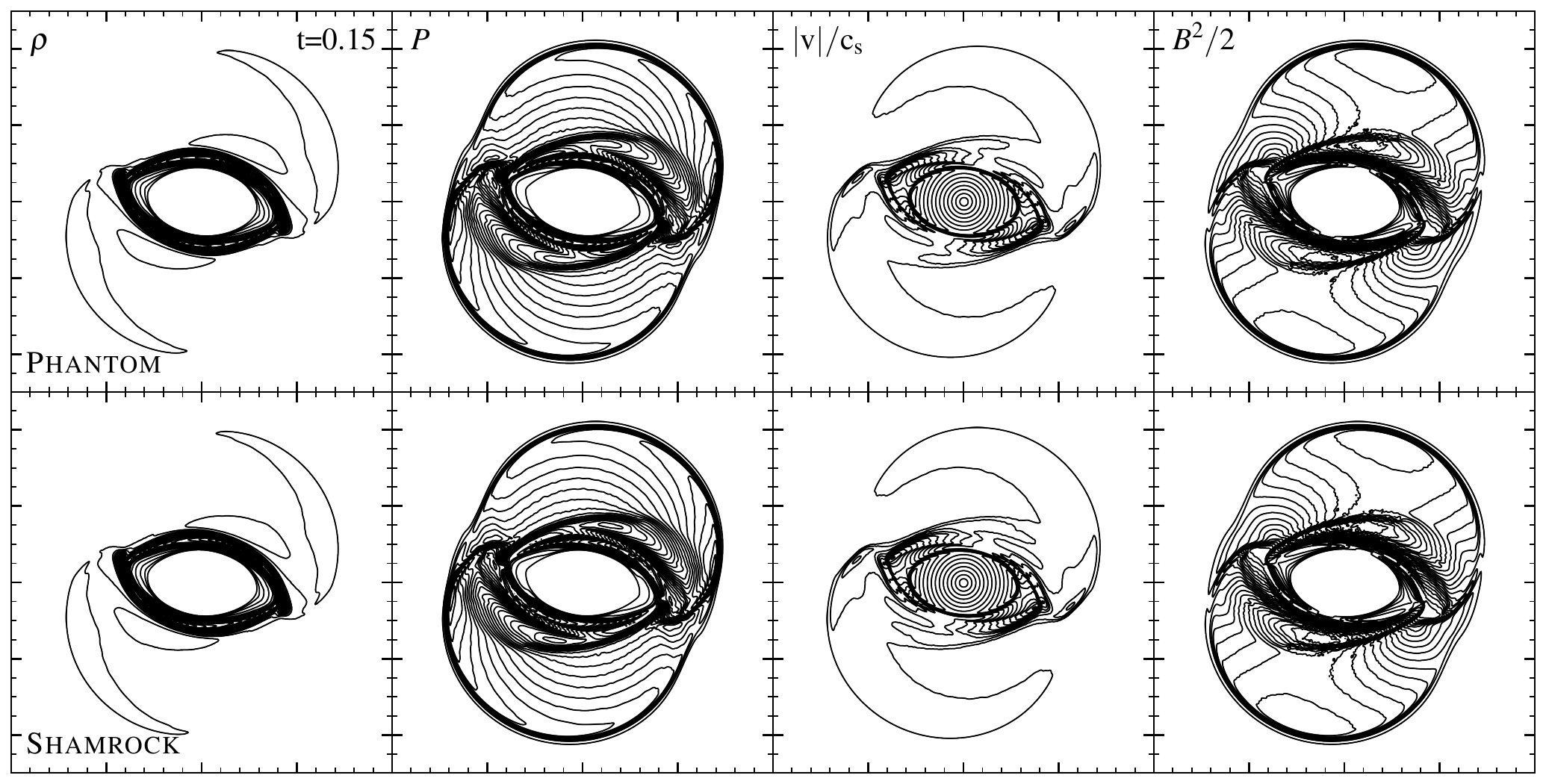}
\caption{MHD rotor test at $t=0.15$, computed with \textsc{Phantom} (\textit{top}) and \textsc{Shamrock} (\textit{bottom}). From left to right: density, gas pressure, Mach number, and magnetic pressure in the $z=0$ plane. For a direct comparison with previous works, we show 30 contours between the limits used by \citet{toth__2000}: $0.483 < \rho < 12.95$, $0.0202 < P < 2.008$, $0 < |v|/c_\mathrm{s} < 8.18$, and $0 < B^2/2 < 2.642$.}
\label{fig:_rotor}
\end{figure}

Figure~\ref{fig:_rotor} shows the density, pressure, Mach number, and magnetic pressure in the $z=0$ plane at $t=0.15$. The two codes produce almost identical results, and both show the structure characteristic of this test \citep{toth__2000, price_phantom_2018, wissing_smoothed_2020, karapiperis_smoothed_2026}.

\subsection{MHD blast wave}

The MHD blast wave tests the treatment of strong MHD shocks in a strongly magnetized medium. We use the setup of \citet{stone_athena_2008}, as in \citet{price_phantom_2018}, in a periodic box of unit size, with $n_x = 256$ particles along the $x$-axis on a close-packed lattice. The gas is initially at rest, with uniform density $\rho = 1$ and $\gamma = 1.4$, and is threaded by a uniform magnetic field $\mathbf{B} = (10/\sqrt{2},\, 0,\, 10/\sqrt{2})$, which lies in the $x$--$z$ plane at $45^\circ$ to the axes. A sphere of radius $R_{\rm blast} = 0.125$ at the centre of the box is over-pressurized, such that the pressure and plasma beta are set to
\begin{equation}
(P,\, \beta) =
\begin{cases}
(100,\, 2), & r < R_{\rm blast}, \\
(1,\, 0.02), & r \geq R_{\rm blast}.
\end{cases}
\end{equation}
The over-pressure drives a blast wave whose expansion is strongly anisotropic: the field resists compression perpendicular to the field lines, so the gas expands mainly along them.

\begin{figure}[t!]
\centering
\includegraphics[width=\textwidth]{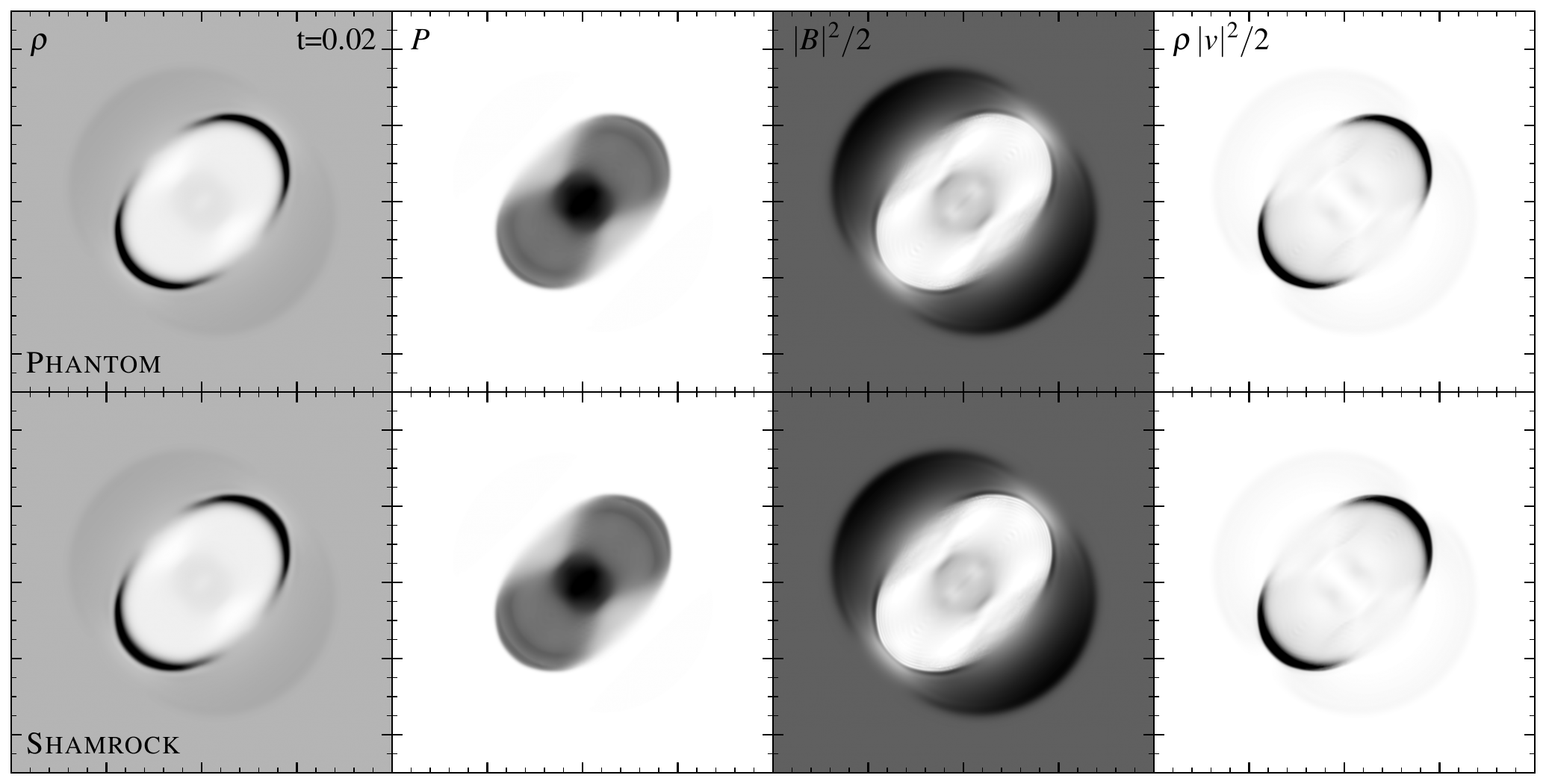}
\caption{MHD blast wave at $t=0.02$, computed with \textsc{Phantom} (\textit{top}) and \textsc{Shamrock} (\textit{bottom}). From left to right: density, gas pressure, magnetic pressure, and kinetic energy density in the $y=0$ plane. For a direct comparison with previous works, we use the same limits as \citet{stone_athena_2008}: $0.19 < \rho < 2.98$, $1 < P < 42.4$, $25.2 < B^2/2 < 64.9$, and $0 < \rho v^2/2 < 33.1$.}
\label{fig:_blast}
\end{figure}

Figure~\ref{fig:_blast} shows the density, gas pressure, magnetic pressure, and kinetic energy density in the $y=0$ plane at $t=0.02$. The two codes produce almost identical results, in agreement with previous works \citep{stone_athena_2008, price_phantom_2018, wissing_smoothed_2020}.

\subsection{Orszag--Tang vortex}

As the last standard test, we use the Orszag--Tang vortex \citep{orszag_small-scale_1979}, which evolves from smooth initial conditions into a network of interacting shocks and, at later times, develops small-scale MHD turbulence. We use a periodic box of unit size in the $x$--$y$ plane and a thin slab in $z$ arranging 12 particles, with $n_x = 512$ particles along the $x$-axis. The density and pressure are initially uniform, $\rho = 25/(36\pi)$ and $P = 5/(12\pi)$, with $\gamma = 5/3$, velocity $\mathbf{v} = (-\sin 2\pi y,\, \sin 2\pi x,\, 0)$, and magnetic field $\mathbf{B} = B_0\,(-\sin 2\pi y,\, \sin 4\pi x,\, 0)$ with $B_0 = 1/\sqrt{4\pi}$, where $x$ and $y$ are measured from the lower-left corner of the box. These values give an initial plasma beta of $10/3$ and a Mach number of unity.

\begin{figure}[t!]
\centering
\includegraphics[width=0.5\textwidth]{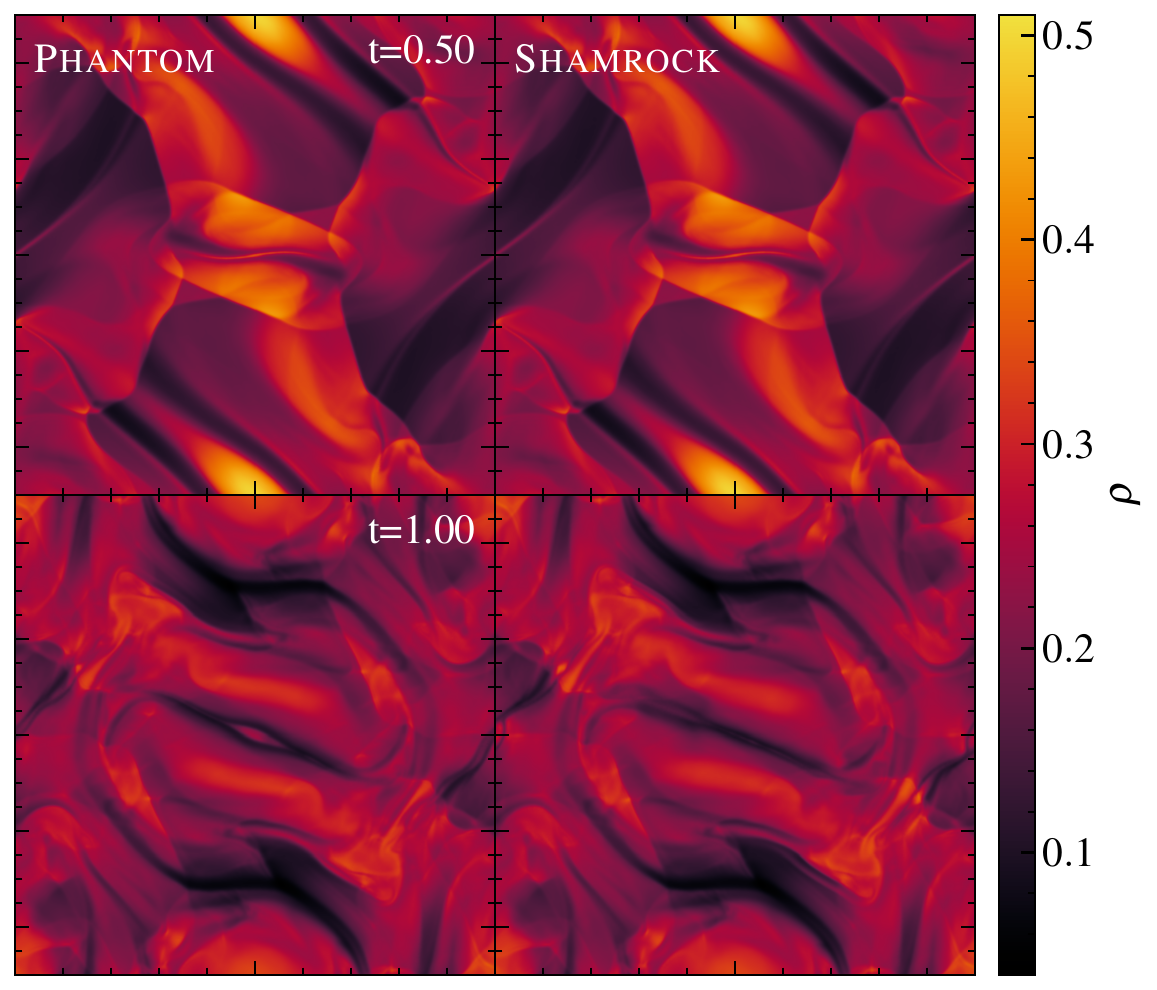}
\caption{Density in the $z=0$ plane for the Orszag--Tang vortex at $t=0.5$ (\textit{top}) and $t=1$ (\textit{bottom}), computed with \textsc{Phantom} (\textit{left}) and \textsc{Shamrock} (\textit{right}).}
\label{fig:orstang}
\end{figure}

Figure~\ref{fig:orstang} shows the density in the $z=0$ plane at $t=0.5$ and $t=1$. At $t=0.5$, the two codes are practically indistinguishable. At $t=1$, small differences appear in the smallest structures, while the large-scale structure remains the same. Similar results are shown in \citet{price_phantom_2018} and \citet{wissing_smoothed_2020}.

\subsection{Accretion Disc: Model 102M\_2aB2}

\begin{figure}[t!]
\centering
\includegraphics[width=0.5\textwidth]{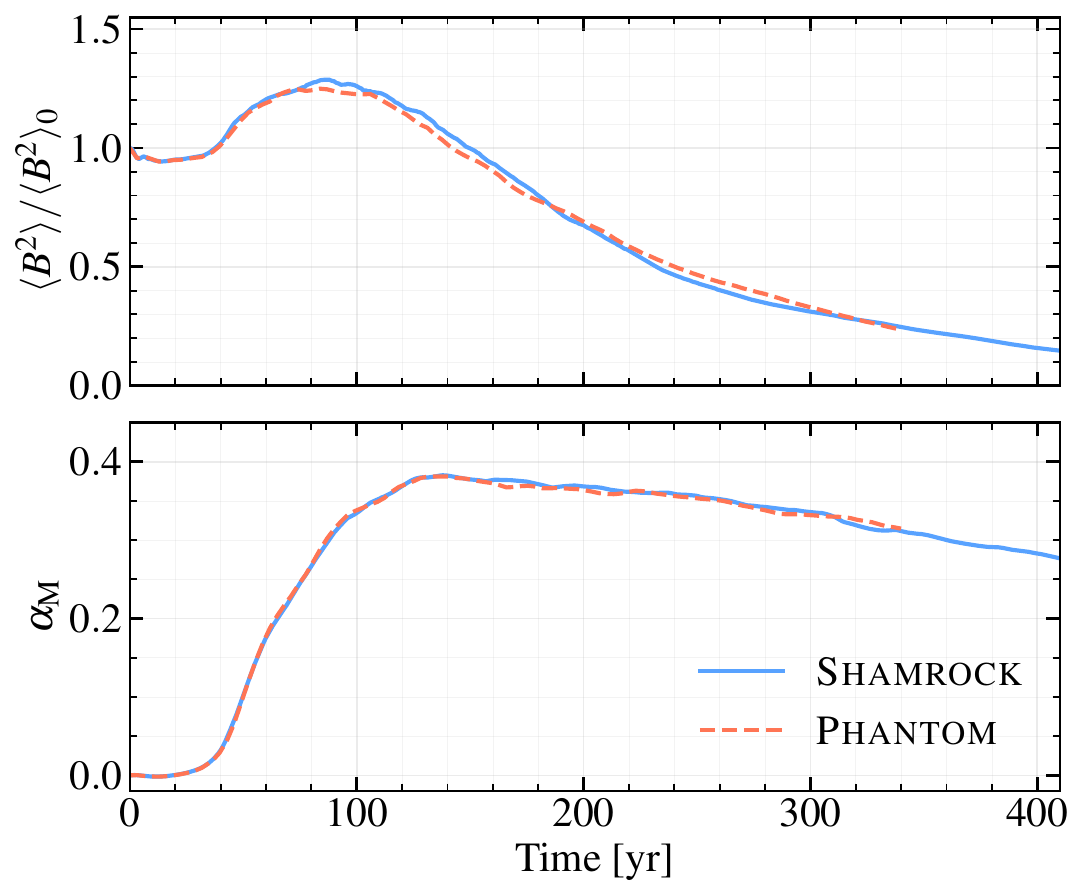}
\caption{Comparison between models S102M\_2aB2 (\textsc{Shamrock}, solid line) and P102M\_2aB2 (\textsc{Phantom}, dashed line). Evolution of the magnetic energy (\textit{top}) and the normalized Maxwell stress (\textit{bottom}). The \textsc{Phantom} run was evolved only up to $t \approx 340$~yr.}
\label{fig:disc_phsh}
\end{figure}

As mentioned in Section~\ref{subsec:simulations}, the implementation in \textsc{Shamrock} differs slightly from that in \textsc{Phantom}, since they are optimized for different parallelization strategies. To test how these differences affect the evolution of the disc, we also ran model S102M\_2aB2 with \textsc{Phantom} (P102M\_2aB2), over a shorter time. Figure~\ref{fig:disc_phsh} shows the time evolution of the magnetic energy (\textit{top}) and the normalized Maxwell stress (\textit{bottom}). We find no significant differences between the two runs. Part of the small differences that remain can also come from restarting the simulations from different dump files. Both codes use a second-order leapfrog integrator, generalized with a velocity predictor and corrector, so the accelerations used at the start of each step are those computed at the end of the previous step with the predicted velocities. These accelerations are not saved in the dump files, and for the case of \textsc{phantom}, it saves $\alpha$, $h$, and $\nabla \cdot v$ as single precision. On restart, these values are recomputed from the stored state, so the run does not resume from exactly the state at which it stopped. Combined with the turbulent nature of flow, this means that runs restarted at different times differ slightly, even with the same code. The turbulence nevertheless keeps the same statistical properties, and the evolution of the disc does not change significantly.

\bibliography{sample701}{}

\begin{thebibliography}{}
\expandafter\ifx\csname natexlab\endcsname\relax\def\natexlab#1{#1}\fi
\providecommand{\url}[1]{\href{#1}{#1}}
\providecommand{\dodoi}[1]{doi:~\href{http://doi.org/#1}{\nolinkurl{#1}}}
\providecommand{\doeprint}[1]{\href{http://ascl.net/#1}{\nolinkurl{http://ascl.net/#1}}}
\providecommand{\doarXiv}[1]{\href{https://arxiv.org/abs/#1}{\nolinkurl{https://arxiv.org/abs/#1}}}

% type= article
\bibitem[{ {Arlt, R.} \&  {Rüdiger, G.}(2001){Arlt, R.} \& {Rüdiger,
  G.}}]{arlt_r_global_2001}
{Arlt, R.}, \& {Rüdiger, G.} 2001, \bibinfo{title}{Global accretion disk
  simulations of magneto-rotational instability,} A\&A, 374, 1035,
  \dodoi{10.1051/0004-6361:20010797}

% type= article
\bibitem[{X.-N. Bai \& J.~M. Stone(2013)Bai \& Stone}]{bai_local_2013}
Bai, X.-N., \& Stone, J.~M. 2013, \bibinfo{title}{Local {Study} of {Accretion}
  {Disks} with a {Strong} {Vertical} {Magnetic} {Field}: {Magnetorotational}
  {Instability} and {Disk} {Outflow},} The Astrophysical Journal, 767, 30,
  \dodoi{10.1088/0004-637X/767/1/30}

% type= article
\bibitem[{S.~A. Balbus \& J.~F. Hawley(1991)Balbus \&
  Hawley}]{balbus_powerful_1991}
Balbus, S.~A., \& Hawley, J.~F. 1991, \bibinfo{title}{A {Powerful} {Local}
  {Shear} {Instability} in {Weakly} {Magnetized} {Disks}. {I}. {Linear}
  {Analysis},} The Astrophysical Journal, 376, 214, \dodoi{10.1086/170270}

% type= article
\bibitem[{S.~A. Balbus \& J.~F. Hawley(1998)Balbus \&
  Hawley}]{balbus_instability_1998}
Balbus, S.~A., \& Hawley, J.~F. 1998, \bibinfo{title}{Instability, turbulence,
  and enhanced transport in accretion disks,} Reviews of Modern Physics, 70, 1,
  \dodoi{10.1103/RevModPhys.70.1}

% type= article
\bibitem[{D.~S. Balsara \& D.~S. Spicer(1999)Balsara \&
  Spicer}]{balsara_staggered_1999}
Balsara, D.~S., \& Spicer, D.~S. 1999, \bibinfo{title}{A {Staggered} {Mesh}
  {Algorithm} {Using} {High} {Order} {Godunov} {Fluxes} to {Ensure}
  {Solenoidal} {Magnetic} {Fields} in {Magnetohydrodynamic} {Simulations},}
  Journal of Computational Physics, 149, 270, \dodoi{10.1006/jcph.1998.6153}

% type= article
\bibitem[{C. Bonnerot {et~al.}(2017)Bonnerot, Price, Lodato, \&
  Rossi}]{bonnerot_magnetic_2017}
Bonnerot, C., Price, D.~J., Lodato, G., \& Rossi, E.~M. 2017,
  \bibinfo{title}{Magnetic field evolution in tidal disruption events,} Monthly
  Notices of the Royal Astronomical Society, 469, 4879,
  \dodoi{10.1093/mnras/stx1210}

% type= article
\bibitem[{A. Brandenburg {et~al.}(1995)Brandenburg, Nordlund, Stein, \&
  Torkelsson}]{brandenburg_dynamo-generated_1995}
Brandenburg, A., Nordlund, A., Stein, R.~F., \& Torkelsson, U. 1995,
  \bibinfo{title}{Dynamo-generated {Turbulence} and {Large}-{Scale} {Magnetic}
  {Fields} in a {Keplerian} {Shear} {Flow},} The Astrophysical Journal, 446,
  741, \dodoi{10.1086/175831}

% type= article
\bibitem[{W. Béthune \& H. Latter(2022)Béthune \&
  Latter}]{bethune_gravitoturbulent_2022}
Béthune, W., \& Latter, H. 2022, \bibinfo{title}{Gravitoturbulent dynamo in
  global simulations of gaseous disks,} Astronomy and Astrophysics, 663, A138,
  \dodoi{10.1051/0004-6361/202243219}

% type= article
\bibitem[{S. Børve {et~al.}(2001)Børve, Omang, \&
  Trulsen}]{borve_regularized_2001}
Børve, S., Omang, M., \& Trulsen, J. 2001, \bibinfo{title}{Regularized
  {Smoothed} {Particle} {Hydrodynamics}: {A} {New} {Approach} to {Simulating}
  {Magnetohydrodynamic} {Shocks},} The Astrophysical Journal, 561, 82,
  \dodoi{10.1086/323228}

% type= article
\bibitem[{S. Børve {et~al.}(2004)Børve, Omang, \&
  Trulsen}]{borve_two-dimensional_2004}
Børve, S., Omang, M., \& Trulsen, J. 2004, \bibinfo{title}{Two-dimensional
  {MHD} {Smoothed} {Particle} {Hydrodynamics} {Stability} {Analysis},} The
  Astrophysical Journal Supplement Series, 153, 447, \dodoi{10.1086/421520}

% type= article
\bibitem[{C. Chen \& C.~J. Nixon(2025)Chen \& Nixon}]{chen_minimizing_2025}
Chen, C., \& Nixon, C.~J. 2025, \bibinfo{title}{Minimizing the numerical
  viscosity in smoothed particle hydrodynamics simulations of discs,} Monthly
  Notices of the Royal Astronomical Society, 540, 2465,
  \dodoi{10.1093/mnras/staf881}

% type= article
\bibitem[{P. Cossins {et~al.}(2009)Cossins, Lodato, \&
  Clarke}]{cossins_characterizing_2009}
Cossins, P., Lodato, G., \& Clarke, C.~J. 2009, \bibinfo{title}{Characterizing
  the gravitational instability in cooling accretion discs,} Monthly Notices of
  the Royal Astronomical Society, 393, 1157,
  \dodoi{10.1111/j.1365-2966.2008.14275.x}

% type= article
\bibitem[{C. Cui \& X.-N. Bai(2022)Cui \& Bai}]{cui_turbulence_2022}
Cui, C., \& Bai, X.-N. 2022, \bibinfo{title}{Turbulence in outer protoplanetary
  discs: {MRI} or {VSI}?} Monthly Notices of the Royal Astronomical Society,
  516, 4660, \dodoi{10.1093/mnras/stac2580}

% type= article
\bibitem[{L. Cullen \& W. Dehnen(2010)Cullen \& Dehnen}]{cullen_inviscid_2010}
Cullen, L., \& Dehnen, W. 2010, \bibinfo{title}{Inviscid smoothed particle
  hydrodynamics,} Monthly Notices of the Royal Astronomical Society, 408, 669,
  \dodoi{10.1111/j.1365-2966.2010.17158.x}

% type= article
\bibitem[{T. David-Cléris {et~al.}(2025)David-Cléris, Laibe, \&
  Lapeyre}]{david-cleris_shamrock_2025}
David-Cléris, T., Laibe, G., \& Lapeyre, Y. 2025, \bibinfo{title}{The
  {SHAMROCK} code: {I} - smoothed particle hydrodynamics on {GPUs},}
  {\textbackslash}mnras, 539, 1, \dodoi{10.1093/mnras/staf444}

% type= article
\bibitem[{S.~W. Davis {et~al.}(2010)Davis, Stone, \&
  Pessah}]{davis_sustained_2010}
Davis, S.~W., Stone, J.~M., \& Pessah, M.~E. 2010, \bibinfo{title}{Sustained
  {Magnetorotational} {Turbulence} in {Local} {Simulations} of {Stratified}
  {Disks} with {Zero} {Net} {Magnetic} {Flux},} The Astrophysical Journal, 713,
  52, \dodoi{10.1088/0004-637X/713/1/52}

% type= article
\bibitem[{W. {Dehnen} \& H. {Aly}(2012){Dehnen} \& {Aly}}]{2012MNRAS.425.1068D}
{Dehnen}, W., \& {Aly}, H. 2012, \bibinfo{title}{{Improving convergence in
  smoothed particle hydrodynamics simulations without pairing instability},}
  \mnras, 425, 1068, \dodoi{10.1111/j.1365-2966.2012.21439.x}

% type= article
\bibitem[{H. Deng {et~al.}(2020)Deng, Mayer, \& Latter}]{deng_global_2020}
Deng, H., Mayer, L., \& Latter, H. 2020, \bibinfo{title}{Global {Simulations}
  of {Self}-gravitating {Magnetized} {Protoplanetary} {Disks},} The
  Astrophysical Journal, 891, 154, \dodoi{10.3847/1538-4357/ab77b2}

% type= article
\bibitem[{H. Deng {et~al.}(2019)Deng, Mayer, Latter, Hopkins, \&
  Bai}]{deng_local_2019}
Deng, H., Mayer, L., Latter, H., Hopkins, P.~F., \& Bai, X.-N. 2019,
  \bibinfo{title}{Local {Simulations} of {MRI} turbulence with {Meshless}
  {Methods},} The Astrophysical Journal Supplement Series, 241, 26,
  \dodoi{10.3847/1538-4365/ab0957}

% type= article
\bibitem[{C.~L. Dobbs {et~al.}(2016)Dobbs, Price, Pettitt, Bate, \&
  Tricco}]{dobbs_magnetic_2016}
Dobbs, C.~L., Price, D.~J., Pettitt, A.~R., Bate, M.~R., \& Tricco, T.~S. 2016,
  \bibinfo{title}{Magnetic field evolution and reversals in spiral galaxies,}
  Monthly Notices of the Royal Astronomical Society, 461, 4482,
  \dodoi{10.1093/mnras/stw1625}

% type= article
\bibitem[{T.~P. Fleming {et~al.}(2000)Fleming, Stone, \&
  Hawley}]{fleming_effect_2000}
Fleming, T.~P., Stone, J.~M., \& Hawley, J.~F. 2000, \bibinfo{title}{The
  {Effect} of {Resistivity} on the {Nonlinear} {Stage} of the
  {Magnetorotational} {Instability} in {Accretion} {Disks},} The Astrophysical
  Journal, 530, 464, \dodoi{10.1086/308338}

% type= article
\bibitem[{M. Flock {et~al.}(2012{\natexlab{a}})Flock, Dzyurkevich, Klahr,
  Turner, \& Henning}]{flock_large-scale_2012}
Flock, M., Dzyurkevich, N., Klahr, H., Turner, N., \& Henning, T.
  2012{\natexlab{a}}, \bibinfo{title}{Large-scale {Azimuthal} {Structures} of
  {Turbulence} in {Accretion} {Disks}: {Dynamo} {Triggered} {Variability} of
  {Accretion},} {\textbackslash}apj, 744, 144,
  \dodoi{10.1088/0004-637X/744/2/144}

% type= article
\bibitem[{M. Flock {et~al.}(2011)Flock, Dzyurkevich, Klahr, Turner, \&
  Henning}]{flock_turbulence_2011}
Flock, M., Dzyurkevich, N., Klahr, H., Turner, N.~J., \& Henning, T. 2011,
  \bibinfo{title}{Turbulence and {Steady} {Flows} in {Three}-dimensional
  {Global} {Stratified} {Magnetohydrodynamic} {Simulations} of {Accretion}
  {Disks},} {\textbackslash}apj, 735, 122, \dodoi{10.1088/0004-637X/735/2/122}

% type= article
\bibitem[{M. Flock {et~al.}(2012{\natexlab{b}})Flock, Henning, \&
  Klahr}]{flock_turbulence_2012}
Flock, M., Henning, T., \& Klahr, H. 2012{\natexlab{b}},
  \bibinfo{title}{Turbulence in {Weakly} {Ionized} {Protoplanetary} {Disks},}
  The Astrophysical Journal, 761, 95, \dodoi{10.1088/0004-637X/761/2/95}

% type= article
\bibitem[{M. Flock {et~al.}(2017)Flock, Nelson, Turner, Bertrang,
  Carrasco-González, Henning, Lyra, \& Teague}]{flock_radiation_2017}
Flock, M., Nelson, R.~P., Turner, N.~J., {et~al.} 2017,
  \bibinfo{title}{Radiation {Hydrodynamical} {Turbulence} in {Protoplanetary}
  {Disks}: {Numerical} {Models} and {Observational} {Constraints},} The
  Astrophysical Journal, 850, 131, \dodoi{10.3847/1538-4357/aa943f}

% type= article
\bibitem[{D. Forgan {et~al.}(2016)Forgan, Price, \&
  Bonnell}]{forgan_fragmentation_2016}
Forgan, D., Price, D.~J., \& Bonnell, I. 2016, \bibinfo{title}{On the
  fragmentation boundary in magnetized self-gravitating discs,} Monthly Notices
  of the Royal Astronomical Society, 466, 3406, \dodoi{10.1093/mnras/stw3314}

% type= article
\bibitem[{S. Fromang {et~al.}(2013)Fromang, Latter, Lesur, \&
  Ogilvie}]{fromang_local_2013}
Fromang, S., Latter, H., Lesur, G., \& Ogilvie, G.~I. 2013,
  \bibinfo{title}{Local outflows from turbulent accretion disks,} Astronomy and
  Astrophysics, 552, A71, \dodoi{10.1051/0004-6361/201220016}

% type= article
\bibitem[{ {Fromang, S.} \&  {Nelson, R. P.}(2006){Fromang, S.} \& {Nelson, R.
  P.}}]{fromang_s_global_2006}
{Fromang, S.}, \& {Nelson, R. P.} 2006, \bibinfo{title}{Global {MHD}
  simulations of stratified and turbulent protoplanetary discs - {I}. {Model}
  properties,} A\&A, 457, 343, \dodoi{10.1051/0004-6361:20065643}

% type= article
\bibitem[{E. Gaburov \& K. Nitadori(2011)Gaburov \&
  Nitadori}]{gaburov_astrophysical_2011}
Gaburov, E., \& Nitadori, K. 2011, \bibinfo{title}{Astrophysical weighted
  particle magnetohydrodynamics,} Monthly Notices of the Royal Astronomical
  Society, 414, 129, \dodoi{10.1111/j.1365-2966.2011.18313.x}

% type= article
\bibitem[{D. Gagnier \& O. Pejcha(2024)Gagnier \&
  Pejcha}]{gagnier_post-dynamical_2024}
Gagnier, D., \& Pejcha, O. 2024, \bibinfo{title}{Post-dynamical inspiral phase
  of common envelope evolution,} A\&A, 683, \dodoi{10.1051/0004-6361/202348383}

% type= article
\bibitem[{C.~F. Gammie(1996)Gammie}]{gammie_layered_1996}
Gammie, C.~F. 1996, \bibinfo{title}{Layered {Accretion} in {T} {Tauri}
  {Disks},} The Astrophysical Journal, 457, 355, \dodoi{10.1086/176735}

% type= article
\bibitem[{T.~A. Gardiner \& J.~M. Stone(2008)Gardiner \&
  Stone}]{gardiner_unsplit_2008}
Gardiner, T.~A., \& Stone, J.~M. 2008, \bibinfo{title}{An unsplit {Godunov}
  method for ideal {MHD} via constrained transport in three dimensions,}
  Journal of Computational Physics, 227, 4123,
  \dodoi{10.1016/j.jcp.2007.12.017}

% type= article
\bibitem[{R.~A. Gingold \& J.~J. Monaghan(1977)Gingold \&
  Monaghan}]{gingold_smoothed_1977}
Gingold, R.~A., \& Monaghan, J.~J. 1977, \bibinfo{title}{Smoothed particle
  hydrodynamics: theory and application to non-spherical stars.,} Monthly
  Notices of the Royal Astronomical Society, 181, 375,
  \dodoi{10.1093/mnras/181.3.375}

% type= article
\bibitem[{X. Guan \& C.~F. Gammie(2011)Guan \& Gammie}]{guan_radially_2011}
Guan, X., \& Gammie, C.~F. 2011, \bibinfo{title}{Radially {Extended},
  {Stratified}, {Local} {Models} of {Isothermal} {Disks},} The Astrophysical
  Journal, 728, 130, \dodoi{10.1088/0004-637X/728/2/130}

% type= article
\bibitem[{X. Guan {et~al.}(2009)Guan, Gammie, Simon, \&
  Johnson}]{guan_locality_2009}
Guan, X., Gammie, C.~F., Simon, J.~B., \& Johnson, B.~M. 2009,
  \bibinfo{title}{Locality of {MHD} {Turbulence} in {Isothermal} {Disks},} The
  Astrophysical Journal, 694, 1010, \dodoi{10.1088/0004-637X/694/2/1010}

% type= article
\bibitem[{A. Harris \& T.~S. Tricco(2023)Harris \&
  Tricco}]{harris_sarracen_2023}
Harris, A., \& Tricco, T.~S. 2023, \bibinfo{title}{Sarracen: a {Python} package
  for analysis and visualization of smoothed particle hydrodynamics data,}
  Journal of Open Source Software, 8, 5263, \dodoi{10.21105/joss.05263}

% type= article
\bibitem[{C.~R. Harris {et~al.}(2020)Harris, Millman, Walt, Gommers, Virtanen,
  Cournapeau, Wieser, Taylor, Berg, Smith, Kern, Picus, Hoyer, Kerkwijk, Brett,
  Haldane, Río, Wiebe, Peterson, Gérard-Marchant, Sheppard, Reddy, Weckesser,
  Abbasi, Gohlke, \& Oliphant}]{harris_array_2020}
Harris, C.~R., Millman, K.~J., Walt, S. J. v.~d., {et~al.} 2020,
  \bibinfo{title}{Array programming with {NumPy},} Nature, 585, 357,
  \dodoi{10.1038/s41586-020-2649-2}

% type= article
\bibitem[{J.~F. Hawley(2000)Hawley}]{hawley_global_2000}
Hawley, J.~F. 2000, \bibinfo{title}{Global {Magnetohydrodynamical}
  {Simulations} of {Accretion} {Tori},} The Astrophysical Journal, 528, 462,
  \dodoi{10.1086/308180}

% type= article
\bibitem[{J.~F. Hawley(2001)Hawley}]{hawley_global_2001}
Hawley, J.~F. 2001, \bibinfo{title}{Global {Magnetohydrodynamic} {Simulations}
  of {Cylindrical} {Keplerian} {Disks},} {\textbackslash}apj, 554, 534,
  \dodoi{10.1086/321348}

% type= article
\bibitem[{J.~F. Hawley \& S.~A. Balbus(1991)Hawley \&
  Balbus}]{hawley_powerful_1991}
Hawley, J.~F., \& Balbus, S.~A. 1991, \bibinfo{title}{A {Powerful} {Local}
  {Shear} {Instability} in {Weakly} {Magnetized} {Disks}. {II}. {Nonlinear}
  {Evolution},} The Astrophysical Journal, 376, 223, \dodoi{10.1086/170271}

% type= article
\bibitem[{J.~F. Hawley {et~al.}(1995)Hawley, Gammie, \&
  Balbus}]{hawley_local_1995}
Hawley, J.~F., Gammie, C.~F., \& Balbus, S.~A. 1995, \bibinfo{title}{Local
  {Three}-dimensional {Magnetohydrodynamic} {Simulations} of {Accretion}
  {Disks},} The Astrophysical Journal, 440, 742, \dodoi{10.1086/175311}

% type= article
\bibitem[{J.~F. Hawley {et~al.}(2011)Hawley, Guan, \&
  Krolik}]{hawley_assessing_2011}
Hawley, J.~F., Guan, X., \& Krolik, J.~H. 2011, \bibinfo{title}{{ASSESSING}
  {QUANTITATIVE} {RESULTS} {IN} {ACCRETION} {SIMULATIONS}: {FROM} {LOCAL} {TO}
  {GLOBAL},} The Astrophysical Journal, 738, 84,
  \dodoi{10.1088/0004-637X/738/1/84}

% type= article
\bibitem[{P.~F. Hopkins \& M.~J. Raives(2016)Hopkins \&
  Raives}]{hopkins_accurate_2016}
Hopkins, P.~F., \& Raives, M.~J. 2016, \bibinfo{title}{Accurate, meshless
  methods for magnetohydrodynamics,} Monthly Notices of the Royal Astronomical
  Society, 455, 51, \dodoi{10.1093/mnras/stv2180}

% type= article
\bibitem[{J.~D. Hunter(2007)Hunter}]{hunter_matplotlib_2007}
Hunter, J.~D. 2007, \bibinfo{title}{Matplotlib: {A} {2D} graphics environment,}
  Computing in Science \& Engineering, 9, 90, \dodoi{10.1109/MCSE.2007.55}

% type= article
\bibitem[{J.~D. Ilee {et~al.}(2020)Ilee, Hall, Walsh, Jiménez-Serra, Pinte,
  Terry, Bourke, \& Hoare}]{ilee_observing_2020}
Ilee, J.~D., Hall, C., Walsh, C., {et~al.} 2020, \bibinfo{title}{Observing
  protoplanetary discs with the {Square} {Kilometre} {Array} - {I}.
  {Characterizing} pebble substructure caused by forming planets,} Monthly
  Notices of the Royal Astronomical Society, 498, 5116,
  \dodoi{10.1093/mnras/staa2699}

% type= article
\bibitem[{T. Jannaud \& H.~N. Latter(2026)Jannaud \&
  Latter}]{jannaud_numerical_2026}
Jannaud, T., \& Latter, H.~N. 2026, \bibinfo{title}{On the numerical
  convergence of {MRI} simulations,} Monthly Notices of the Royal Astronomical
  Society, 545, staf1966, \dodoi{10.1093/mnras/staf1966}

% type= article
\bibitem[{O.~A. Karapiperis {et~al.}(2026)Karapiperis, Schaller, Shchutskyi,
  Stasyszyn, \& Elion}]{karapiperis_smoothed_2026}
Karapiperis, O.~A., Schaller, M., Shchutskyi, N., Stasyszyn, F.~A., \& Elion,
  M. 2026, \bibinfo{title}{Smoothed particle magnetohydrodynamics for
  simulations of galaxy and cosmic structure formation,} arXiv e-prints,
  arXiv:2608.06978, \dodoi{10.48550/arXiv.2608.06978}

% type= article
\bibitem[{G. Lesur {et~al.}(2014)Lesur, Kunz, \&
  Fromang}]{lesur_thanatology_2014}
Lesur, G., Kunz, M.~W., \& Fromang, S. 2014, \bibinfo{title}{Thanatology in
  protoplanetary discs. {The} combined influence of {Ohmic}, {Hall}, and
  ambipolar diffusion on dead zones,} Astronomy and Astrophysics, 566, A56,
  \dodoi{10.1051/0004-6361/201423660}

% type= article
\bibitem[{G.~R.~J. Lesur(2021)Lesur}]{lesur_magnetohydrodynamics_2021}
Lesur, G. R.~J. 2021, \bibinfo{title}{Magnetohydrodynamics of protoplanetary
  discs,} Journal of Plasma Physics, 87, 205870101,
  \dodoi{10.1017/S0022377820001002}

% type= article
\bibitem[{G. Lodato \& D.~J. Price(2010)Lodato \&
  Price}]{lodato_diffusive_2010}
Lodato, G., \& Price, D.~J. 2010, \bibinfo{title}{On the diffusive propagation
  of warps in thin accretion discs,} Monthly Notices of the Royal Astronomical
  Society, 405, 1212, \dodoi{10.1111/j.1365-2966.2010.16526.x}

% type= article
\bibitem[{L.~B. Lucy(1977)Lucy}]{lucy_numerical_1977}
Lucy, L.~B. 1977, \bibinfo{title}{A numerical approach to the testing of the
  fission hypothesis.,} The Astronomical Journal, 82, 1013,
  \dodoi{10.1086/112164}

% type= article
\bibitem[{K. Menou \& E. Quataert(2001)Menou \&
  Quataert}]{menou_ionization_2001}
Menou, K., \& Quataert, E. 2001, \bibinfo{title}{Ionization,
  {Magnetorotational}, and {Gravitational} {Instabilities} in {Thin}
  {Accretion} {Disks} {Around} {Supermassive} {Black} {Holes},} The
  Astrophysical Journal, 552, 204, \dodoi{10.1086/320466}

% type= article
\bibitem[{K.~A. Miller \& J.~M. Stone(2000)Miller \&
  Stone}]{miller_formation_2000}
Miller, K.~A., \& Stone, J.~M. 2000, \bibinfo{title}{The {Formation} and
  {Structure} of a {Strongly} {Magnetized} {Corona} above a {Weakly}
  {Magnetized} {Accretion} {Disk},} The Astrophysical Journal, 534, 398,
  \dodoi{10.1086/308736}

% type= article
\bibitem[{J.~J. Monaghan(1997)Monaghan}]{monaghan_sph_1997}
Monaghan, J.~J. 1997, \bibinfo{title}{{SPH} and {Riemann} {Solvers},} Journal
  of Computational Physics, 136, 298, \dodoi{10.1006/jcph.1997.5732}

% type= article
\bibitem[{J.~J. Monaghan(2005)Monaghan}]{monaghan_smoothed_2005}
Monaghan, J.~J. 2005, \bibinfo{title}{Smoothed particle hydrodynamics,} Reports
  on Progress in Physics, 68, 1703, \dodoi{10.1088/0034-4885/68/8/R01}

% type= article
\bibitem[{S.~C. Noble {et~al.}(2010)Noble, Krolik, \&
  Hawley}]{noble_dependence_2010}
Noble, S.~C., Krolik, J.~H., \& Hawley, J.~F. 2010,
  \bibinfo{title}{{DEPENDENCE} {OF} {INNER} {ACCRETION} {DISK} {STRESS} {ON}
  {PARAMETERS}: {THE} {SCHWARZSCHILD} {CASE},} The Astrophysical Journal, 711,
  959, \dodoi{10.1088/0004-637X/711/2/959}

% type= article
\bibitem[{S.~A. Orszag \& C.-M. Tang(1979)Orszag \&
  Tang}]{orszag_small-scale_1979}
Orszag, S.~A., \& Tang, C.-M. 1979, \bibinfo{title}{Small-scale structure of
  two-dimensional magnetohydrodynamic turbulence,} Journal of Fluid Mechanics,
  90, 129, \dodoi{10.1017/S002211207900210X}

% type= article
\bibitem[{E.~R. Parkin \& G.~V. Bicknell(2013{\natexlab{a}})Parkin \&
  Bicknell}]{parkin_equilibrium_2013}
Parkin, E.~R., \& Bicknell, G.~V. 2013{\natexlab{a}},
  \bibinfo{title}{{EQUILIBRIUM} {DISKS}, {MAGNETOROTATIONAL} {INSTABILITY}
  {MODE} {EXCITATION}, {AND} {STEADY}-{STATE} {TURBULENCE} {IN} {GLOBAL}
  {ACCRETION} {DISK} {SIMULATIONS},} The Astrophysical Journal, 763, 99,
  \dodoi{10.1088/0004-637X/763/2/99}

% type= article
\bibitem[{E.~R. Parkin \& G.~V. Bicknell(2013{\natexlab{b}})Parkin \&
  Bicknell}]{parkin_global_2013}
Parkin, E.~R., \& Bicknell, G.~V. 2013{\natexlab{b}}, \bibinfo{title}{Global
  simulations of magnetorotational turbulence – {I}. {Convergence} and the
  quasi-steady state,} Monthly Notices of the Royal Astronomical Society, 435,
  2281, \dodoi{10.1093/mnras/stt1450}

% type= article
\bibitem[{D.~J. Price(2012)Price}]{price_smoothed_2012}
Price, D.~J. 2012, \bibinfo{title}{Smoothed particle hydrodynamics and
  magnetohydrodynamics,} Special Issue: Computational Plasma Physics, 231, 759,
  \dodoi{10.1016/j.jcp.2010.12.011}

% type= article
\bibitem[{D.~J. Price \& J.~J. Monaghan(2004{\natexlab{a}})Price \&
  Monaghan}]{price_smoothed_2004a}
Price, D.~J., \& Monaghan, J.~J. 2004{\natexlab{a}}, \bibinfo{title}{Smoothed
  {Particle} {Magnetohydrodynamics} – {I}. {Algorithm} and tests in one
  dimension,} Monthly Notices of the Royal Astronomical Society, 348, 123,
  \dodoi{10.1111/j.1365-2966.2004.07345.x}

% type= article
\bibitem[{D.~J. Price \& J.~J. Monaghan(2004{\natexlab{b}})Price \&
  Monaghan}]{price_smoothed_2004b}
Price, D.~J., \& Monaghan, J.~J. 2004{\natexlab{b}}, \bibinfo{title}{Smoothed
  {Particle} {Magnetohydrodynamics} – {II}. {Variational} principles and
  variable smoothing-length terms,} Monthly Notices of the Royal Astronomical
  Society, 348, 139, \dodoi{10.1111/j.1365-2966.2004.07346.x}

% type= article
\bibitem[{D.~J. Price {et~al.}(2018)Price, Wurster, Tricco, Nixon, Toupin,
  Pettitt, Chan, Mentiplay, Laibe, Glover, Dobbs, Nealon, Liptai, Worpel,
  Bonnerot, Dipierro, Ballabio, Ragusa, Federrath, Iaconi, Reichardt, Forgan,
  Hutchison, Constantino, Ayliffe, Hirsh, \& Lodato}]{price_phantom_2018}
Price, D.~J., Wurster, J., Tricco, T.~S., {et~al.} 2018,
  \bibinfo{title}{Phantom: {A} {Smoothed} {Particle} {Hydrodynamics} and
  {Magnetohydrodynamics} {Code} for {Astrophysics},} {\textbackslash}pasa, 35,
  e031, \dodoi{10.1017/pasa.2018.25}

% type= article
\bibitem[{A. Reboul-Salze {et~al.}(2025)Reboul-Salze, Held, \&
  Kiuchi}]{reboul-salze_mri-driven_2025}
Reboul-Salze, A., Held, L.~E., \& Kiuchi, K. 2025, \bibinfo{title}{{MRI}-driven
  {PLACEHOLDERalphaPLACEHOLDEROmega} dynamo at high {Pm} numbers,} arXiv
  e-prints, arXiv:2501.05393, \dodoi{10.48550/arXiv.2501.05393}

% type= article
\bibitem[{O. Regev \& O.~M. Umurhan(2008)Regev \&
  Umurhan}]{regev_viability_2008}
Regev, O., \& Umurhan, O.~M. 2008, \bibinfo{title}{On the viability of the
  shearing box approximation for numerical studies of {MHD} turbulence in
  accretion disks,} Astronomy and Astrophysics, 481, 21,
  \dodoi{10.1051/0004-6361:20078413}

% type= article
\bibitem[{A. Riols \& H. Latter(2018)Riols \&
  Latter}]{riols_magnetorotational_2018}
Riols, A., \& Latter, H. 2018, \bibinfo{title}{Magnetorotational instability
  and dynamo action in gravito-turbulent astrophysical discs,} Monthly Notices
  of the Royal Astronomical Society, 474, 2212, \dodoi{10.1093/mnras/stx2455}

% type= article
\bibitem[{A. Riols \& H. Latter(2019)Riols \&
  Latter}]{riols_gravitoturbulent_2019}
Riols, A., \& Latter, H. 2019, \bibinfo{title}{Gravitoturbulent dynamos in
  astrophysical discs,} Monthly Notices of the Royal Astronomical Society, 482,
  3989, \dodoi{10.1093/mnras/sty2804}

% type= article
\bibitem[{M.~J.~O. Roberts {et~al.}(2025)Roberts, Latter, \&
  Lesur}]{roberts_global_2025}
Roberts, M. J.~O., Latter, H.~N., \& Lesur, G. 2025, \bibinfo{title}{Global
  magnetohydrodynamic simulations of the inner regions of protoplanetary discs.
  {I}. {Zero}-net flux regime,} Monthly Notices of the Royal Astronomical
  Society, 544, 1284, \dodoi{10.1093/mnras/staf1672}

% type= article
\bibitem[{P.~E. Rodman \& C.~S. Reynolds(2024)Rodman \&
  Reynolds}]{rodman_evolution_2024}
Rodman, P.~E., \& Reynolds, C.~S. 2024, \bibinfo{title}{Evolution of the
  {Magnetic} {Field} in {High}- and {Low}-β {Disks} with {Initially}
  {Toroidal} {Fields},} The Astrophysical Journal, 960, 97,
  \dodoi{10.3847/1538-4357/ad0384}

% type= article
\bibitem[{G. Salvesen {et~al.}(2016)Salvesen, Simon, Armitage, \&
  Begelman}]{salvesen_accretion_2016}
Salvesen, G., Simon, J.~B., Armitage, P.~J., \& Begelman, M.~C. 2016,
  \bibinfo{title}{Accretion disc dynamo activity in local simulations spanning
  weak-to-strong net vertical magnetic flux regimes,} Monthly Notices of the
  Royal Astronomical Society, 457, 857, \dodoi{10.1093/mnras/stw029}

% type= article
\bibitem[{T. Sano {et~al.}(2004)Sano, Inutsuka, Turner, \&
  Stone}]{sano_angular_2004}
Sano, T., Inutsuka, S.-i., Turner, N.~J., \& Stone, J.~M. 2004,
  \bibinfo{title}{Angular {Momentum} {Transport} by {Magnetohydrodynamic}
  {Turbulence} in {Accretion} {Disks}: {Gas} {Pressure} {Dependence} of the
  {Saturation} {Level} of the {Magnetorotational} {Instability},} The
  Astrophysical Journal, 605, 321, \dodoi{10.1086/382184}

% type= article
\bibitem[{N.~I. Shakura \& R.~A. Sunyaev(1973)Shakura \&
  Sunyaev}]{shakura_black_1973}
Shakura, N.~I., \& Sunyaev, R.~A. 1973, \bibinfo{title}{Black holes in binary
  systems. {Observational} appearance.,} Astronomy and Astrophysics, 24, 337.
\newblock \url{https://ui.adsabs.harvard.edu/abs/1973A&A....24..337S}

% type= article
\bibitem[{J.~B. Simon {et~al.}(2012)Simon, Beckwith, \&
  Armitage}]{simon_emergent_2012}
Simon, J.~B., Beckwith, K., \& Armitage, P.~J. 2012, \bibinfo{title}{Emergent
  mesoscale phenomena in magnetized accretion disc turbulence,} Monthly Notices
  of the Royal Astronomical Society, 422, 2685,
  \dodoi{10.1111/j.1365-2966.2012.20835.x}

% type= article
\bibitem[{J.~B. Simon {et~al.}(2011)Simon, Hawley, \&
  Beckwith}]{simon_resistivity-driven_2011}
Simon, J.~B., Hawley, J.~F., \& Beckwith, K. 2011,
  \bibinfo{title}{Resistivity-driven {State} {Changes} in {Vertically}
  {Stratified} {Accretion} {Disks},} The Astrophysical Journal, 730, 94,
  \dodoi{10.1088/0004-637X/730/2/94}

% type= article
\bibitem[{K.~A. Sorathia {et~al.}(2012)Sorathia, Reynolds, Stone, \&
  Beckwith}]{sorathia_global_2012}
Sorathia, K.~A., Reynolds, C.~S., Stone, J.~M., \& Beckwith, K. 2012,
  \bibinfo{title}{Global {Simulations} of {Accretion} {Disks}. {I}.
  {Convergence} and {Comparisons} with {Local} {Models},} {\textbackslash}apj,
  749, 189, \dodoi{10.1088/0004-637X/749/2/189}

% type= article
\bibitem[{J. Squire {et~al.}(2025)Squire, Quataert, \&
  Hopkins}]{squire_rapid_2025}
Squire, J., Quataert, E., \& Hopkins, P.~F. 2025, \bibinfo{title}{Rapid,
  strongly magnetized accretion in the zero-net-vertical-flux shearing box,}
  The Open Journal of Astrophysics, 8, 39, \dodoi{10.33232/001c.136467}

% type= article
\bibitem[{U.~P. Steinwandel {et~al.}(2020)Steinwandel, Dolag, Lesch, Moster,
  Burkert, \& Prieto}]{steinwandel_origin_2020}
Steinwandel, U.~P., Dolag, K., Lesch, H., {et~al.} 2020, \bibinfo{title}{On the
  origin of magnetic driven winds and the structure of the galactic dynamo in
  isolated galaxies,} Monthly Notices of the Royal Astronomical Society, 494,
  4393, \dodoi{10.1093/mnras/staa817}

% type= article
\bibitem[{J.~M. Stone {et~al.}(2008)Stone, Gardiner, Teuben, Hawley, \&
  Simon}]{stone_athena_2008}
Stone, J.~M., Gardiner, T.~A., Teuben, P., Hawley, J.~F., \& Simon, J.~B. 2008,
  \bibinfo{title}{Athena: {A} {New} {Code} for {Astrophysical} {MHD},} The
  Astrophysical Journal Supplement Series, 178, 137, \dodoi{10.1086/588755}

% type= article
\bibitem[{J.~M. Stone {et~al.}(1996)Stone, Hawley, Gammie, \&
  Balbus}]{stone_three-dimensional_1996}
Stone, J.~M., Hawley, J.~F., Gammie, C.~F., \& Balbus, S.~A. 1996,
  \bibinfo{title}{Three-dimensional {Magnetohydrodynamical} {Simulations} of
  {Vertically} {Stratified} {Accretion} {Disks},} The Astrophysical Journal,
  463, 656, \dodoi{10.1086/177280}

% type= article
\bibitem[{T.~K. Suzuki(2023)Suzuki}]{suzuki_mhd_2023}
Suzuki, T.~K. 2023, \bibinfo{title}{{MHD} in a {Cylindrical} {Shearing} {Box}.
  {II}. {Intermittent} {Bursts} and {Substructures} in {MRI} {Turbulence},} The
  Astrophysical Journal, 957, 99, \dodoi{10.3847/1538-4357/acfb88}

% type= article
\bibitem[{T.~S. Tricco(2015)Tricco}]{tricco_simulating_2015}
Tricco, T.~S. 2015, \bibinfo{title}{Simulating {Astrophysical} {Magnetic}
  {Fields} with {Smoothed} {Particle} {Magnetohydrodynamics},} arXiv e-prints,
  arXiv:1505.04494, \dodoi{10.48550/arXiv.1505.04494}

% type= article
\bibitem[{T.~S. Tricco(2023)Tricco}]{tricco_smoothed_2023}
Tricco, T.~S. 2023, \bibinfo{title}{Smoothed particle magnetohydrodynamics,}
  Frontiers in Astronomy and Space Sciences, 10,
  \dodoi{10.3389/fspas.2023.1288219}

% type= article
\bibitem[{T.~S. Tricco \& D.~J. Price(2012)Tricco \&
  Price}]{tricco_constrained_2012}
Tricco, T.~S., \& Price, D.~J. 2012, \bibinfo{title}{Constrained hyperbolic
  divergence cleaning for smoothed particle magnetohydrodynamics,} Journal of
  Computational Physics, 231, 7214, \dodoi{10.1016/j.jcp.2012.06.039}

% type= article
\bibitem[{T.~S. Tricco {et~al.}(2016{\natexlab{a}})Tricco, Price, \&
  Bate}]{tricco_constrained_2016}
Tricco, T.~S., Price, D.~J., \& Bate, M.~R. 2016{\natexlab{a}},
  \bibinfo{title}{Constrained hyperbolic divergence cleaning in smoothed
  particle magnetohydrodynamics with variable cleaning speeds,} Journal of
  Computational Physics, 322, 326, \dodoi{10.1016/j.jcp.2016.06.053}

% type= article
\bibitem[{T.~S. Tricco {et~al.}(2016{\natexlab{b}})Tricco, Price, \&
  Federrath}]{tricco_comparison_2016}
Tricco, T.~S., Price, D.~J., \& Federrath, C. 2016{\natexlab{b}},
  \bibinfo{title}{A comparison between grid and particle methods on the
  small-scale dynamo in magnetized supersonic turbulence,} Monthly Notices of
  the Royal Astronomical Society, 461, 1260, \dodoi{10.1093/mnras/stw1280}

% type= article
\bibitem[{Y. Tsukamoto(2026)Tsukamoto}]{tsukamoto_adjoint_2026}
Tsukamoto, Y. 2026, \bibinfo{title}{An {Adjoint} {Projection} {Formulation} for
  {Enforcing} the divergence-free {Constraint} in {Smoothed} {Particle}
  {Magnetohydrodynamics},} arXiv e-prints, arXiv:2606.28197,
  \dodoi{10.48550/arXiv.2606.28197}

% type= article
\bibitem[{G. Tóth(2000)Tóth}]{toth__2000}
Tóth, G. 2000, \bibinfo{title}{The ∇· {B}=0 {Constraint} in
  {Shock}-{Capturing} {Magnetohydrodynamics} {Codes},} Journal of Computational
  Physics, 161, 605, \dodoi{10.1006/jcph.2000.6519}

% type= misc
\bibitem[{E. van~der Velden(2021)van~der Velden}]{van_der_velden_cmasher_2021}
van~der Velden, E. 2021, {CMasher}: {Scientific} colormaps for making
  accessible, informative plots,

% type= article
\bibitem[{S. Vanaverbeke {et~al.}(2014)Vanaverbeke, Keppens, \&
  Poedts}]{vanaverbeke_gradspmhd_2014}
Vanaverbeke, S., Keppens, R., \& Poedts, S. 2014, \bibinfo{title}{{GRADSPMHD}:
  {A} parallel {MHD} code based on the {SPH} formalism,} Computer Physics
  Communications, 185, 1053, \dodoi{https://doi.org/10.1016/j.cpc.2013.11.006}

% type= article
\bibitem[{J.~W. Wadsley {et~al.}(2017)Wadsley, Keller, \&
  Quinn}]{wadsley_gasoline2_2017}
Wadsley, J.~W., Keller, B.~W., \& Quinn, T.~R. 2017, \bibinfo{title}{Gasoline2:
  a modern smoothed particle hydrodynamics code,} Monthly Notices of the Royal
  Astronomical Society, 471, 2357, \dodoi{10.1093/mnras/stx1643}

% type= article
\bibitem[{R. Wissing \& S. Shen(2020)Wissing \& Shen}]{wissing_smoothed_2020}
Wissing, R., \& Shen, S. 2020, \bibinfo{title}{Smoothed particle
  magnetohydrodynamics with the geometric density average force expression,}
  A\&A, 638, \dodoi{10.1051/0004-6361/201936739}

% type= article
\bibitem[{R. Wissing \& S. Shen(2023)Wissing \& Shen}]{wissing_numerical_2023}
Wissing, R., \& Shen, S. 2023, \bibinfo{title}{Numerical dependencies of the
  galactic dynamo in isolated galaxies with {SPH},} åp, 673, A47,
  \dodoi{10.1051/0004-6361/202244753}

% type= article
\bibitem[{R. Wissing {et~al.}(2022)Wissing, Shen, Wadsley, \&
  Quinn}]{wissing_magnetorotational_2022}
Wissing, R., Shen, S., Wadsley, J., \& Quinn, T. 2022,
  \bibinfo{title}{Magnetorotational instability with smoothed particle
  hydrodynamics,} åp, 659, A91, \dodoi{10.1051/0004-6361/202141206}

% type= article
\bibitem[{J. Wurster \& B.~T. Lewis(2020{\natexlab{a}})Wurster \&
  Lewis}]{wurster_non-ideal_2020a}
Wurster, J., \& Lewis, B.~T. 2020{\natexlab{a}}, \bibinfo{title}{Non-ideal
  magnetohydrodynamics versus turbulence - {I}. {Which} is the dominant process
  in protostellar disc formation?} Monthly Notices of the Royal Astronomical
  Society, 495, 3795, \dodoi{10.1093/mnras/staa1339}

% type= article
\bibitem[{J. Wurster \& B.~T. Lewis(2020{\natexlab{b}})Wurster \&
  Lewis}]{wurster_non-ideal_2020b}
Wurster, J., \& Lewis, B.~T. 2020{\natexlab{b}}, \bibinfo{title}{Non-ideal
  magnetohydrodynamics versus turbulence {II}: {Which} is the dominant process
  in stellar core formation?} Monthly Notices of the Royal Astronomical
  Society, 495, 3807, \dodoi{10.1093/mnras/staa1340}

% type= article
\bibitem[{J. Wurster {et~al.}(2016)Wurster, Price, \& Bate}]{wurster_can_2016}
Wurster, J., Price, D.~J., \& Bate, M.~R. 2016, \bibinfo{title}{Can non-ideal
  magnetohydrodynamics solve the magnetic braking catastrophe?} Monthly Notices
  of the Royal Astronomical Society, 457, 1037, \dodoi{10.1093/mnras/stw013}

% type= article
\bibitem[{Z. Zhu \& J.~M. Stone(2018)Zhu \& Stone}]{zhu_global_2018}
Zhu, Z., \& Stone, J.~M. 2018, \bibinfo{title}{Global {Evolution} of an
  {Accretion} {Disk} with a {Net} {Vertical} {Field}: {Coronal} {Accretion},
  {Flux} {Transport}, and {Disk} {Winds},} The Astrophysical Journal, 857, 34,
  \dodoi{10.3847/1538-4357/aaafc9}

% type= article
\bibitem[{O. Zier \& V. Springel(2022)Zier \& Springel}]{zier_simulating_2022}
Zier, O., \& Springel, V. 2022, \bibinfo{title}{Simulating the
  magnetorotational instability on a moving mesh with the shearing box
  approximation,} Monthly Notices of the Royal Astronomical Society, 517, 2639,
  \dodoi{10.1093/mnras/stac2831}

\end{thebibliography}
\bibliographystyle{aasjournalv7}

\end{document}